\documentclass[twocolumn]{aastex631}

\newcommand\hst{\textit{HST}}

\newcommand {\A}[1]{And~{\sc #1}}

\usepackage[caption=false]{subfig}
\usepackage{threeparttablex}
\usepackage{mathtools}
\usepackage{amsmath}

\begin{document}

\title{The Hubble Space Telescope Survey of M31 Satellite Galaxies IV. \\ Survey Overview and Lifetime Star Formation Histories}

\email{asavino@berkeley.edu}
\author[0000-0002-1445-4877]{Alessandro Savino}
\affiliation{Department of Astronomy, University of California, Berkeley, Berkeley, CA, 94720, USA}

\author[0000-0002-6442-6030]{Daniel R. Weisz}
\affiliation{Department of Astronomy, University of California, Berkeley, Berkeley, CA, 94720, USA}

\author{Andrew E. Dolphin}
\affiliation{Raytheon, 1151 E. Hermans Road, Tucson, AZ 85756, USA}
\affiliation{Steward Observatory, University of Arizona, 933 N. Cherry Avenue, Tucson, AZ 85719, USA}

\author[0000-0001-7531-9815]{Meredith J. Durbin}
\affiliation{Department of Astronomy, University of California, Berkeley, Berkeley, CA, 94720, USA}

\author[0000-0002-3204-1742]{Nitya Kallivayalil}
\affiliation{Department of Astronomy, University of Virginia, 530 McCormick Road, Charlottesville, VA 22904, USA}

\author[0000-0003-0603-8942]{Andrew Wetzel}
\affiliation{Department of Physics and Astronomy, University of California, Davis, CA 95616, USA}

\author{Jay Anderson}
\affiliation{Space Telescope Science Institute, 3700 San Martin Drive, Baltimore, MD 21218, USA}

\author[0000-0003-0715-2173]{Gurtina Besla}
\affiliation{Department of Astronomy, University of Arizona, 933 North Cherry Avenue, Tucson, AZ 85721, USA}

\author[0000-0002-9604-343X]{Michael Boylan-Kolchin}
\affiliation{Department of Astronomy, The University of Texas at Austin, 2515 Speedway, Stop C1400, Austin, TX 78712, USA}

\author{Thomas M. Brown}
\affiliation{Space Telescope Science Institute, 3700 San Martin Drive, Baltimore, MD 21218, USA}

\author{James S. Bullock}
\affiliation{Department of Physics and Astronomy, University of California, Irvine, CA 92697 USA}

\author[0000-0003-0303-3855]{Andrew A. Cole}
\affiliation{School of Natural Sciences, University of Tasmania, Private Bag 37, Hobart, Tasmania 7001, Australia}

\author[0000-0002-1693-3265]{Michelle L.M. Collins}
\affiliation{Physics Department, University of Surrey, Guildford GU2 7XH, UK}

\author[0000-0003-1371-6019]{M. C. Cooper}
\affiliation{Department of Physics and Astronomy, University of California, Irvine, CA 92697 USA}

\author[0000-0001-6146-2645]{Alis J. Deason}
\affiliation{Institute for Computational Cosmology, Department of Physics, Durham University, Durham DH1 3LE, UK}

\author{Aaron L. Dotter}
\affiliation{Department of Physics and Astronomy, Dartmouth College, 6127 Wilder Laboratory, Hanover, NH 03755, USA}

\author{Mark Fardal}

\affiliation{Eureka Scientific, 2452 Delmer St., Suite 100, Oakland, CA 96402, USA}

\author{Annette M. N. Ferguson}
\affiliation{Institute for Astronomy, University of Edinburgh, Royal Observatory, Blackford Hill, Edinburgh, EH9 3HJ, UK}

\author[0000-0002-3122-300X]{Tobias K. Fritz}
\affiliation{ School of Data Science, University of Virginia, 1919 Ivy Road, Charlottesville, VA 22903, USA}

\author[0000-0002-7007-9725]{Marla C. Geha}
\affiliation{Department of Astronomy, Yale University, New Haven, CT 06520, USA}

\author[0000-0003-0394-8377]{Karoline M. Gilbert}

\affiliation{Space Telescope Science Institute, 3700 San Martin Drive, Baltimore, MD 21218, USA}
\affiliation{The William H. Miller III Department of Physics \& Astronomy, Bloomberg Center for Physics and Astronomy, Johns Hopkins University, 3400 N. Charles Street, Baltimore, MD 21218}

\author{Puragra Guhathakurta}
\affiliation{UCO/Lick Observatory, Department of Astronomy \& Astrophysics, University of California Santa Cruz, 1156 High Street, Santa Cruz, California 95064, USA}

\author{Rodrigo Ibata}
\affiliation{Observatoire de Strasbourg, 11, rue de l’Universite, F-67000 Strasbourg, France}

\author[0000-0002-2191-9038]{Michael J. Irwin}
\affiliation{Institute of Astronomy, University of Cambridge, Cambridge CB3 0HA, UK}

\author{Myoungwon Jeon}
\affiliation{School of Space Research, Kyung Hee University, 1732 Deogyeong-daero, Yongin-si, Gyeonggi-do 17104, Republic of Korea}

\author[0000-0001-6196-5162]{Evan N. Kirby}
\affiliation{Department of Physics \& Astronomy, University of Notre Dame, Notre Dame, IN 46556, USA}

\author[0000-0003-3081-9319]{Geraint F. Lewis}
\affiliation{Sydney Institute for Astronomy, School of Physics, A28,
The University of Sydney, NSW 2006, Australia}

\author[0000-0002-6529-8093]{Dougal Mackey}
\affiliation{Research School of Astronomy and Astrophysics, Australian National
University, Canberra 2611, ACT, Australia}

\author{Steven R. Majewski}
\affiliation{Department of Astronomy, University of Virginia, 530 McCormick Road, Charlottesville, VA 22904, USA}

\author[0000-0002-1349-202X]{Nicolas Martin}
\affiliation{Universit\'e de Strasbourg, CNRS, Observatoire astronomique de Strasbourg, UMR 7550, F-67000 Strasbourg, France}
\affiliation{Max-Planck-Institut f\"{u}r Astronomie, K\"{o}nigstuhl 17, D-69117 Heidelberg, Germany}

\author{Alan McConnachie}
\affiliation{NRC Herzberg Astronomy and Astrophysics, 5071 West Saanich Road, Victoria, BC V9E 2E7, Canada}

\author[0000-0002-9820-1219]{Ekta~Patel}\thanks{NASA Hubble Fellow} \affiliation{Department of Physics and Astronomy, University of Utah, 115 South 1400 East, Salt Lake City, Utah 84112, USA}

\author[0000-0003-0427-8387]{R. Michael Rich}
\affiliation{Department of Physics and Astronomy, UCLA, 430 Portola Plaza, Box 951547, Los Angeles, CA 90095-1547, USA}

\author[0000-0003-0605-8732]{Evan D. Skillman}
\affiliation{University of Minnesota, Minnesota Institute for Astrophysics, School of Physics and Astronomy, 116 Church Street, S.E., Minneapolis,
MN 55455, USA}

\author[0000-0002-4733-4994]{Joshua D. Simon}
\affiliation{Observatories of the Carnegie Institution for Science, 813 Santa Barbara Street, Pasadena, CA 91101, USA}

\author[0000-0001-8368-0221]{Sangmo Tony Sohn}
\affiliation{Space Telescope Science Institute, 3700 San Martin Drive, Baltimore, MD 21218, USA}
\affiliation{Dept. of Astronomy \& Space Science, Kyung Hee University, Gyeonggi-do 17104, Republic of Korea}

\author{Erik J. Tollerud}
\affiliation{Space Telescope Science Institute, 3700 San Martin Drive, Baltimore, MD 21218, USA}

\author[0000-0001-7827-7825]{Roeland P. van der Marel}
\affiliation{Space Telescope Science Institute, 3700 San Martin Drive, Baltimore, MD 21218, USA}
\affiliation{Center for Astrophysical Sciences,
  The William H. Miller III Department of Physics \& Astronomy,
  Johns Hopkins University, Baltimore, MD 21218, USA}

\vspace{5mm}
\begin{abstract}
From $>1000$ orbits of HST imaging, we present deep homogeneous resolved star color-magnitude diagrams that reach the oldest main sequence turnoff and uniformly measured star formation histories (SFHs) of 36 dwarf galaxies ($-6 \ge M_V \ge -17$) associated with the M31 halo, and for 10 additional fields in M31, M33, and the Giant Stellar Stream.
From our SFHs we find: i) the median stellar age and quenching epoch of M31 satellites correlate with galaxy luminosity and galactocentric distance. Satellite luminosity and present-day distance from M31 predict the satellite quenching epoch to  within $1.8$~Gyr at all epochs. This tight relationship highlights the fundamental connection between satellite halo mass, environmental history, and star formation duration. ii) There is no difference between the median SFH of galaxies on and off the great plane of Andromeda satellites. iii) $\sim50$\% of our M31 satellites show prominent ancient star formation ($>12$~Gyr ago) followed by delayed quenching ($8-10$~Gyr ago), which is not commonly observed among the MW satellites. iv) A comparison with 
TNG50 and FIRE-2 simulated satellite dwarfs around M31-like hosts show that some of these trends (dependence of SFH on satellite luminosity) are reproduced in the simulations while others (dependence of SFH on galactocentric distance, presence of the delayed-quenching population) are weaker or absent. We provide all photometric catalogs and SFHs as High-Level Science Products on MAST.

\end{abstract}



\section{Introduction} \label{sec:intro}

Our knowledge of low-mass galaxy formation has long been anchored by Milky Way (MW) satellite galaxies.  Their demographics (i.e. numbers, luminosities, spatial configuration), kinematics, stellar abundance patterns, orbital histories, and star formation histories (SFHs) provide stringent tests of Cold Dark Matter (CDM), the physics of galaxy formation on the smallest scales, and the relationship between cosmic reionization and low-mass galaxies \citep[e.g.,][]{Mateo98, Simon07, Tolstoy09, Kirby13, Brown14, Weisz14a, Frebel15, Bullock17, Simon19, Patel20, Sacchi21, Sales22}.

The progenitors of present-day MW satellite galaxies likely occupied a broadly representative volume in the very early Universe ($z\sim7$), but now the MW satellites reside in a highly biased region \citep[e.g.,][]{Boylan-Kolchin16}.  It remains unclear if the insights learned from MW satellites, and their particular formation pathways, are applicable to other satellite systems and low-mass galaxies in general \citep[e.g.,][]{Gandhi24}.   

On the one hand, some present-day properties (e.g. luminosity function, quenched fraction) of the brightest MW satellites are similar to other satellite systems around MW-mass hosts in the Local Volume \citep[e.g.,][]{Chiboucas13, Carlsten22, Mao21, Danieli23, Mao24, Geha24}.  On the other hand, there are well-documented differences in the properties (e.g. numbers, morphologies, spatial configurations, sizes, and SFHs) of satellites of the MW compared to those of M31, the closest $L_{\star}$ galaxy and the only other satellite system which can be observed to a comparable level of detail and faintness as the MW \citep[e.g.,][]{McConnachie06a, Brasseur11, Tollerud12, Monachesi12, Conn13, Ibata13, Geha15, Martin16, Geha17, Skillman17, D'Souza18, Greco18, Muller18, Pawlowski18, Smercina18, Pawlowski19, DolivaDolinsky23,Savino22,Savino23}.  Whether these differences are the result of variance in the intrinsic nature of low-mass galaxy formation, due to the specific accretion histories of the MW and M31 \citep[e.g.,][]{D'Souza21, Engler21}, or some other factor(s) is an open question that can only be resolved through systematic studies of low-mass satellite systems.

M31 and its satellites are an important frontier for exploring the formation and evolution of low-mass satellites over cosmic time.  With characteristic distances of $\sim800$~kpc \citep[e.g.,][]{deGrijs14,Savino22}, it is possible to resolve M31 and its satellites into individual stars with our most powerful ground- and space-based facilities.  Resolved star studies of the M31 satellites enable measurements of SFHs, stellar abundance patterns, and proper motions to a level of precision comparable to MW satellites \citep[e.g.,][]{Williams09,Hidalgo11, Barker11, Ho12, Monachesi12,Tollerud12,Ho15, Collins13, Vargas14, Weisz14c, Geha15,  Monelli16, Skillman17, VanderMarel19,Kirby20, McQuinn23, Sohn20, Savino23,Fu24b}. 

For the foreseeable future (e.g. until large-aperture UV/optical space telescopes, such as the Habitable Worlds Observatory, are deployed), M31 is the only satellite system that can be studied in comparable detail and completeness as the MW satellites. The next closest satellite systems around $L_{\star}$ hosts have distances of $\sim$3-4~Mpc.  At such distances, measuring SFHs from the oldest main sequence turnoff (oMSTO; $m_I\gtrsim31$) is prohibitive (crowding, faintness) even with JWST \citep[e.g.,][]{Savino24}, abundances of the red giant branch (RGB) stars are beyond even our most powerful spectrographs currently in operation (e.g., Keck, VLT)\footnote{At such distances stellar abundances of the brightest resolved RGB stars may be tractable with ELTs and JWST spectroscopy \citep[e.g.,][]{Sandford20}.}, and proper motions will require multi-decade time baselines.  

The enormous scientific potential for resolved star studies of the M31 satellites has been recognized for decades.  Since the discovery of the first faint M31 satellites by \citet{VandenBergh72, VandenBergh74}, there have been increasingly dedicated efforts to identify and characterize sub-structures (i.e., satellites and streams) around M31 through imaging \citep[e.g.,][]{Ibata01,Ferguson02, McConnachie03, Zucker04a, Zucker04b, Martin06,McConnachie06a, McConnachie06b, Ibata07,Majewski07,Zucker07, Irwin08, McConnachie08, Martin09, McConnachie09, Bell11, Richardson11,Slater11,Bernard15a,  Martin16, DolivaDolinsky22, DolivaDolinsky23}.  In particular, the Pan-Andromeda Archaeological Survey (PAndAS) has been foundational for our knowledge of the M31 system.  Using the CFHT, PAndAS undertook a deep, wide-area imaging survey out to a projected distance of 150~kpc from M31, resulting in the discovery of 16 M31 satellites as faint as $M_V\sim-6$, along with prominent streams, shells, globular clusters, and other stellar sub-structures (see \citealt{McConnachie18} and references therein).  The discovery of new M31 satellites continues apace with fainter, lower surface brightness, and/or more distant M31 satellites recently being reported \citep[e.g.,][]{Martin13a,Martin13b,Collins22a, Martinez-Delgado22, McQuinn23, Collins24}. 

There has been substantial effort to characterize the stellar populations and kinematics of the M31 satellites.  First, the Keck telescope has played a central role in spectroscopic studies of the entire M31 system.  With its excellent sensitivity and moderate resolution, large investments of Keck/DEIMOS time have provided velocities for thousands of stars throughout the M31 systems, along with chemical abundances for a sizable fraction of these stars \citep[e.g.,][]{Majewski07,Kalirai09,Kalirai10,Collins10,Tollerud12,Howley13,Vargas14,Ho15,Gilbert19,Kirby20,Wojno20,Escala21}.  These incredible spectroscopic datasets have provided dark matter halo mass estimates and radial velocities for most known M31 satellites \citep[e.g.,][]{Tollerud12,Collins13}, along with metallicities and detailed abundance patterns for the brighter systems \citep{Vargas14,Ho15,Kirby20,Wojno20}.  Fainter M31 satellites are not as well characterized spectroscopically, primarily owing to the paucity of RGB stars in each system.  More dedicated spectroscopy is required and/or alternative approaches to acquiring stellar chemistry are required for these faint systems \citep[e.g.,][]{Fu22,Fu24b}.

Second, the exquisite sensitivity and high angular resolution of HST has been required for measuring SFHs and orbital histories.  On the SFH side, only HST has had the ability to construct color-magnitude diagrams (CMDs) that reach the oldest main sequence turnoff (oMSTO).  Initially, shallower HST data revealed the presence of asymptotic giant branch stars (AGB) and red horizontal branches in many M31 satellites, suggestive of intermediate age star formation in many of these systems \citep[e.g.,][]{Dacosta96,Dacosta00,Dacosta02,Martin17}.  Given the significant integration times involved in reaching the oMSTO at $\sim800$~kpc, the measurement of `gold standard' SFHs was a slow process with an early focus on M31, M33, and the bright elliptical systems not found around the MW (M32, NGC 147, and NGC 185; \citealt{Brown06, Brown08, Williams09, Barker11, Bernard12,  Monachesi12,Bernard15a, Bernard15b, Geha15}) and an expansion to other, fainter systems in the mid-2010s.  

The first SFHs of faint M31 satellites based on MSTO-depth CMDs were from the Initial Star formation and Lifetimes of Andromeda Satellites (ISLAndS) program \citep{Weisz14c, Monelli16, Skillman17}.  This $\sim$100 orbit HST program surveyed six M31 satellites that spanned a factor of $\sim$100 in luminosity.  Perhaps the most remarkable finding from ISLAndS was the high degree of similarity in the SFHs of all six systems.  All formed $\sim$50\% of their stellar mass $\gtrsim12$~Gyr ago, while the remainder formed at intermediate ages, followed by nearly simultaneous quenching $\sim5-6$~Gyr ago.  These trends were qualitatively different from those of MW satellites with similar properties (e.g., comparable luminosity, distance from host galaxy).  However, the limited statistical power of six galaxies prohibited broader conclusions.  A shallower HST survey of the remainder of the M31 satellites \citep{Martin17} hinted at similar trends across the M31 system, but the limited depth of the imaging resulted in large systematic uncertainties on the formal SFH measurements \citep{Weisz19}.

HST imaging of M31 satellites that reaches the oMSTO is also well-suited for proper motion measurements.  The astrometric stability, sensitivity, and angular resolution (for high fidelity point spread function characterization and star-galaxy differentiation) of HST (and now JWST) is uniquely suited for these measurements.  However, to date, only M31, NGC~147, NGC~185, \A{III}, and \A{VII} have proper motions based on HST imaging \citep{Sohn12,Sohn20,Warfield23,Casetti-Dinescu24}; proper motions based on other observations (e.g., Gaia, VLBI) have been published for a handful of galaxies associated with the M31 system \citep{Brunthaler05,Brunthaler07,VanderMarel19,McConnachie21,Battaglia22,Bennet24}. Several HST and JWST programs are targeting the ISLAndS galaxies for PM measurements.  As discussed at several points in this paper, a complete census of orbital histories for the M31 satellites is essential for unraveling the likely complex formation of the M31 system as we see it today.

The HST Survey of M31 Satellites Treasury program is a Cycle 27 program (GO-15902, PI:Weisz) that was designed to acquire oMSTO imaging for all known M31 satellites that lacked such data.  The primary goals of the program are to (a) uniformly measure SFHs of the entire M31 satellite system; (b) determine homogenous RR Lyrae-based distances to all known M31 satellites; (c) provide first epoch proper motion imaging for the faint M31 satellites; and (d) provide the community with a uniformly reduced HST legacy photometric dataset for all new and archival observations of the M31 satellites.  

To date, this program has produced several significant results.  First, \citet{Savino22} uniformly measured RR Lyrae distances to nearly all M31 satellites and anchored them to the Gaia distance scale.  This work resulted in a revision to the 3D geometry of the M31 satellite systems, affirmed the anisotropic spatial distribution of the satellites, and added new complexity into questions about the putative plane of M31 satellites.  Second, \citet{Savino23} presented SFHs of six ultra-faint dwarfs (UFDs) orbiting M31, the first oMSTO SFH of UFDs outside the MW halo.  They show that while most systems are predominantly ancient, similar to MW UFDs, they tend to have longer periods of low-level star formation post-reionization and later quenching epochs.  \citet{Savino23} also present the SFH of And~XIII, which formed a remarkable 75\% of its stellar mass in a rapid burst $\sim10$~Gyr ago.

In this paper, we present an overview of our HST Survey of M31 Satellites, and describe and release the photometric data from $>1000$ orbits of new and archival HST imaging. We then present the lifetime SFHs of all 36 satellites galaxies studied as part of this program along with redeterminations of SFHs for M31, M33, and the Giant Stellar Stream (GSS). The high-level science products (HLSPs) produced by this program are hosted on MAST and can be accessed here: \url{ https://archive.stsci.edu/hlsp/m31-satellites}. We provide a detailed description of the data products and their content in Appendix~\ref{App:Products}.

This paper is organized as follows. We describe the observational strategy in \S \ref{sec:obs}. We detail our photometry and artificial star tests (ASTs), and present the CMDs of all galaxies in \S \ref{Sec:Reduction}.  We summarize our approach to SFH measurements in \S \ref{sec:cmd_model}.  We present the SFHs, and undertake a variety of science analyses with them, in \S \ref{sec:sfhs}.  We present our conclusions in \S \ref{sec:conclusions}.

\begin{table*}\centering
\caption{Observation summary for the 36 dwarf galaxies in our sample. We report RA and Dec of the primary ACS field, the total exposure time across all frames, the filter combination of the dataset, the approximate magnitudes (apparent and absolute) of 50\% completeness, the fraction of target stellar light contained in the primary ACS field ($f_{\star}$), and the program IDs the data belong to. }
\hspace*{-2cm}\begin{threeparttable}


\begin{tabular}{lrrccccll}


\toprule
Galaxy ID\tnote{a} &RA&Dec&$t_{exp}$& Filters  &$m_{50\%}$ &$M_{50\%}$&$f_{\star}$\tnote{b}& \hst\ Proposal ID \\
&deg&deg&s&&mag&mag&&\\
\toprule

         \underline{M32}&10.73300& 40.84732&59,036&F606W, F814W&26.55, 25.95&1.94, 1.40&0.05&9392, 15658\\

         \underline{NGC147}&8.21250& 48.37778&98,464&F606W, F814W&28.85, 27.90&4.10, 3.29&0.02&10794, 11724, 14769\\

         \underline{NGC185}&9.79458& 48.44383&78,782&F606W, F814W&29.10, 28.30&4.48, 3.87&0.02&10794, 11724, 14769\\

         NGC205&10.07441& 41.74730&21,958&F606W, F814W&26.45, 25.85&1.76, 1.18&0.04&\textbf{15902}\\

         \underline{And~{\sc I}}&11.42833& 38.03967&51,964&F475W, F814W&29.05, 28.05&4.37, 3.49&0.19&13739\\

         \underline{And~{\sc II}}&19.09918& 33.43487&40,268&F475W, F814W&29.05, 27.90&4.81, 3.72&0.12&13028\\

         \underline{And~{\sc III}}&8.87779& 36.50394&51,964&F475W, F814W&29.00, 27.90&4.49, 3.50&0.43&13739\\

         And~{\sc V}&17.57237& 47.63018&21,989&F606W, F814W&28.50, 27.65&3.79, 3.04&0.58&\textbf{15902}\\

         And~{\sc VI}&357.95009& 24.59728&22,070&F606W, F814W&28.65, 27.90&3.83, 3.15&0.41&\textbf{15902}\\

         And~{\sc VII}&351.62917& 50.69194&24,425&F606W, F814W&28.65, 27.90&3.68, 3.06&0.23&\textbf{15902}\\

         And~{\sc IX}&13.22000& 43.19972&24,362&F606W, F814W&28.65, 27.85&4.17, 3.45&0.47&13699, \textbf{15902}\\

         And~{\sc X}&16.64625& 44.80861&19,907&F606W, F814W&28.55, 27.65&4.16, 3.39&0.77&13699, \textbf{15902}\\

         And~{\sc XI}&11.58520& 33.79880&22,065&F606W, F814W&28.40, 27.60&3.77, 3.05&0.96 (0.85)&\textbf{15902}\\

         And~{\sc XII}&11.86020& 34.37750&21,896&F606W, F814W&28.65, 27.85&3.89, 3.26&0.46&\textbf{15902}\\

         And~{\sc XIII}&12.96030& 33.00240&17,595&F606W, F814W&28.40, 27.70&3.43, 2.87&0.86 (0.82)&\textbf{15902}\\

         And~{\sc XIV}&12.89722& 29.66901&22,065&F606W, F814W&28.45, 27.85&3.76, 3.24&0.50&\textbf{15902}\\

         \underline{And~{\sc XV}}&18.57792& 38.11750&40,216&F475W, F814W&28.95, 27.85&4.39, 3.39&0.69&13739\\

         \underline{And~{\sc XVI}}&14.88474& 32.39414&30,816&F475W, F814W&28.85, 27.80&5.01, 4.10&0.67 (0.64)&13028\\

         And~{\sc XVII}&9.27554& 44.32576&37,564&F606W, F814W&28.60, 28.70&4.00, 4.17&0.63&13699, \textbf{15902}\\

         \underline{And~{\sc XIX}}&4.89725& 35.07382&40,505&F606W, F814W&28.95, 28.25&4.26, 3.61&0.02&15302\\

         And~{\sc XX}&1.86976& 35.13206&20,096&F606W, F814W&28.30, 27.65&3.73, 3.15&0.99 (0.85)&13699, \textbf{15902}\\

         And~{\sc XXI}&358.67062& 42.48853&28,777&F606W, F814W&28.80, 28.15&4.05, 3.51&0.15&13699, \textbf{15902}\\

         And~{\sc XXII}&21.91871& 28.08953&35,454&F606W, F814W&28.40, 27.75&3.73, 3.17&0.80 (0.76)&13699, \textbf{15902}\\

         And~{\sc XXIII}&22.33708& 38.72444&24,321&F606W, F814W&28.45, 27.75&3.95, 3.30&0.12&13699, \textbf{15902}\\

         And~{\sc XXIV}&19.61540& 46.37167&17,875&F606W, F814W&28.35, 27.55&4.15, 3.44&0.32&13699, \textbf{15902}\\

         And~{\sc XXV}&7.56628& 46.87198&26,746&F606W, F814W&28.65, 27.85&3.99, 3.28&0.28&13699, \textbf{15902}\\

         And~{\sc XXVI}&5.93973& 47.91706&31,096&F606W, F814W&28.55, 27.60&3.82, 2.95&0.80 (0.78)&13699, \textbf{15902}\\

         \underline{And~{\sc XXVIII}}&338.17167& 31.21617&47,240&F475W, F814W&28.90, 27.85&4.13, 3.29&0.76&13739\\

         And~{\sc XXIX}&359.73156& 30.75567&26,184&F606W, F814W&28.35, 27.55&4.00, 3.23&0.54&13699, \textbf{15902}\\

         And~{\sc XXX}&9.14417& 49.64639&19,976&F606W, F814W&28.35, 28.55&4.27, 4.59&0.60&13699, \textbf{15902}\\

         And~{\sc XXXI}&344.54982& 41.34282&28,834&F606W, F814W&28.65, 27.95&3.95, 3.37&0.08&13699, \textbf{15902}\\

         And~{\sc XXXII}&8.98667& 51.61403&33,400&F606W, F814W&28.65, 27.90&3.60, 3.03&0.04&\textbf{15902}\\

         And~{\sc XXXIII}&45.34833& 40.98833&24,364&F606W, F814W&28.50, 27.90&3.95, 3.46&0.54&13699, \textbf{15902}\\

         \underline{Psc {\sc I}}&15.96112& 21.88757&58,416&F475W, F814W&29.15, 28.70&5.09, 4.72&0.38&10505, 13738\\

         \underline{Peg DIG}&352.15417& 14.73444&71,670&F475W, F814W&28.45, 27.75&3.48, 2.90&0.42&13768\\
         
         \underline{IC~1613}&16.11750& 2.16014&58,608&F475W, F814W&29.40, 28.40&5.00, 4.04&0.03&10505\\
         \toprule

\end{tabular}
\begin{tablenotes}
    \item [a] Galaxies for which the deep photometry is from the archive are underlined.
    \item [b] If $f_{\star}$ changes after the $2 r_h$ spatial cut (\S~\ref{Sec:SpatialCut}), the new value is listed in parentheses.
\end{tablenotes}

\label{Tab:Sample}
\end{threeparttable}

\end{table*}


\section{Observations}
\label{sec:obs}
\subsection{Sample Selection}
Our program was designed to target any known dwarf galaxy that, as of early 2019, fulfilled the following criteria: i) it is not in an advanced state of disruption (e.g., \A{XXVII}, \citealt{Collins13,Preston19,Preston24}); ii) it is located within $\sim500$~kpc (deprojected) of M31; iii) it does not already have deep HST imaging from other programs, though we do include those archival observations in our analysis (see \S~\ref{Sec:Archival}). This selection resulted in a sample of 24 targets, from which we excluded the galaxy IC~10, as its high line-of-sight extinction ($A_V = 4.5$, \citealt{Green19}) makes its oMSTO unreachable with HST. As of writing, there are a handful of additional galaxies, such as Peg~W, Tri~III, and \A{XXXIV}, that are known to be associated with the M31 system \citep[Smith et al. submitted]{Collins22b,Martinez-Delgado22,McQuinn23,Collins24}. However, these systems were discovered after our program was executed and were therefore not targeted by our survey. The final sample of 23 galaxies observed by our program, along with a summary of the observations, is listed in Table~\ref{Tab:Sample}.
\subsection{General Observing Strategy}
\label{sec:obs_strategy}

For all new observations, we used the F606W and F814W filters on ACS/WFC and WFC3/UVIS.  This filter combination has a higher throughput, but reduced sensitivity to metallicity and effective temperature, compared to the F475W and F814W filters used by the LCID and ISLAndS programs.  We designed our program assuming we would use informative priors on the age-metallicity relationship (AMR), which have been commonly used in SFH measurements of galaxies with CMDs of all depths, including those that include the oMSTO \citep[e.g.,][]{Weisz11,Weisz14a,McQuinn15,Skillman17,Albers19,Mcquinn24a}. These priors mitigate some of the reduced sensitivity to AMRs introduced by our filter choice, and, even in the case of the F475W and F814W filters, they help to reduce the effects of including poorly modeled foreground/background contamination in the SFH measurement. 

Motivated by the observing strategy of \citet{Cole14}, our observations are designed to reach the subgiant branch (SGB; $M_{V}+3.2$, $M_I \sim +2.7$) of an ancient (12 Gyr), metal-poor ($\sim$[Fe/H] = -1.5) population with a SNR$\sim10$ in both filters.  This provides a slightly lower SNR ($\sim8$) for the corresponding MSTO. Our simulations showed it to be sufficient for accurate SFH recovery, consistent with \citet{Cole14}. Simultaneously, we optimized the observational cadence for the detection of RR Lyrae in order to measure high-fidelity distances as discussed in \citet{Savino22}.  Finally, the number of exposures and dithers used, as detailed in the public Phase II file, satisfied the astrometric requirements for first epoch proper motions.

We used ACS/WFC as our primary instrument and placed it near the center of each galaxy.  WFC3/UVIS operated in parallel and sampled the outskirts, either providing a useful sampling of local background for modeling the CMDs of faint, compact galaxies or providing a second deep CMD for population gradient studies.  We discuss the field placement more in \S \ref{sec:primary} and \S \ref{sec:parallel}.

Because of the large nature of the program, we were required to follow the SCHED100 orbit duration protocol, which reduces the length of each orbit to accommodate HST scheduling pressure.  Including this time constraint, we found that 244 prime and 244 parallel orbits would achieve our depth and cadence requirements.  We note that adopting F475W instead of F606W, for the same target depth, would have increased the total exposure time, and number of orbits, required to complete our survey by a factor of $\sim3$.

\subsection{Archival Dwarf Galaxies Data}
\label{Sec:Archival}
There are 13 dwarf galaxies within approximately 500~kpc of M31 that had HST imaging that reaches the oMSTO.  We did not acquire new data for these systems and instead, we homogeneously re-analyze these data and include it in our final sample. With the exception of IC~10 and of very recent discoveries noted above, which do not possess deep HST photometry, the combination of our observations and the archival data comprises the entire known satellite galaxy system of M31. Here, we describe the archival data, which is also summarized in Table~\ref{Tab:Sample}.

For \A{I}, \A{II}, \A{III}, \A{XV}, \A{XVI}, and \A{XXVIII} we use ACS/WFC3 and WFC3/UVIS imaging from the ISLAndS program \citep[GO-13038, GO-13739, PI: Skillman;][]{Skillman17}. For \A{XIX}, we use the ACS/WFC imaging from GO-15302 (PI: Collins, \citealt{Collins22b}). For NGC~147 and NGC~185, we use ACS/WFC and WFC3/UVIS imaging from GO-10794 and GO-11724 (PI: Geha, \citealt{Geha15}), and additional observations from GO-14769 (PI: Sohn, \citealt{Sohn20}). For M32 we use ACS/WFC imaging from GO-9392 (PI: Mateo) and GO-15658 (PI: Sohn). For IC~1613 and Psc~{\sc I} (also known as LGS3) we use ACS/WFC imaging from the LCID program (GO-1505, PI: Gallart, \citealt{Hidalgo11,Skillman14,Gallart15}) and from GO-13738 (PI: Shaya). Finally, for the Pegasus dwarf irregular galaxy we use ACS/WFC imaging from GO-13768 (PI: Weisz, \citealt{Cole17}).

Additionally, 15 galaxies in our sample have shallow ACS/WFC imaging from GO-13699 (PI: Martin, \citealt{Martin17}). Although these observations only consist of one orbit of data per galaxy, all but one of these fields are spatially coincident with our deep ACS/WFC fields, and use the same filters, so we include those data in our reduction.

\begin{figure*}
        \centering
    \includegraphics[width=0.9\textwidth]{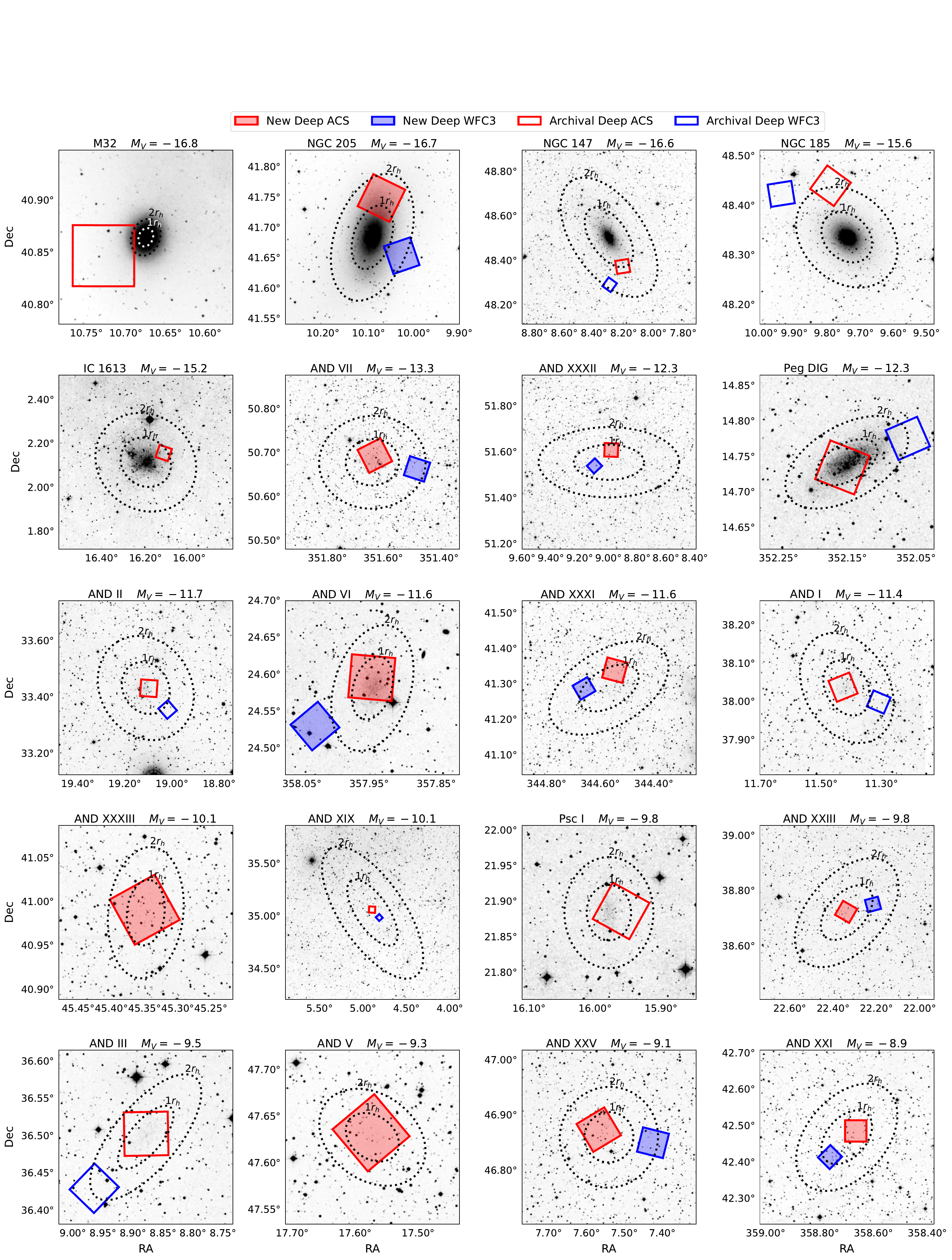}
    \caption{The HST pointings for the 36 dwarf galaxies in our sample, overlaid on DSS archival imaging. The galaxies are ordered by luminosity. The ACS/WFC and WFC3/UVIS fields are shown in red and blue, respectively. The new deep imaging acquired by our program is illustrated as a filled footprint, while archival deep fields are shown as empty contours. We also report the $1 r_h$ and $2 r_h$ ellipses (dotted lines).}
    \label{Fig:Footprint}
\end{figure*}

\begin{figure*}
        \centering
        \ContinuedFloat
    \includegraphics[width=0.9\textwidth]{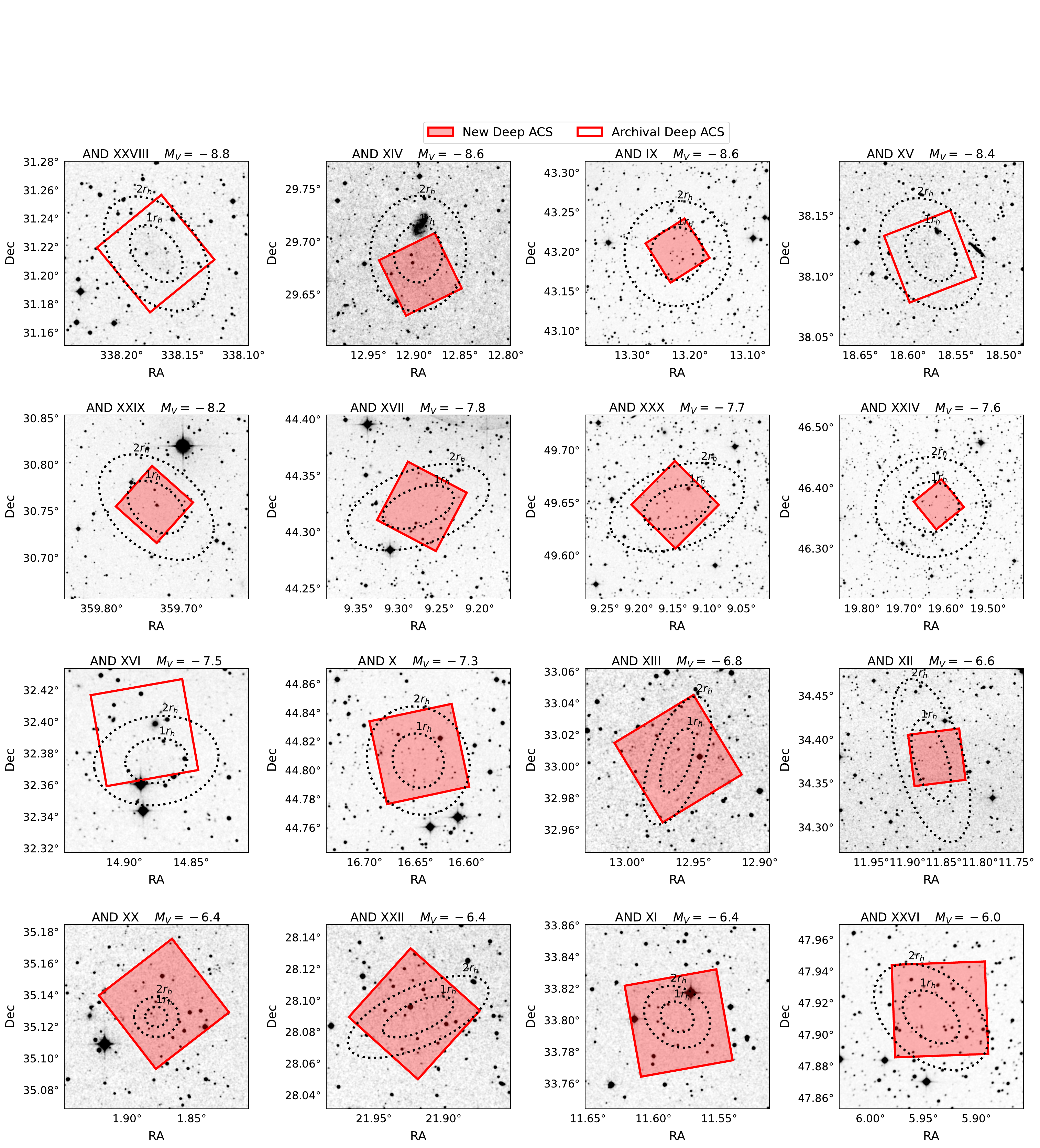}
    \caption{Continued.}

\end{figure*}

\subsection{Primary ACS Fields}
\label{sec:primary}
Figure~\ref{Fig:Footprint} shows the placement of our primary ACS photometric fields on the 36 target dwarf galaxies in this paper (which we refer to as our primary sample). Most of the primary ACS fields observed by our program are placed to cover the photometric center of the target galaxy. The exceptions are NGC~205, \A{XXXI}, and \A{XXXII}, for which the field is placed off-center, due to crowding concerns. Additionally, the archival ACS fields of NGC~147, NGC~185, M32, and IC~1613 are also offset from the galaxy center, to avoid the highly crowded inner regions. 

Due to the different apparent sizes of our targets and to the different field
placements, the primary ACS pointings sample different fractions of the galaxy total stellar population. For each target, we have therefore calculated the fraction of total stellar light contained in our ACS field. We do this by integrating a 2D exponential surface density profile (e.g., as described in \citealt{Martin16}) over our ACS footprint. The only exceptions are NGC~147, NGC~185, NGC~205, and M32, for which we use a S\'ersic profile \citep{Sersic63}. We construct our profile using structural parameters (i.e., half-light radius, ellipticity, position angle and, where relevant, S\'ersic index) from the literature \citep{Lee95,Choi02,McConnachie06b,Bell11,Slater11,McConnachie12,Martin13b,Martin13c,Crnojevic14,Martin16}. We refer to \citet{Savino22} for further details.

The fraction of the total stellar light sampled by the ACS fields, which we call $f_{\star}$, is listed in Table~\ref{Tab:Sample}. This varies from 99\% for our most compact galaxy (\A{XX}) to 2\% for extended galaxies such as NGC~147, NGC~185 or \A{XIX}.

\begin{figure*}
    \centering
    \plottwo{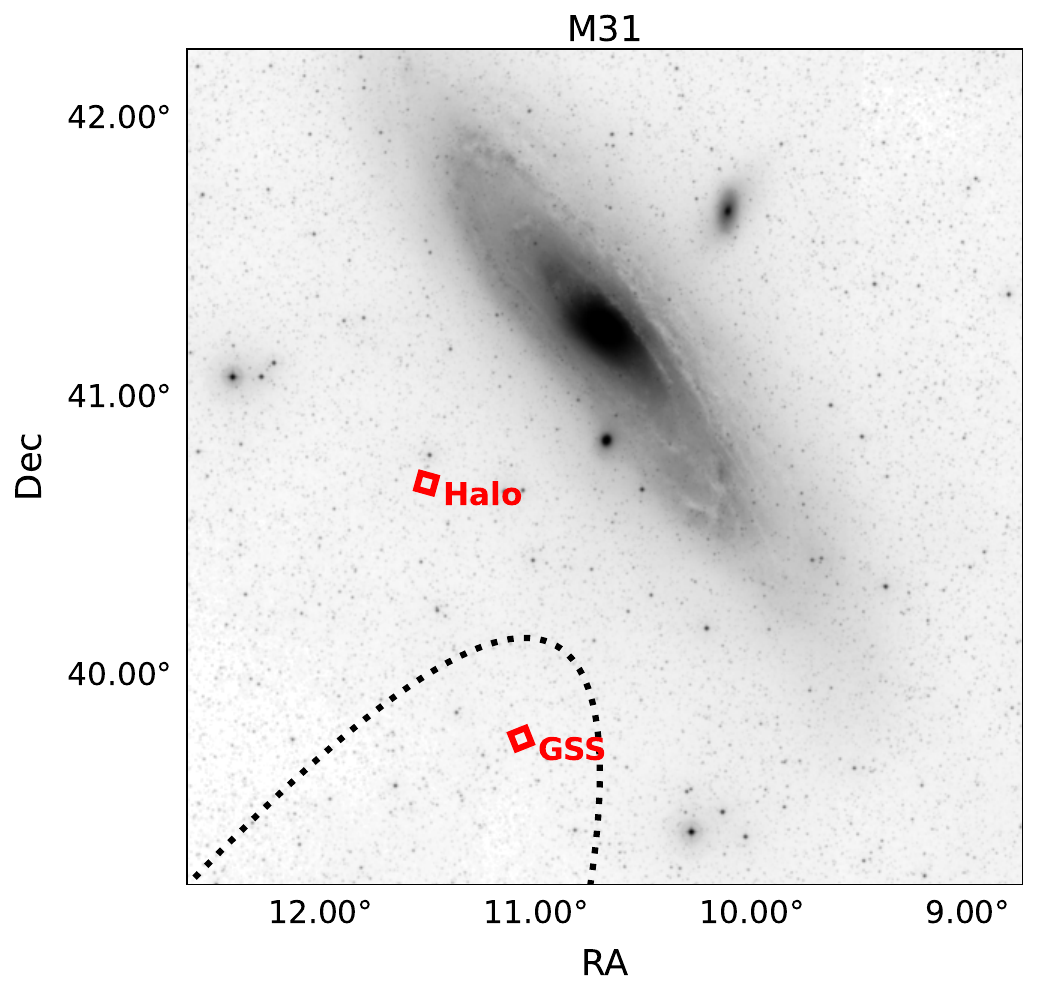}{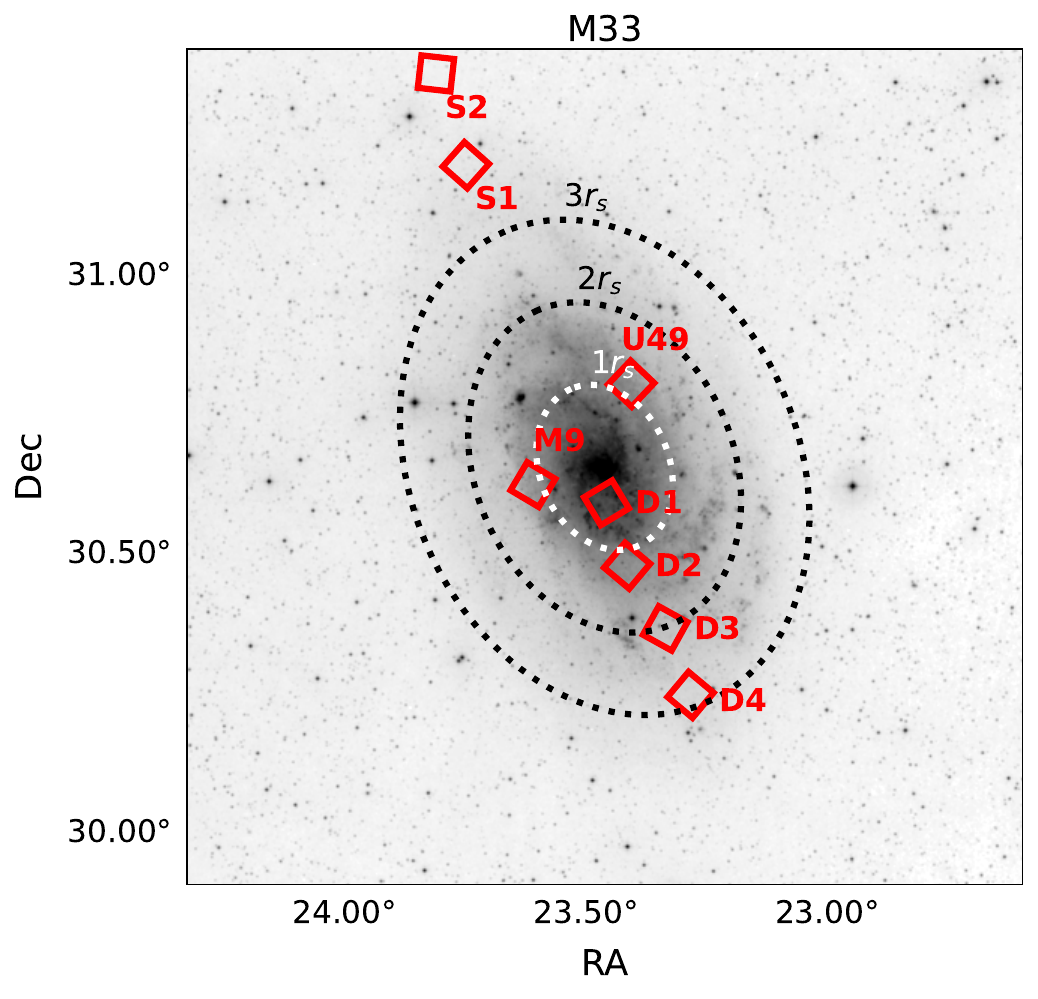}
    \caption{The archival ACS/WFC pointings we use for M31 (left) and M33 (right), overlaid on DSS archival imaging.  For M31, the approximate outline of the region in which the GSS dominates the stellar density is also outlined (dotted line). For M33, ellipses corresponding to 1, 2, and 3 disc scale lengths are shown (dotted lines).
}
    \label{Fig:M31_Footprint}
\end{figure*}
\subsection{Parallel UVIS Fields}
\label{sec:parallel}
All the new observations of our program include parallel WFC3/UVIS fields, mirroring the same F606W/F814W exposures of the primary ACS field. In addition, most of the dwarf galaxies with archival data have a deep parallel UVIS field (the exceptions being M32, IC 1613, and Psc I). However, due to the different apparent sizes of our sample galaxies, the parallel fields capture a variable amount of the targets' stellar light, ranging from very populated fields in our large and luminous targets (such as NGC~205) to fields showing few, if any, galaxy member stars in our most compact galaxies (such as \A{XX}).

Figure~\ref{Fig:Footprint} shows the placement of the parallel UVIS field in each target for which the images contain a meaningful amount of member stars (see \S~\ref{Sec:CMDs}). Overall, we detect the target stellar population in the parallel field of 15 of the dwarf galaxies in our sample. In addition, the parallel field of \A{IX} also contains a significant stellar population but this entirely belongs to the M31 halo, as discussed in \citet{Skillman17}. 

While in this paper we present the observations and the photometry from the UVIS parallels, the analysis of the stellar populations in these fields, and a discussion of stellar population gradients in our target galaxies is deferred to a future paper (Garling et al, in prep.).

\begin{table*}
\centering
\caption{Observation summary for 10 auxiliary fields in M31 and M33. We report RA and Dec of the ACS field, the total exposure time across all frames, the filter combination of the dataset, the approximate magnitudes (apparent and absolute) of 50\% completeness, and the program IDs the data belong to. }
\begin{tabular}{lllccccl}
\toprule
Galaxy ID &RA&Dec&$t_{exp}$& Filters  &$m_{50\%}$&$M_{50\%}$& \hst\ Proposal ID \\

&d&d&s&&mag&&\\
\toprule
         M31 (Halo)&11.53285& 40.70991&141,305&F606W, F814W&30.00, 29.10&5.33, 4.50&9453\\
         M31 (GSS)&11.07544& 39.79409&130,880&F606W, F814W&29.85, 29.25&5.13, 4.58&10265\\
         M33 (D1)&23.45785& 30.59670&13,004&F606W, F814W&25.75, 25.05&0.97, 0.31&10190\\
         M33 (D2)&23.41542& 30.48333&47,680&F606W, F814W&26.85, 26.25&2.07, 1.51&10190\\
         M33 (D3)&23.33542& 30.37083&47,680&F606W, F814W&27.85, 27.15&3.07, 2.41&10190\\
         M33 (D4)&23.28292& 30.25194&47,680&F606W, F814W&28.65, 27.95&3.87, 3.21&10190\\
         M33 (M9)&23.62632& 30.63687&31,242&F606W, F814W&26.95, 26.25&2.17, 1.51&9873\\
         M33 (U49)&23.41855& 30.79969&31,242&F606W, F814W&26.95, 26.25&2.17, 1.51&9873\\
         M33 (S1)&23.75000& 31.20000&31,005&F606W, F814W&29.05, 28.55&4.27, 3.81&9837\\
         M33 (S2)&23.81250& 31.36667&31,005&F606W, F814W&28.90, 28.40&4.12, 3.66&9837\\
         \toprule

\end{tabular}

\label{Tab:M31}
\end{table*}

\subsection{Archival M31/M33 Data}
We complement our dwarf galaxy data by reducing and analyzing selected deep archival HST observations of the two large spiral galaxies in the M31 system, namely M33 and M31 itself, which we refer to as the auxiliary sample.  Our motivation to include them is to provide homogeneous photometry and SFHs of these important central galaxies alongside the satellite population.  The spatial position of the M33 and M31 fields is shown in Figure~\ref{Fig:M31_Footprint} and the observations are summarized in Table~\ref{Tab:M31}.

There are several deep HST fields targeting different regions of the M33 galaxy \citep[e.g.,][]{Barker07a,Williams09,Barker11, Bernard12}. We reduced and analyzed ACS/WFC imaging from eight pointings observed as part of GO-9837 (PI: Ferguson), GO-9873 (PI: Sarajedini) and GO-10190 (PI: Garnett). These observations sample the M33 disc at different radii, from 0.5 to 5 disc scale lengths from the center. Due to the large size and complex morphology of M33, jointly analyzing these data is the best way we have to obtain a global SFH that is more representative of M33 as a whole.

For M31, we choose to analyze two deep ACS fields in the stellar halo. As stellar halos are understood to carry the fossil imprint of past accretion events \citep[e.g.,][]{Bullock05,Belokurov06a,Belokurov18,Helmi18,Naidu20a,Deason24}, these data will provide an instructive comparative benchmark to the properties of the satellites. 

The first field, observed by GO-9453 (PI: Brown), targets the inner smooth halo of M31, at a projected distance of 11~kpc from its center. We only analyze roughly half of the 250 images acquired by ACS on this field. Even with this reduced dataset, this is the deepest photometric field in our sample. 

The other field, observed by GO-10265 (PI: Brown), targets the GSS, which is the most prominent tidal feature of the M31 halo, hosts a stellar population that is representative of most of the inner halo substructures \citep[e.g.,][]{Bernard15a}, and is one of the main manifestations of M31's recent accretion history \citep[e.g.,][]{McConnachie18}. As a caveat, we stress that both M31's smooth halo and the GSS have very large apparent sizes and are known to have significant stellar population gradients \citep[e.g.,][]{Ferguson02,Gilbert14,Conn16, Cohen18,Escala20,Escala21,Escala23}. An extensive analysis of the stellar populations in these two components is impractical with the small FoV of ACS and it is not within the scope of this paper. Nevertheless, we deem useful to include these two datasets to enable a first-order comparison between the SFH of the satellites and that of the M31 spheroid.

\begin{figure*}
        \centering
    \includegraphics[width=\textwidth]{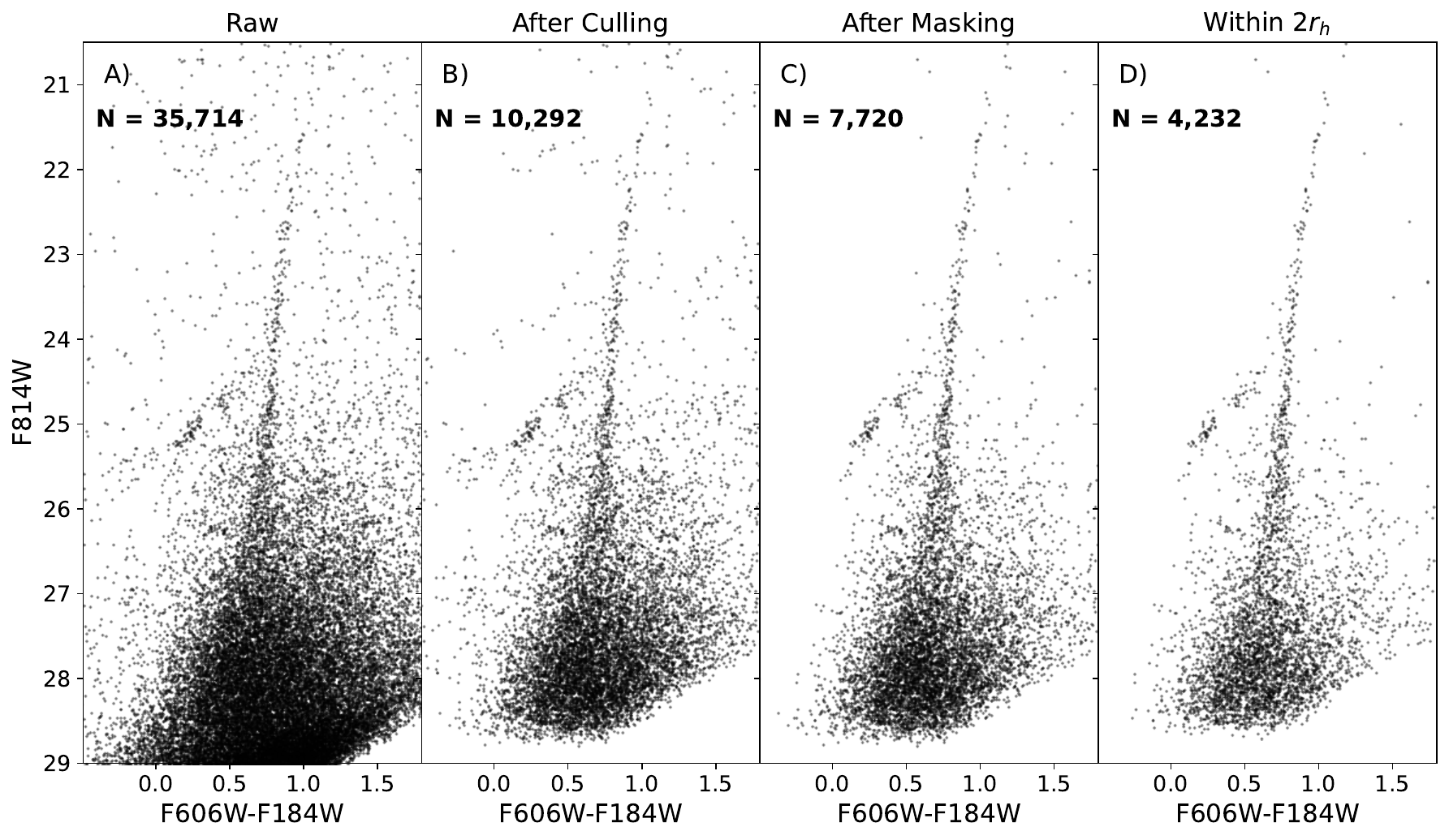}
    \caption{The ACS CMD of \A{XI} ($M_V = -6.4$), at different steps of our contaminant cleaning procedure. A) Raw photometric catalog, as output from DOLPHOT. B) Photometric catalog, after the quality cuts of \S~\ref{Sec:Culling} have been applied. C) Photometric catalog, after the bright-star mask (\S~\ref{Sec:Masks}) has been applied. D) Final photometric catalog, after the spatial cut of \S~\ref{Sec:SpatialCut} has been applied. }
    \label{Fig:Culling}
\end{figure*}

\begin{figure}
    \centering
    \includegraphics[width=0.49\textwidth]{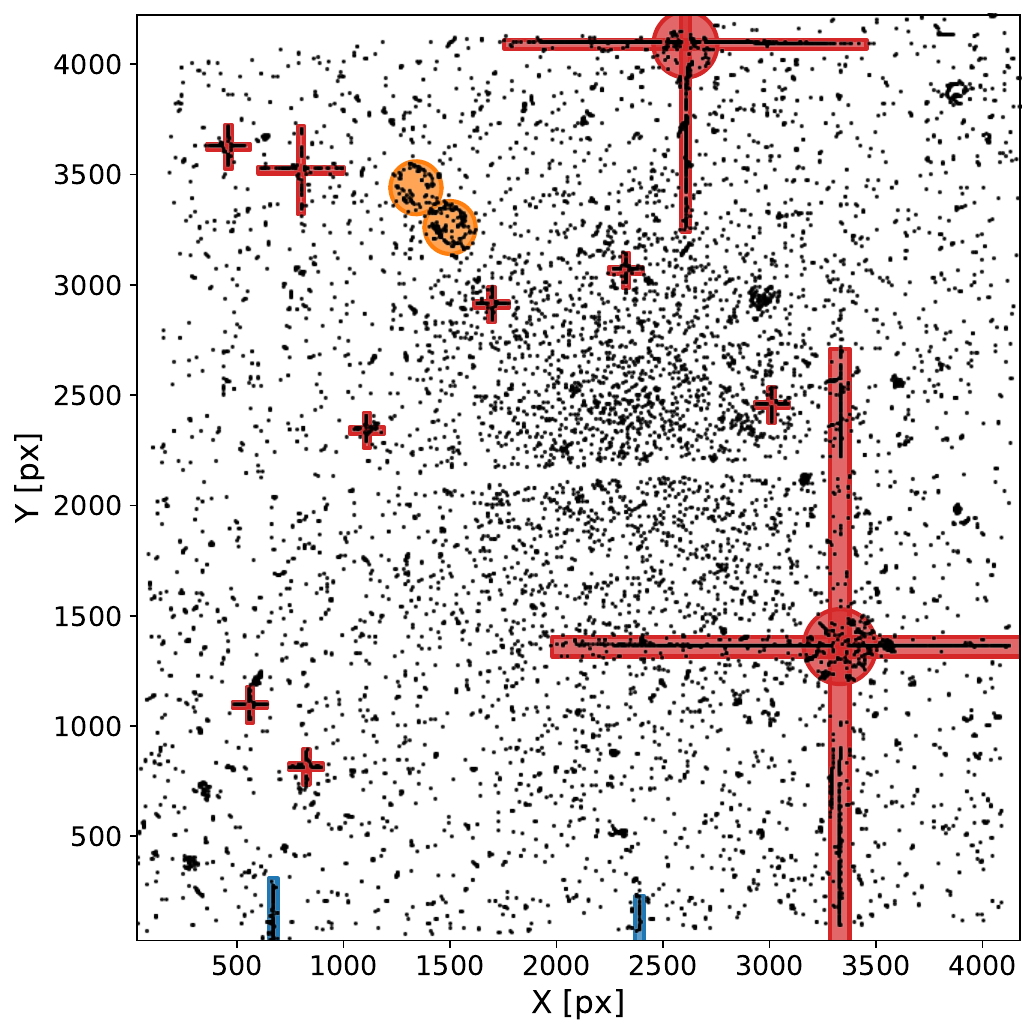}
    \caption{Spatial distribution of sources in the \A{XI} catalog. The shaded regions illustrate our spatial masks used to exclude diffraction spikes (red). In the upper side of this field, a bright stray light artifact is also masked (orange). At the bottom of this field, also visible are two masked diffraction spikes from bright stars just outside our field of view (blue).}
    \label{Fig:Masks}
\end{figure}

\section{Photometry}
\label{Sec:Reduction}
\subsection{Photometric Catalogs}
\label{Sec:Culling}
We start the photometric reduction from the pre-reduced FLC frames, as retrieved from MAST. We reduce the images using \texttt{DOLPHOT} \citep{Dolphin00,Dolphin16a,Weisz24}, a popular point-spread-function (PSF) photometry package that has been used on many HST studies of nearby galaxies \citep[e.g.,][]{Holtzman06,Dalcanton09,McQuinn10a,Monelli10a,Radburn-Smith11,Dalcanton12,Weisz14a,Williams14,Williams21}. We use the same \texttt{DOLPHOT} parameter set up established by the PHAT project \citep{Dalcanton12,Williams14}, which is well-optimized for crowded HST fields. While our dwarf fields are not as crowded as the PHAT observations of M31's disk, testing on our images, and independent testing on a range of crowding regimes \citep[e.g.,][]{Weisz24}, confirmed that this set-up is optimal to analyze our data. The only exception is the PSF-PhotIt parameter, which we set to 2. This choice adds a second iteration to the PSF photometry solution, refining the noise estimates and increasing completeness for faint sources.

After reduction, we cull the resulting photometric catalogs, to mitigate contamination from background galaxies, cosmic rays, hot pixels, and other artifacts. This is done by selecting good sources on a number of photometry quality metrics. Namely we keep photometric sources with:
\begin{itemize}
    \setlength\itemsep{0.001em}
    \item $S/N \geq 4$
    \item $Sharp^2 \leq 0.2$
    \item $Crowd \leq 0.75$
    \item $Round \leq 3$
\end{itemize}

The quality metrics cuts are imposed on both filters simultaneously. The resulting photometric catalogs have a substantially higher level of purity. We illustrate this in Fig.~\ref{Fig:Culling}, using \A{XI} as an example. The comparison between the raw DOLPHOT catalog (panel A) and the culled catalog (panel B) shows that our selection criteria drastically reduces the amount of contaminant sources in the CMD.

\subsection{Bright star masking}
\label{Sec:Masks}
After culling the catalog using the quality metrics described above, a major remaining source of photometric contaminants consists of spurious sources detected in the diffraction spikes of bright stars and in stray-light artifacts over the detector. We mitigate this effect by filtering our data through a series of spatial masks that cover the most obvious artifacts (see Fig.~\ref{Fig:Masks} for an example). 

For each one of our fields, we obtain a list of bright stars by querying the Gaia database for all sources with $G<18$, that are inside or immediately adjacent to our HST field of view. We inspect each source and we create a spatial mask that covers both the saturated core and any visible diffraction spike. The location, orientation, and size of the mask is manually adjusted for each star. We also mask diffraction spikes that arise from sources just outside our field of view. Finally, in some cases, there are bright stray-light artifacts obviously identifiable in our images, which we also mask. 

In Fig.~\ref{Fig:Masks} we provide an example of our masks for the \A{XI} field, where all the features described above are visible. By removing the sources inside the masked regions, we further increase the purity of our photometric catalogs, as demonstrated in Fig.~\ref{Fig:Culling} (comparison between panel B and panel C).

\subsection{Spatial Cuts}
\label{Sec:SpatialCut}
For most sources in our sample, the target's stellar population is the dominant contributor to source counts across the entire primary ACS FoV. However, for some of our most compact target galaxies, the bulk of the stellar population is concentrated in the central regions of the ACS detector, while the outer regions primarily contain contaminant sources. For the CMD analysis of this paper, we therefore impose a spatial cut on our catalogs, including only sources within 2$r_h$ from the target's photometric center (see Fig.~\ref{Fig:Footprint} for a visual representation). We exclude M32, NGC~147, NGC~185, and NGC~205 from this criterion, because their large stellar mass means that even regions at large radii contain a large population of galaxy member stars.

For most galaxies in our sample, their large angular size and the central positioning of the ACS fields mean that this spatial cut has no effect. However, for some of our faintest targets ($M_V > -9$) these cuts significantly increase the purity of the stellar catalog, while only marginally decreasing the stellar counts from the target's stellar population. This is shown for the case of \A{XI} ($M_V = -6.4$) by comparing panels C and D in Fig.~\ref{Fig:Culling}. For those galaxies for which we reject sources at $r>2 r_h$, we have calculated updated $f_{\star}$ values and listed them (in parentheses) in Table~\ref{Tab:Sample}.

\begin{figure*}
    \centering
    \includegraphics[width=\textwidth]{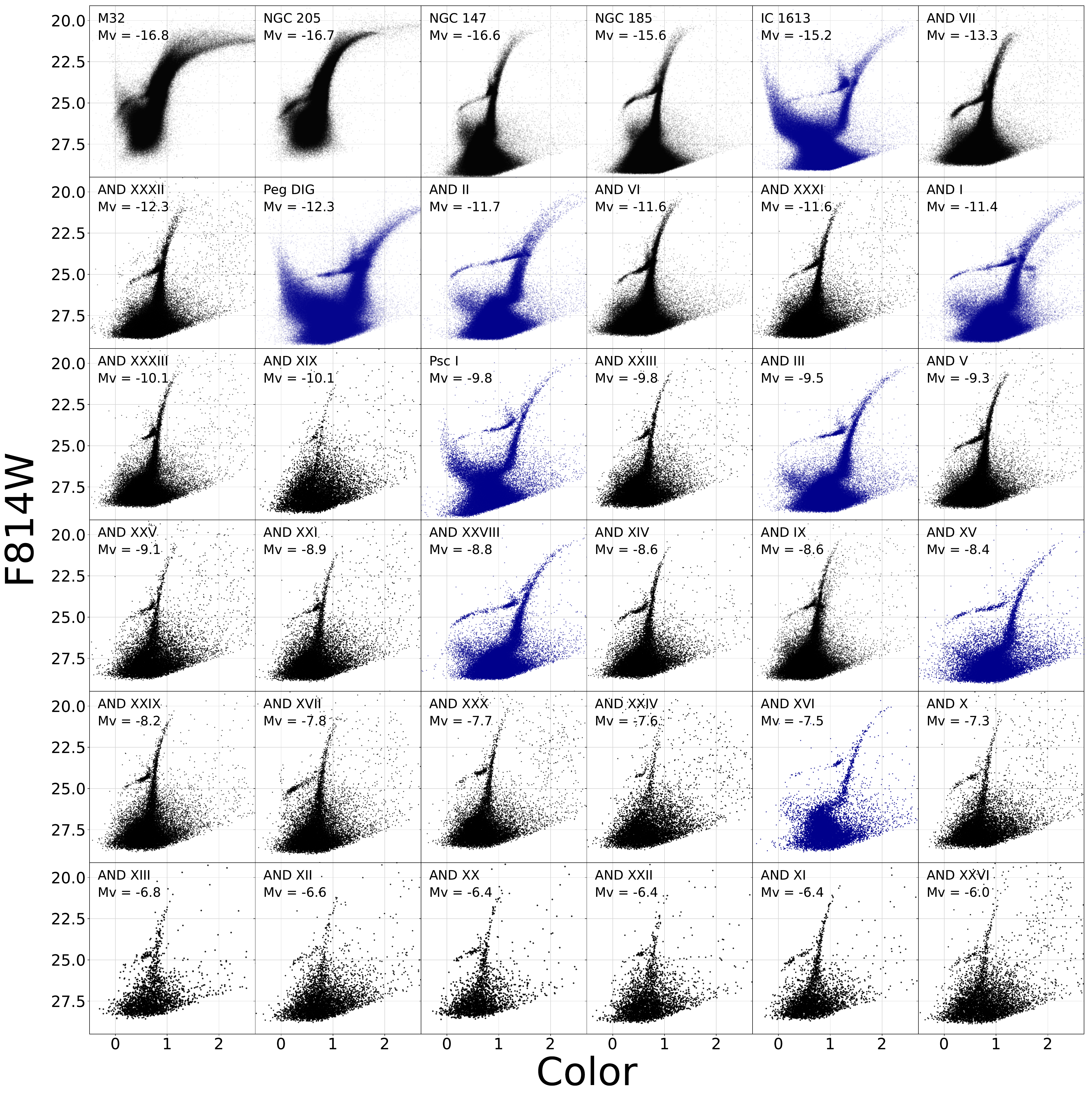}
    \caption{ACS CMDs for the 36 dwarf galaxies in our primary sample. The galaxies are sorted in order of decreasing luminosity. Panels with black points show F814W vs (F606W-F814W) CMDs, while panels with blue points show F814W vs (F475W-F814W) CMDs. The CMDs are not corrected for distance or extinction.}
    \label{Fig:CMDs}
\end{figure*}

\begin{figure*}
    \centering
    \includegraphics[width=\textwidth]{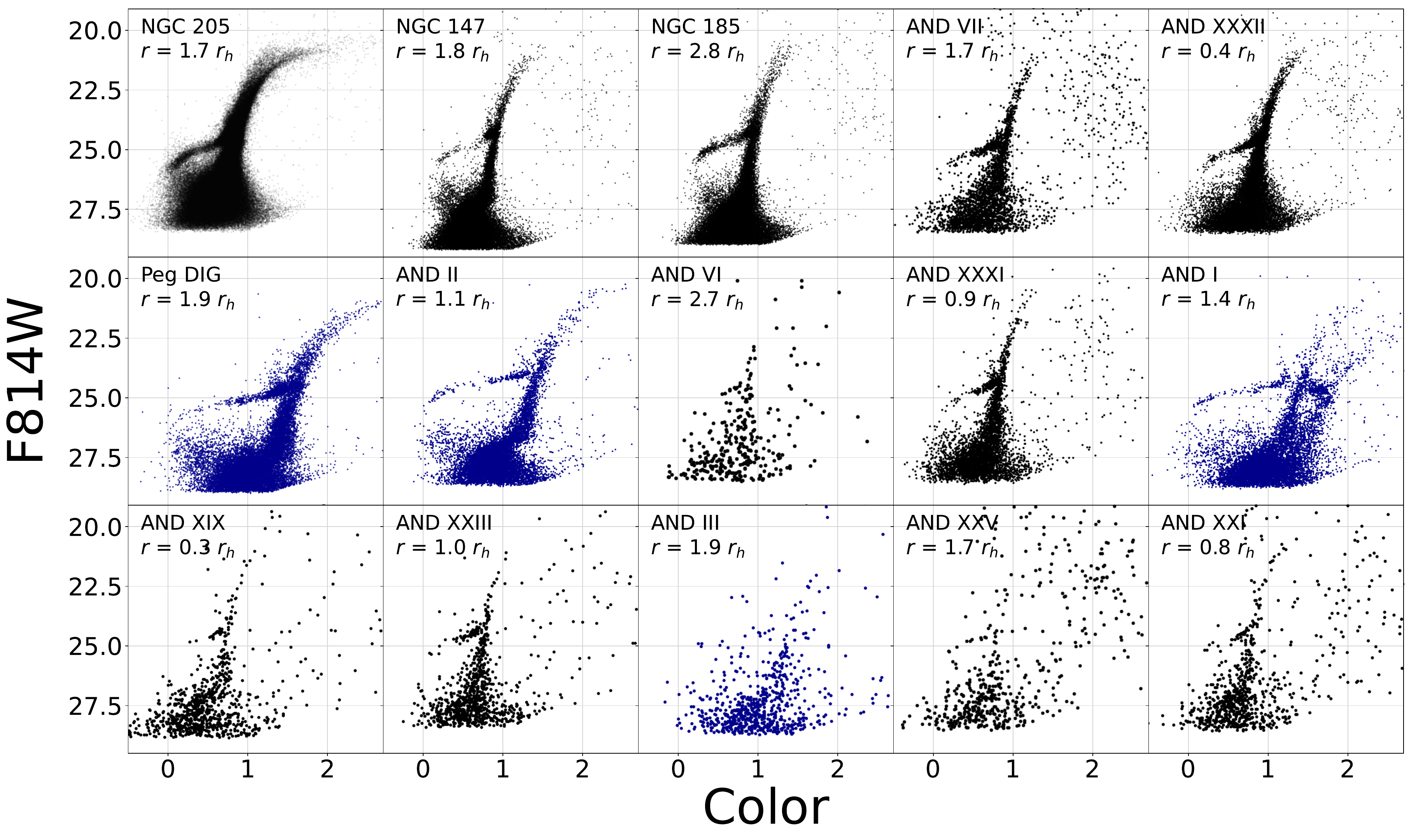}
    \caption{Same as for Fig.~\ref{Fig:CMDs}, but for the 15 UVIS parallel fields that contain member stars from our dwarf galaxy sample. To highlight the stellar populations in our sparsest fields, only stars with $Sharp^2 \leq 0.015$ are shown. The position of the UVIS aperture center is also reported, in unit of $r_h$.}
    \label{Fig:CMDs_Parallel}
\end{figure*}

\begin{figure*}
    \centering
    \includegraphics[width=\textwidth]{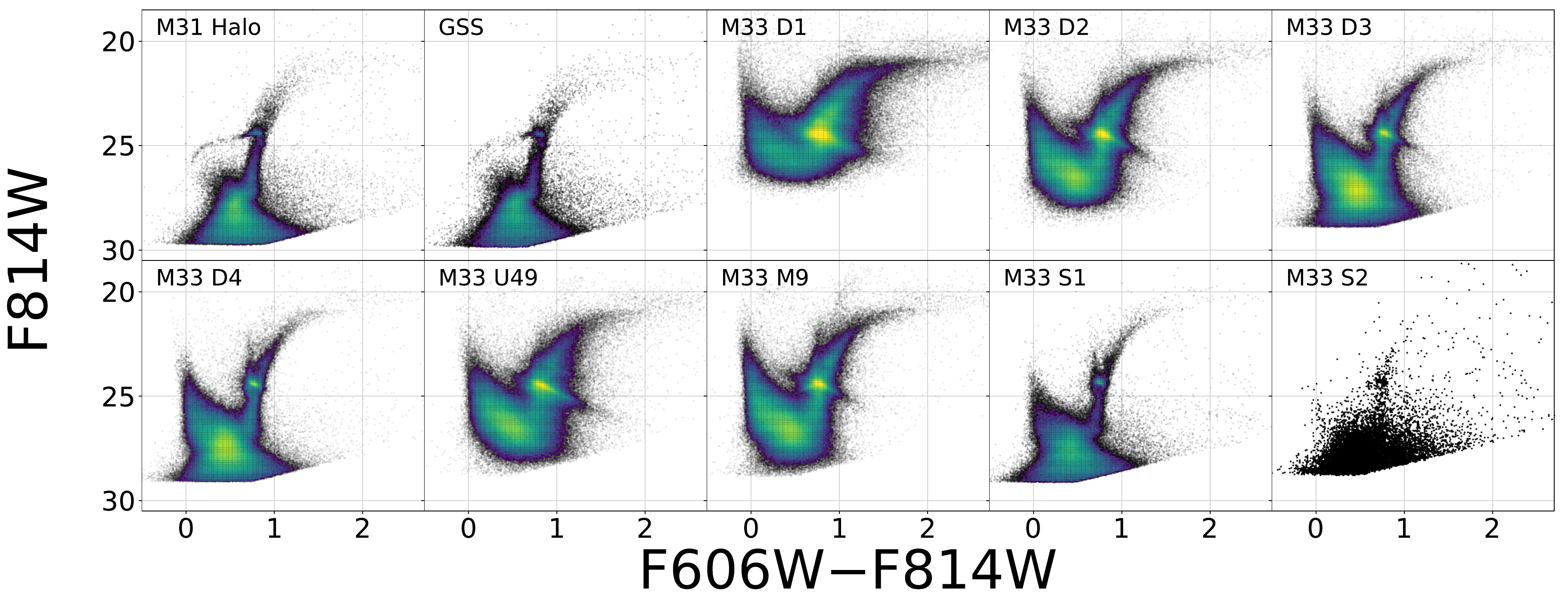}
    \caption{Same as for Fig.~\ref{Fig:CMDs}, but for the 10 ACS fields belonging to M31 and M33. Due to the large number of stars in the CMD, the denser CMD regions are shown  as a density map. All CMDs in this figure are based on F606W/F814W photometry.}
    \label{Fig:CMDs_Extra}
\end{figure*}

\subsection{Color-Magnitude Diagrams}
\label{Sec:CMDs}
Fig.~\ref{Fig:CMDs} shows the ACS CMDs in our primary dwarf galaxy sample and highlights the depth and diversity of our dataset.  In almost every target, the photometry reaches below the SGB and oMSTO per program construction.  The galaxies show a remarkable diversity of CMDs, which imply a variety of formation histories, as we discuss in the context of formal SFH measurements.

In terms of depth, two of the shallower datasets belong to M32 (archival) and NGC~205 (acquired through our program). In these two galaxies, despite long integration times, the extreme crowding limits our photometric depth to F606W $\sim 28$. While we still formally detect the SGB of a 13~Gyr stellar population, the incompleteness and photometric uncertainties at these magnitudes are considerably larger than for the rest of our targets. This limitation should be kept in mind when interpreting the oldest SFH of these two targets.

Fig.~\ref{Fig:CMDs_Parallel} shows the CMDs of the 15 UVIS parallel fields for which we detect dwarf galaxy member stars. They are not quite as deep as the ACS fields, as UVIS is less sensitive in the F814W band.  Moreover, they exhibit a broad range of stellar densities. NGC~205 has the most populated UVIS CMD due to its high luminosity, large size, and the more central location of the UVIS field. The UVIS catalog of NGC~205 contains $\sim$345,000 sources compared to the $\sim$435,000 sources in the ACS field. Conversely, in the lower-luminosity \A{III}, our UVIS field centered roughly at $2 {r_h}$ only contains a few hundred member stars, compared to $\sim$70,000 sources in the ACS field. With the exception of the \A{IX} UVIS field, which contains a large population of M31 halo stars, all the UVIS fields not shown in Figure~\ref{Fig:CMDs_Parallel} are essentially empty, as they are dominated by background galaxies and sparse populations of MW and/or M31 stars.

Finally, Fig.~\ref{Fig:CMDs_Extra} shows the CMDs of the M31 and M33 fields. This figure shows the heterogeneity of these archival datasets. Both the M31 fields (Halo and GSS) are extremely deep, reaching more than 1 magnitude below the oMSTO. In M33, the more external photometric fields (D3, D4, S1, and S2) are also sufficiently deep to detect the oMSTO. However, as we move to inner regions (M9, U49, D2, and D1), crowding increasingly limits the photometric depth achievable. Also noticeable is the change in differential reddening, which is essentially absent for the outermost fields, and progressively increases when moving towards M33's center.

\subsection{Artificial Star Tests}
\label{Sec:ASTs}
We characterize the photometric completeness and the error distribution as a function of position on the CMD through the use of artificial star tests (ASTs). This technique is the `gold standard' to model photometric uncertainties in resolved stellar population studies, and is absolutely essential in crowded fields \citep[e.g.,][]{Stetson87,Stetson88,Dalcanton09,Weisz14b,Williams14,Savino24}.

For each of our ACS and WFC3 fields, we inject $\sim5\times 10^5$ synthetic stars in our images. Our input artificial stars are uniformly distributed across the observed CMD from  $\sim1$ magnitude above the tip of the RGB to $\sim2$ mag below the detection limit. For the 36 dwarf galaxies in our primary sample, the ASTs are spatially distributed following the same 2D density profiles used to calculate $f_\star$. This choice allows us to capture the spatial variation in observational errors and completeness due to crowding gradients. The ASTs for the M31 and M33 are uniformly distributed across the ACS fields. 

We then repeat the DOLPHOT reduction procedure to measure the recovered photometric properties of our synthetic sources. To avoid modifying the crowding properties of the image, the injection and recovery steps are performed on one artificial star at a time. After reduction, the AST catalogs are processed through the exact cuts used for the photometric catalogs (i.e., quality culling, bright star masking, and spatial cuts).

\section{CMD modeling set-up}
\label{sec:cmd_model}
For each of our ACS photometric fields, we derive the SFH of the underlying stellar population through detailed modeling of the CMD density distribution, i.e., Hess diagrams. We do this using \texttt{MATCH} \citep{Dolphin02}, which is a software package designed to measure the SFH of a resolved stellar population by forward modeling the CMD with a combination of simple stellar population models. \texttt{MATCH} is a well-tested software that has been used in numerous studies of nearby stellar populations \citep[e.g.,][]{Skillman03,Williams09,McQuinn10a,Monelli10a,Weisz11,Weisz14a,Lewis15,McQuinn15,Skillman17}. We briefly describe the specific assumptions and configuration of our CMD models and refer the reader to the extensive tests carried out in \citet{Savino23} regarding the effect of many of these assumption on measured SFHs.

\subsection{Stellar Models}
We create stellar population models using stellar evolutionary tracks from the BaSTI theoretical grid \citep{Hidalgo18,Pietrinferni21,Pietrinferni24}. The stellar models we use include the effect of atomic diffusion, convective core overshooting, and use a Reimer's mass-loss efficiency of $\eta=0.3$. We use models with a scaled-solar $\alpha$-element distribution, which is typical of CMD analysis  in dwarf galaxies \citep{Monelli10a,Skillman17}. Although spectroscopic investigations of dwarfs in the M31 system revealed large variations in stellar $\alpha$-enhancement, both within individual dwarfs and across the satellite population \citep{Vargas14,Kirby20}, it has long been established that the choice of chemical mixture has a minor effect on the SFH of old stellar populations, especially if red broadband filters are used \citep[e.g.,][]{Salaris93,Cassisi04,Savino23}. The exact $\alpha$-element distribution adopted will have little impact on our SFHs.

\subsection{Model Grid}
We generate stellar population models over a grid spanning $7.50 \le \log_{10}(t/yr) \le 10.15$, with a 0.05 dex resolution, and $-3.0 < [Fe/H] < 0$, with a 0.1 dex resolution. We draw stars from a Kroupa initial mass function \citep{Kroupa01}, normalized between 0.08 $M_{\odot}$ and 120 $M_{\odot}$, and use an unresolved binary fraction of 0.35, with secondary masses drawn from a uniform mass ratio distribution. As demonstrated in \citet{Savino23}, all these assumptions do not have a strong impact on the recovered SFH. Consistent with previous nearby dwarf analyses \citep[e.g.,][]{Weisz14a,Skillman17}, we impose a physically motivated prior on the age-metallicity relation. During fitting, we impose that the mean metallicity must increase monotonically with time and we allow a modest metallicity dispersion (0.15~dex) in each age bin. This is a common choice to mitigate the age-metallicity degeneracy on the MSTO, and it is especially useful for the F606W/F814W filter pair, which has limited sensitivity to metallicity (see Appendix~B of \citealt{Savino23} for additional discussion on this assumption).

\subsection{Distance and Extinction}
We adopt homogeneous RR Lyrae distances from \citet{Savino22}. We set the MW foreground extinction using the dust maps of \citet{Green19}. The only exceptions are M33 and IC~1613, which are not covered by the \citet{Green19} map, and for which we use extinction values from \citet{Schlafly11}.

For galaxies that have had recent star formation (i.e., Psc~I, Peg~DIG, IC~1613, and the photometric fields in M33), we apply two more extinction terms, in addition to the MW foreground. The first one is an internal reddening term that applies to all stars in the stellar population model. This term is intended to capture the diffuse dust content within the dwarf galaxy itself. We explore values of internal reddening ranging from 0 to $dA_V = 2$~mag, in steps of 0.1~mag, and adopt the value that results in the best CMD goodness-of-fit parameter. Only the M33 fields are found to require a $dA_V$ greater than zero and the best fit values are listed in Table~\ref{tab:M33}.

The second term is a uniformly distributed differential extinction term that only applies to stars younger than 100~Myr and is intended to capture the extinction of the dense natal environments of these stars. Following \citet{Dolphin03}, this second extinction term is set to a differential extinction of $dA_V=0$ at 100~Myr and linearly increased to a maximum value of $dA_V=0.5$~mag for stars younger than 40~Myr. Given that most of our targets have virtually no amount of recent star formation, the impact of this assumption is expected to be negligible in most cases. For the few actively star-forming galaxies in our sample, choices on this extra extinction term only meaningfully affect the SFH within the last 500 Myr. Even for star-forming dwarfs, only 1-3\% of the total stellar mass is formed in this time span (see \S~\ref{sec:sfhs}), so this assumption has a small impact unless in applications where the details of the most recent SFH are of particular interest.

\begin{figure*}
    \centering
    \includegraphics[width=\textwidth]{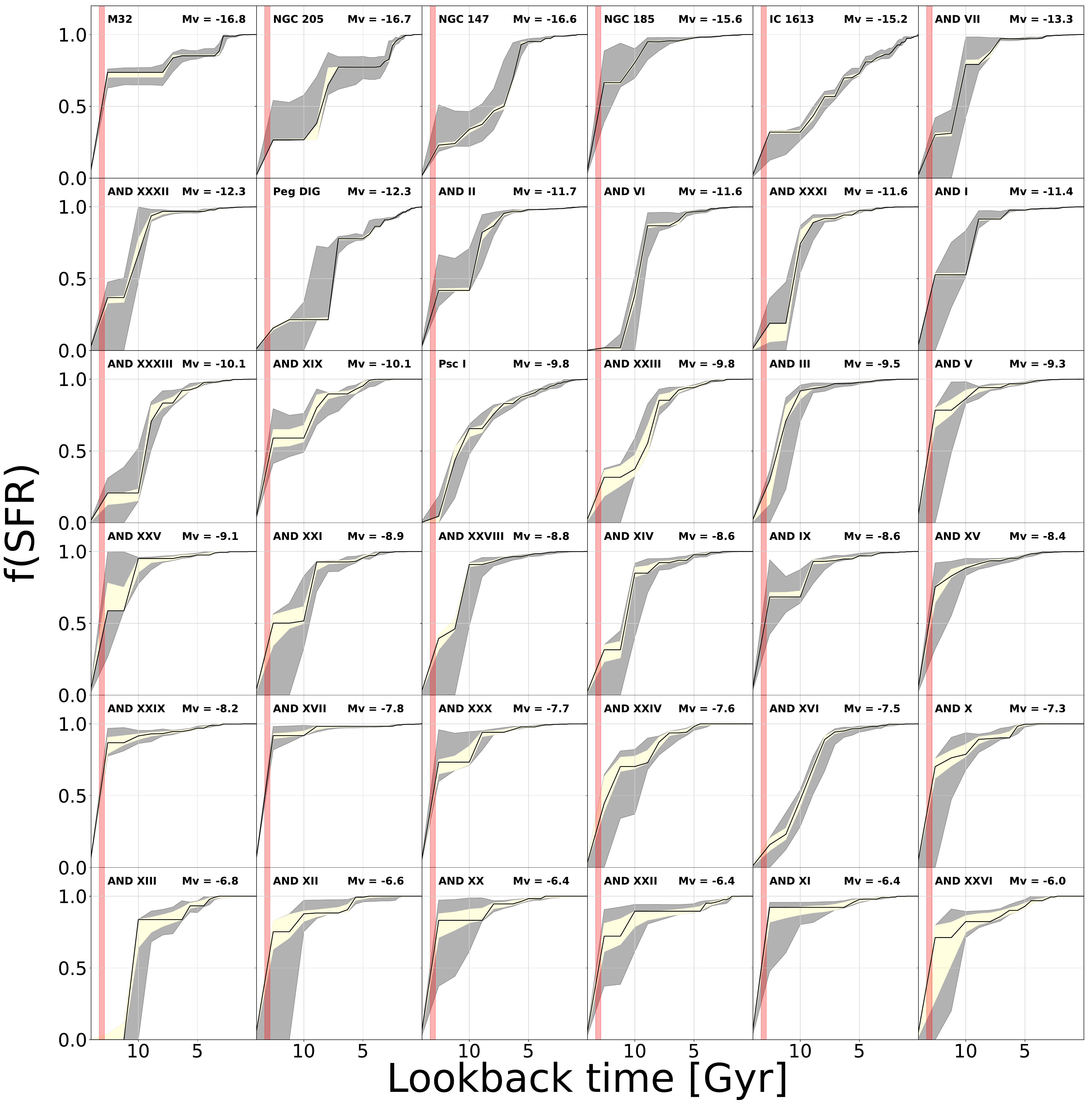}
    \caption{Cumulative SFHs for the 36 dwarf galaxies in our sample, measured from the ACS fields. The black line shows the best-fit SFH, the yellow region show the statistical uncertainties and the grey region show the systematic uncertainties. The epoch of reionization is highlighted by the red shaded region \citep[$6 < z < 10$, corresponding to $13.33~Gyr < t < 12.87~Gyr$ with Planck cosmological parameters,][]{Planck20,Robertson22}. The galaxies are ordered by absolute luminosity.}
    \label{Fig:SFHs}
\end{figure*}

\subsection{CMD Fit}
The stellar population models are convolved with observational effects (photometric errors and incompleteness) measured by the ASTs described in \S~\ref{Sec:ASTs}. Both observed and model CMDs are converted to density maps using bins of 0.05~mag in color  and 0.1~mag in apparent magnitude. We fit the CMD from approximately 1~mag above the tip of the RGB, to a magnitude that corresponds to the 50\% completeness level as listed in Tabs.~\ref{Tab:Sample} and \ref{Tab:M31}. The only exceptions are those fields for which the 50\% completeness limit does not reach the oMSTO and for which we extend our modeling to fainter magnitudes. In particular, for Peg~DIG and for the D2, D3, D4, M9, and U49 fields of M33 we model down to the 30\% completeness level. For the severely crowded fields of M32, NGC~205 and the D1 field of M33, we model the entirety of the CMD down to the detection limit. In all of our targets, we exclude the CMD region around the horizontal branch, to mitigate the accuracy issues of models of this evolutionary phase.  As mentioned above, the oldest age SFHs measured from CMDs that extend below the 50\% completeness limit should be treated with appropriate caution.

\subsection{Contamination Model}
While the procedure outlined in \S~\ref{Sec:Reduction} lets us reject a large fraction of contaminant sources, our photometric catalogs still contain sources that are not genuine member stars of our targets. In particular, unresolved background galaxies are a major source of contamination at the magnitude of the oMSTO \citep[e.g.,][]{Warfield23}. We address this issue by building a contamination model that is added to our stellar population models during the CMD fit.

Ideally, each ACS field has a WFC3/UVIS parallel that could be used as a control field to build the contamination model. In practice, most UVIS parallels are not suitable either because they contain a substantial number of member stars (in our most extended targets) or because they are so sparse that the reduction does not yield reliable photometry.

We instead choose representative, well-reduced, UVIS fields to build a contamination model for all the galaxies in our sample. For the F606W/F814W data, we use the parallel field of \A{XXX}, while for the F475W/F814W data we use the parallel field of \A{XV}. For these two fields, we convert the CMD to a density map, which is smoothed with a $5\times5$ bin kernel. This density field is added to the stellar population model density during the CMD fit.  As part of the CMD optimization process, we allow the code to fit a scale factor for the background model along with the stellar population parameters.

The only exceptions to this approach are the CMDs of \A{I}, \A{IX}, and M32.  They each contain substantial contamination from M31 halo stars. For these galaxies, we create a custom contamination model. The contamination model for \A{IX} is built using the whole UVIS field from its own parallel, which only contains M31 stars. Similar to \citet{Skillman17}, for \A{I} we only use a cut-out ($r>3r_h$) of the UVIS parallel field, chosen to minimize the number of stars from \A{I} itself. For M32, we use an ACS control field, adjacent to M32, observed as part of program GO-9392.

\begin{figure}
    \centering
    \includegraphics[width=0.45\textwidth]{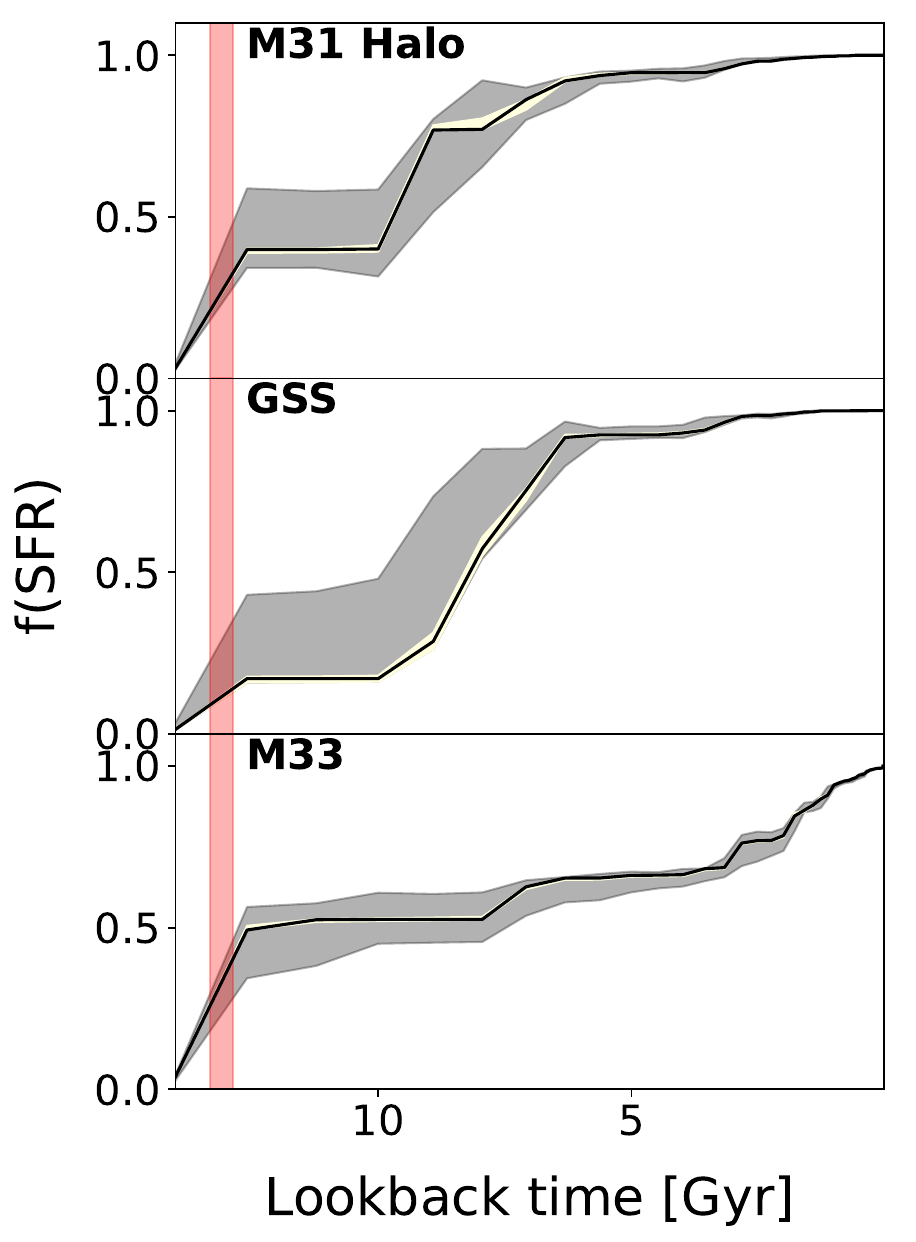}
    \caption{The cumulative SFHs for the M31 Halo field, the GSS field, and M33. The M33 SFH is the combination of the 8 fields from Table\ref{Tab:M31}. Lines and colors have the same meaning as in Figure~\ref{Fig:SFHs}}
    \label{Fig:SFHs_Extra}
\end{figure}

\subsection{Uncertainties}
Once we obtain the best-fit SFH, we calculate the uncertainties following the prescription of \citet{Dolphin12} and \citet{Dolphin13}, which is a common choice in studies of nearby galaxies \citep[e.g.,][]{Weisz14a,Skillman17}. The statistical uncertainties on the star formation rates are obtained by sampling the solution parameter space with a Hamiltonian Monte Carlo algorithm \citep{Duane87}. The systematic uncertainties are estimated by perturbing the stellar model library with shifts in effective temperature and luminosity. We produce 100 perturbed stellar model grids and use the dispersion in the SFH solutions to calculate the systematic uncertainties. As demonstrated in \citet{Savino23}, not only does this method provide a good representation of the systematics associated with the stellar model library, but it also captures the SFH variation associated with other common sources of uncertainties (e.g., distance, extinction, $\alpha$-enhancement).

\begin{table*}
        \caption{Star formation timescales of our sample targets. We list the absolute luminosity of the galaxy, the de-projected distance from M31, the value of distance and extinction used in the CMD fitting, and the values of $\tau_{50}$, $\tau_{80}$, and $\tau_{90}$ (see \S~\ref{Sec:Quenching}). For the latter, uncertainties are only statistical (outside parentheses) and statistical plus systematics (inside parentheses).}
    \centering
    \begin{threeparttable}
    \begin{tabular}{lrrrrrrr}

    \toprule
    Galaxy & $M_V$\tnote{a}&$D_{M31}$\tnote{a}&$(m-M)_0$\tnote{a}&E(B-V)\tnote{b}& $\tau_{50}$ & $\tau_{80}$ & $\tau_{90}$\\
    &mag&kpc&mag&mag& Gyr & Gyr & Gyr\\
    \toprule
    M32 & $-16.8\pm0.1$&$20.6^{+21}_{-13}$&24.44&0.06&$13.1_{-0.5~(-0.5)}^{+1.1~(+1.1)}$&$7.4_{-0.3~(-1.0)}^{+0.6~(+0.6)}$&$3.1_{-0.3~(-0.3)}^{+0.1~(+1.5)}$\\
    
    NGC~147&$-16.6\pm0.07$&$107.0^{+15}_{-8.0}$&24.33&0.15&$7.1_{-0.1~(-0.8)}^{+0.9~(+5.5)}$&$6.0_{-0.4~(-0.4)}^{+0.3~(+1.2)}$&$5.7_{-0.1~(-0.4)}^{+0.6~(+0.9)}$\\
    
    NGC~185&$-15.6\pm0.07$&$154.1^{+23}_{-21}$&24.06&0.20&$13.0_{-0.4~(-1.0)}^{+1.2~(+1.2)}$&$10.0_{-0.1~(-0.9)}^{+1.2~(+2.7)}$&$9.3_{-0.4~(-1.6)}^{+0.7~(+3.0)}$\\
    
    NGC~205&$-16.7\pm0.1$&$58.0^{+30}_{-29}$&24.61&0.03&$8.5_{-0.5~(-0.5)}^{+0.4~(+4.2)}$&$3.3_{-0.1~(-0.4)}^{+0.3~(+5.1)}$&$2.6_{-0.1~(-0.2)}^{+0.2~(+0.5)}$\\
    
    \A{I}&$-11.4\pm0.2$&$48.0^{+10}_{-3.2}$&24.45&0.06&$12.7_{-0.1~(-2.6)}^{+1.5~(+1.5)}$&$9.2_{-0.3~(-0.3)}^{+0.8~(+1.3)}$&$9.0_{-0.0~(-0.9)}^{+1.0~(+1.0)}$\\
    
    \A{II}&$-11.7\pm0.2$&$168.9^{+19}_{-16}$&24.12&0.03&$9.8_{-0.9~(-0.9)}^{+0.2~(+3.2)}$&$9.0_{-1.0~(-1.0)}^{+0.0~(+1.0)}$&$7.6_{-0.5~(-0.5)}^{+0.3~(+1.5)}$\\
    
    \A{III}&$-9.5\pm0.2$&$84.9^{+19}_{-14}$&24.29&0.06&$11.9_{-0.7~(-1.4)}^{+0.7~(+0.7)}$&$10.7_{-0.7~(-1.2)}^{+0.6~(+0.7)}$&$10.1_{-1.2~(-1.2)}^{+0.1~(+1.1)}$\\
    
    \A{V}&$-9.3\pm0.2$&$110.5^{+7.0}_{-3.5}$&24.40&0.11&$13.1_{-0.5~(-2.0)}^{+1.0~(+1.0)}$&$11.0_{-0.3~(-1.0)}^{+1.6~(+1.6)}$&$9.5_{-0.6~(-1.2)}^{+1.0~(+2.3)}$\\
    
    \A{VI}&$-11.6\pm0.2$&$281.6^{+8.6}_{-7.1}$&24.60&0.08&$9.7_{-0.8~(-0.8)}^{+0.3~(+0.3)}$&$9.1_{-0.1~(-1.0)}^{+0.9~(+0.9)}$&$6.4_{-0.8~(-0.3)}^{+0.1~(+2.7)}$\\
    
    \A{VII}&$-13.3\pm0.3$&$230.8^{+8.4}_{-6.5}$&24.58&0.14&$10.7_{-0.7~(-1.0)}^{+0.5~(+0.5)}$&$8.8_{-0.1~(-0.9)}^{+2.4~(+1.6)}$&$7.7_{-0.6~(-0.6)}^{+0.3~(+2.5)}$\\
    
    \A{IX}&$-8.6\pm0.3$&$82.0^{+26}_{-24}$&24.23&0.09&$13.0_{-0.4~(-1.1)}^{+1.1~(+1.1)}$&$9.5_{-0.6~(-0.8)}^{+0.5~(+3.3)}$&$9.0_{-0.1~(-1.6)}^{+1.0~(+3.6)}$\\
    
    \A{X}&$-7.3\pm0.3$&$162.2^{+25}_{-24}$&24.00&0.14&$13.0_{-0.4~(-2.0)}^{+1.1~(+1.1)}$&$9.9_{-0.2~(-1.0)}^{+1.7~(+2.3)}$&$7.5_{-0.2~(-1.0)}^{+1.7~(+3.8)}$\\
    
    \A{XI}&$-6.4\pm0.4$&$104.2^{+11}_{-4.2}$&24.38&0.09&$13.3_{-0.7~(-0.9)}^{+0.9~(+0.9)}$&$12.8_{-0.2~(-2.8)}^{+1.3~(+1.3)}$&$12.6_{-0.2~(-5.5)}^{+1.3~(+1.5)}$\\
    
    \A{XII}&$-6.6\pm0.5$&$107.7^{+20}_{-13}$&24.28&0.17&$13.1_{-0.5~(-2.7)}^{+1.0~(+1.0)}$&$10.7_{-0.7~(-1.3)}^{+1.9~(+1.9)}$&$6.5_{-0.7~(-0.3)}^{+1.9~(+4.4)}$\\
    
    \A{XIII}&$-6.8\pm0.4$&$126.4^{+16}_{-8.0}$&24.57&0.14&$10.5_{-0.5~(-1.3)}^{+0.7~(+0.7)}$&$10.1_{-2.5~(-3.5)}^{+0.0~(+1.2)}$&$5.8_{-2.5~(-1.0)}^{+0.0~(+2.7)}$\\
    
    \A{XIV}&$-8.6\pm0.3$&$160.8^{+3.8}_{-4.2}$&24.44&0.09&$10.8_{-0.8~(-1.1)}^{+0.4~(+0.4)}$&$10.1_{-0.1~(-1.7)}^{+1.1~(+1.1)}$&$8.2_{-0.3~(-1.9)}^{+0.8~(+1.8)}$\\
    
    \A{XV}&$-8.4\pm0.3$&$95.8^{+12}_{-4.8}$&24.37&0.05&$13.1_{-0.5~(-1.6)}^{+1.0~(+1.0)}$&$11.7_{-0.5~(-1.6)}^{+0.9~(+1.0)}$&$9.4_{-0.5~(-1.8)}^{+0.8~(+3.2)}$\\
    
    \A{XVI}&$-7.5\pm0.3$&$280.0^{+26}_{-27}$&23.57&0.07&$9.9_{-0.9~(-0.9)}^{+0.2~(+0.4)}$&$8.4_{-0.4~(-1.1)}^{+0.5~(+0.5)}$&$7.8_{-0.4~(-1.4)}^{+0.5~(+0.3)}$\\
    
    \A{XVII}&$-7.8\pm0.3$&$49.9^{+17}_{-5.8}$&24.40&0.07&$13.3_{-0.7~(-0.7)}^{+0.9~(+0.9)}$&$12.8_{-0.2~(-0.2)}^{+1.3~(+1.3)}$&$12.6_{-0.2~(-2.2)}^{+1.3~(+1.5)}$\\
    
    \A{XIX}&$-10.1\pm0.3$&$113.3^{+18}_{-6.9}$&24.55&0.05&$12.8_{-0.2~(-2.9)}^{+1.3~(+1.3)}$&$8.9_{-0.4~(-2.1)}^{+1.1~(+1.0)}$&$6.3_{-0.5~(-0.7)}^{+2.1~(+2.9)}$\\
    
    \A{XX}&$-6.4\pm0.4$&$128.4^{+12}_{-5.5}$&24.35&0.08&$13.2_{-0.6~(-2.4)}^{+0.9~(+0.9)}$&$12.6_{-2.3~(-3.6)}^{+0.1~(+1.5)}$&$8.3_{-2.3~(-0.8)}^{+0.1~(+4.4)}$\\
    
    \A{XXI}&$-8.9\pm0.3$&$124.4^{+5.1}_{-3.8}$&24.44&0.11&$12.6_{-2.6~(-3.1)}^{+0.2~(+1.5)}$&$9.2_{-0.3~(-0.9)}^{+0.8~(+0.9)}$&$9.0_{-0.8~(-2.8)}^{+1.0~(+1.0)}$\\
    
    \A{XXII}&$-6.4\pm0.4$&$216.8^{+5.7}_{-5.6}$&24.39&0.10&$13.0_{-0.5~(-2.4)}^{+1.1~(+1.1)}$&$10.7_{-1.0~(-1.7)}^{+2.0~(+2.1)}$&$4.4_{-1.0~(-0.4)}^{+2.0~(+8.2)}$\\
    
    \A{XXIII}&$-9.8\pm0.2$&$128.1^{+10}_{-4.9}$&24.36&0.05&$9.2_{-0.4~(-0.4)}^{+0.8~(+1.4)}$&$8.1_{-0.2~(-0.8)}^{+0.8~(+0.9)}$&$6.6_{-0.2~(-0.3)}^{+0.8~(+1.7)}$\\
    
    \A{XXIV}&$-7.6\pm0.3$&$194.5^{+25}_{-24}$&23.92&0.10&$12.3_{-0.3~(-2.8)}^{+1.8~(+0.6)}$&$8.4_{-0.5~(-0.8)}^{+0.9~(+2.9)}$&$7.6_{-0.5~(-1.5)}^{+0.9~(+1.4)}$\\
    
    \A{XXV}&$-9.1_{-0.2}^{+0.3}$&$85.2^{+12}_{-4.4}$&24.38&0.10&$12.8_{-0.2~(-1.2)}^{+1.3~(+1.3)}$&$10.5_{-0.5~(-0.8)}^{+0.7~(+2.4)}$&$10.2_{-0.9~(-1.8)}^{+1.1~(+2.6)}$\\
    
    \A{XXVI}&$-6.0_{-0.5}^{+0.7}$&$104.6^{+6.8}_{-3.5}$&24.48&0.09&$13.0_{-1.7~(-2.5)}^{+1.1~(+1.1)}$&$10.2_{-1.3~(-2.1)}^{+2.1~(+2.3)}$&$6.3_{-1.3~(-0.9)}^{+2.1~(+5.0)}$\\
    
    \A{XXVIII}&$-8.8_{-1.0}^{+0.4}$&$368.8^{+7.8}_{-7.3}$&24.36&0.11&$11.1_{-1.1~(-1.2)}^{+0.5~(+0.5)}$&$10.3_{-0.3~(-1.3)}^{+0.9~(+0.9)}$&$10.0_{-0.6~(-2.1)}^{+1.2~(+1.2)}$\\
    
    \A{XXIX}&$-8.2\pm0.4$&$189.1^{+12}_{-8.8}$&24.26&0.03&$13.2_{-0.6~(-0.6)}^{+0.9~(+0.9)}$&$12.7_{-0.3~(-1.1)}^{+1.4~(+1.4)}$&$10.4_{-0.8~(-2.1)}^{+2.2~(+2.3)}$\\
    
    \A{XXX}&$-7.7_{-0.2}^{+0.3}$&$238.6^{+24}_{-24}$&23.74&0.12&$13.1_{-0.5~(-0.5)}^{+1.1~(+1.1)}$&$9.6_{-0.7~(-0.7)}^{+1.2~(+3.2)}$&$9.1_{-0.7~(-1.0)}^{+1.2~(+3.6)}$\\
    
    \A{XXXI}&$-11.6\pm0.7$&$261.4^{+6.9}_{-5.9}$&24.36&0.12&$10.5_{-0.5~(-0.5)}^{+0.7~(+0.7)}$&$9.6_{-0.7~(-0.9)}^{+0.5~(+0.6)}$&$8.6_{-0.2~(-1.5)}^{+1.4~(+1.0)}$\\
    
    \A{XXXII}&$-12.3\pm0.7$&$146.8^{+7.8}_{-4.2}$&24.52&0.19&$10.7_{-0.7~(-0.7)}^{+0.6~(+0.8)}$&$9.5_{-0.5~(-0.5)}^{+0.5~(+1.0)}$&$9.1_{-0.1~(-0.2)}^{+0.9~(+1.2)}$\\
    
    \A{XXXIII}&$-10.1\pm0.7$&$340.3^{+10}_{-8.7}$&24.24&0.11&$9.3_{-0.4~(-0.4)}^{+0.7~(+0.8)}$&$8.2_{-0.3~(-1.0)}^{+0.8~(+0.9)}$&$6.5_{-0.5~(-0.6)}^{+0.6~(+1.7)}$\\
    
    Psc~{\sc I}&$-9.8\pm0.1$&$292.1^{+17}_{-16}$&23.91&0.04&$10.8_{-0.8~(-1.0)}^{+0.4~(+0.5)}$&$7.4_{-0.3~(-1.1)}^{+0.5~(+0.9)}$&$4.7_{-0.2~(-0.6)}^{+0.4~(+0.7)}$\\
    
    Peg~DIG&$-12.3\pm0.2$&$458.2^{+11}_{-9.4}$&24.74&0.06&$7.5_{-0.4~(-0.4)}^{+0.5~(+2.0)}$&$4.6_{-0.2~(-0.2)}^{+0.4~(+1.8)}$&$3.2_{-0.0~(-0.1)}^{+0.0~(+1.2)}$\\
    
    IC~1613&$-15.2\pm0.1$&$511.1^{+10}_{-9.8}$&24.32&0.02&$8.4_{-0.5~(-0.8)}^{+0.5~(+0.5)}$&$4.5_{-0.1~(-0.5)}^{+0.5~(+0.5)}$&$2.1_{-0.0~(-0.3)}^{+0.0~(+0.6)}$\\
    
    M31 (Halo)&-&-&24.45&0.08&$9.7_{-0.8~(-0.8)}^{+0.3~(+3.1)}$&$7.7_{-0.6~(-0.6)}^{+0.5~(+1.3)}$&$6.6_{-0.3~(-0.8)}^{+0.5~(+1.5)}$\\
    
    M31 (GSS)&-&$53.4^{+30}_{-26}$&24.58&0.05&$8.2_{-0.2~(-0.2)}^{+0.7~(+1.7)}$&$6.9_{-0.5~(-0.5)}^{+0.2~(+1.6)}$&$6.4_{-0.1~(-0.7)}^{+0.7~(+0.7)}$\\
    
    M33 (Combined)&$-19.1\pm0.1$&$226.7^{+15}_{-11}$&24.67&0.04&$12.3_{-0.1~(-4.8)}^{+1.9~(+1.9)}$&$1.9_{-0.2~(-0.2)}^{+0.1~(+0.2)}$&$1.2_{-0.0~(-0.1)}^{+0.0~(+0.1)}$\\
    \toprule

    \end{tabular}
    \begin{tablenotes}
        \item [a] From \citet{Savino22}.
        \item [b] From \citet{Green19} and \citet{Schlafly11}.
    \end{tablenotes}
    \end{threeparttable}
    \label{Tab:SFHs}
\end{table*}

\section{Star formation histories}
\label{sec:sfhs}
Fig.~\ref{Fig:SFHs} and Fig.~\ref{Fig:SFHs_Extra} show the cumulative SFHs for our primary sample of 36 dwarfs and for our M31/M33 fields, respectively. The corresponding star formation rates vs.\ lookback time are shown in Appendix~\ref{App:SFR}. We report the SFH of the M31 Halo field and that of the GSS separately. We obtain the SFH of M33 by an unweighted sum of the SFRs of the eight M33 fields (which corresponds to a mass-weighted sum of the cumulative SFHs). SFHs for the individual M33 fields are shown in Appendix~\ref{App:M33}.

For most targets, our measurements are the first SFHs obtained from oMSTO photometry. However, roughly 30\% of our sample (roughly corresponding to the archival fields) have previous SFH determination from deep photometry \citep[e.g.,][]{Skillman03,Brown06,Williams09,Hidalgo11,Bernard12,Monachesi12,Geha15,Skillman17,Collins22a}. In Appendix~\ref{App:Lit}, we provide a comparison between our SFHs and some of these previous literature results. Overall, there is good agreement between our SFHs and literature SFHs, with minor discrepancies due to choices in stellar models, distance, and/or extinction.

We now take advantage of our large and homogeneous SFH compilation to analyze the star formation timescales of the M31 satellites on a population level, both as a function of other satellite properties and in comparison to the MW satellite population.

\subsection{The Effect of Spatial Coverage}
\label{sec:Coverage}
HST's field of view only covers part of most galaxies.  In this analysis, we assume that the SFHs measured from each ACS field are representative of the whole galaxy. It is important to acknowledge the potential limitations of this assumption. Our observations only cover a certain fraction of the galaxy's stellar mass (approximately equal to $f_{\star}$ in Table~\ref{Tab:Sample}, assuming small radial variations in mass-to-light ratio), typically close to the galaxy's center. 

Both theory \citep[e.g.,][]{El-badry16,Graus19} and observations \citep[e.g.,][]{Harbeck01,Tolstoy04,Battaglia06, Hidalgo13, Savino19a,Taibi22,Fu24a,Fu24b} find stellar population gradients in many low-mass galaxies.  SFHs measured from a limited field of view may therefore be biased relative to a SFH of the entire galaxy \citep{Graus19}. Given that our fields target the galaxy centers, and that the most recent star formation is typically more spatially concentrated in low-mass galaxies, the effect of incomplete spatial coverage is expected to be more important in our measurement involving old ages (e.g., median star formation epochs), and less impactful on estimates of the quenching epoch \citep{Graus19}. There is no clear way to eliminate this source of bias without a prohibitive investment of HST/JWST time.  Measuring SFHs over the full spatial extent of nearby galaxies is better suited for wide field-of-view space telescopes such as Euclid or Roman.

The effect of limited spatial coverage varies substantially across the sample. For 15 of our 36 dwarf galaxy targets, our ACS field contains more than 50\% of the total stellar mass, meaning that, while stellar population gradients can still impact our measurements, our SFH can still be taken as reasonably representative of the whole galaxy. For spatially larger galaxies, our ACS fields can contain as little as a few percent of the total stellar mass and stellar population gradients can have a larger impact. Many of our most massive galaxies have parallel WFC3/UVIS observations that can be used to constrain the presence of age gradients, which we plan to explore in  a forthcoming paper (Garling et al., in prep.).

Our limited spatial coverage is most impactful in our three auxiliary targets: M33, the GSS, and the M31 halo. These stellar systems have large angular sizes, can host important stellar population gradients \citep[e.g., ][see Appendix~\ref{App:M33}]{Williams09,Gilbert14,Conn16,Escala20,Escala21,Escala23}, and, in the case of M33, have a complex morphology. The SFHs presented in this paper should not be interpreted as a characterization of the global SFH of these systems. Rather, they represent the best that is possible given the piecemeal nature of archival HST data.  They are included to provide a homogeneously measured point of comparison to the properties of the satellite population.

\subsection{Identifying the Quenching Epoch}
\label{Sec:Quenching}
A particularly important timescale in a galaxy's SFH is its quenching epoch, i.e., the lookback time at which star formation activity ceased completely. However, extracting this information from CMD-based SFHs is surprisingly challenging \citep[e.g.,][]{Skillman17,Savino23}. Beyond physical mechanisms that initiate galaxy quenching, there are a number of observational contaminants that can introduce low levels of late star formation in CMD-based SFHs, which affect estimation of the quenching epoch.

\begin{figure*}[t]
    \centering
    \includegraphics[width=\textwidth]{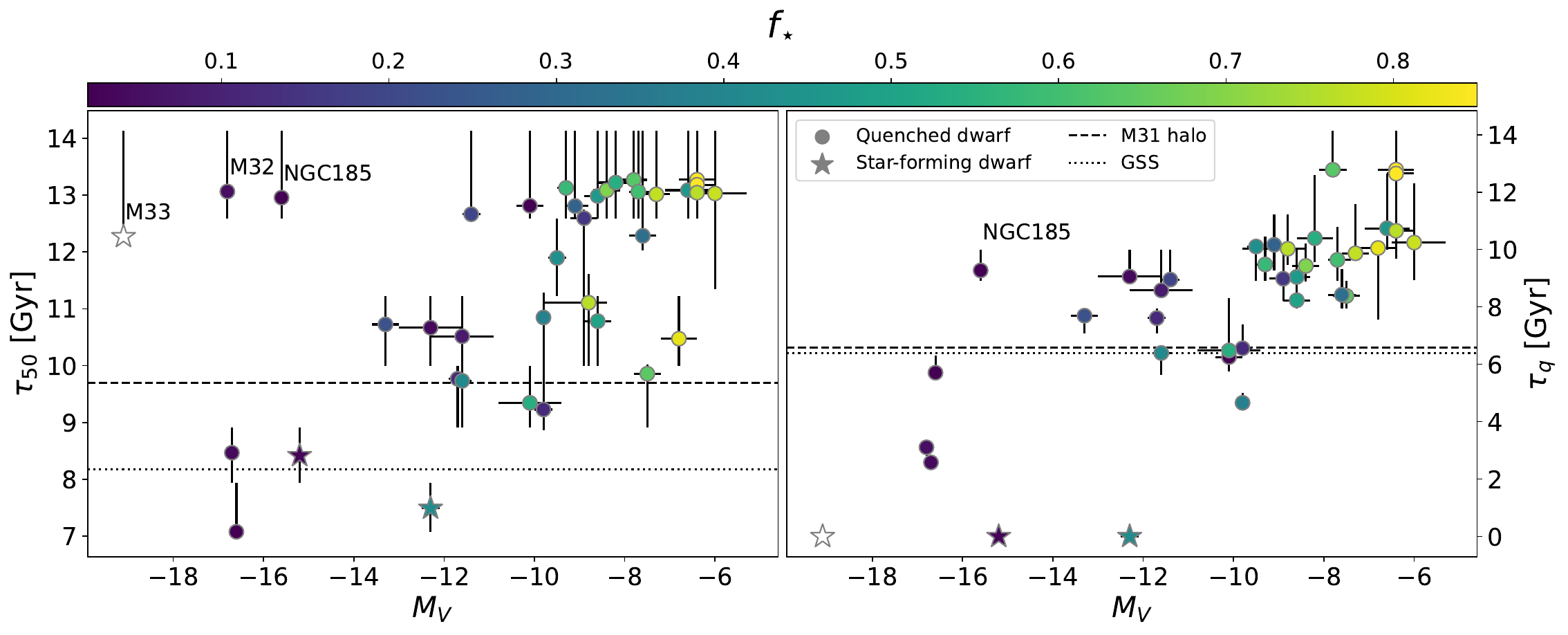}
    \caption{Median star formation epoch ($\tau_{50}$, left), and quenching epoch ($\tau_{q}$, right) as function of galaxy absolute magnitude (taken from \citealt{Savino22}). The symbols are color-coded by the amount of stellar light sampled by the ACS field ($f_{\star}$, reported in Table\ref{Tab:Sample}). M33 is shown as a white symbol. Star forming galaxies are shown as star symbols. Notable outliers from the general trend are highlighted by their name. The dashed and dotted line show the measured values for the M31 Halo and the GSS fields, respectively. }
    \label{Fig:Tq_Mv}
\end{figure*}

One such source are blue-straggler stars. On an optical CMD, they can mimic recent star formation activity at the level of a few percent of the total stellar mass \citep[e.g., ][]{Mapelli07,Momany07,Monelli12}. Other effects, such as imperfect modeling of the foreground/background population and residual photometric artifacts, can also introduce noise in the SFH, pushing the complete end of star formation to younger ages.

For these reasons, it is common practice to use metrics such as $\tau_{90}$ (i.e., the lookback time at which 90\% of the total star formation occurs), to approximately trace star formation quenching while mitigating the above concerns about contamination \citep[e.g.,][]{Weisz14b,Skillman17,Weisz19}. While this might be an appropriate choice for relatively luminous dwarfs, there is the possibility that it is not sufficiently conservative for low-mass dwarfs, as their sparse CMDs are more susceptible to contaminants such as background galaxies, especially if the SNR at the oMSTO is relatively low, as is the case for some of our faint systems.

To mitigate this, we also calculate and adopt the more conservative metric $\tau_{80}$ (i.e., the lookback time at which 80\% of the total star formation occurs).  Other studies have shown this to be a more robust tracer of quenching in the faintest galaxies such as UFDs \citep[e.g.,][]{Savino23,McQuinn24b}. The trade-off of a more conservative metric is that it can miss genuine late star formation in brighter dwarfs, which also tend to have more extended SFHs. 

Given that our galaxy sample spans $11$~magnitudes in luminosity, it is challenging to adopt a single quenching metric for all systems.  Instead, we  adopt the quenching metric $\tau_q$ defined as: i) $\tau_{90}$ for quenched dwarfs with $M_V<-8$, ii) $\tau_{80}$ for quenched dwarfs with $M_V>-8$, iii) 0 for those galaxies with active star formation (Peg~DIG, IC~1613, and M33). We find this to be a reasonable compromise to balance the purity versus completeness concerns outlined above. We realize that this choice introduces a degree of heterogeneity that might complicate comparisons with other results. Accordingly, in appendix~\ref{App:T8090} we show our results in terms of only $\tau_{80}$ or $\tau_{90}$. We provide all the star formation timescales discussed in this paper in Table~\ref{Tab:SFHs}.

\begin{figure*}
    \centering
    \includegraphics[width=\textwidth]{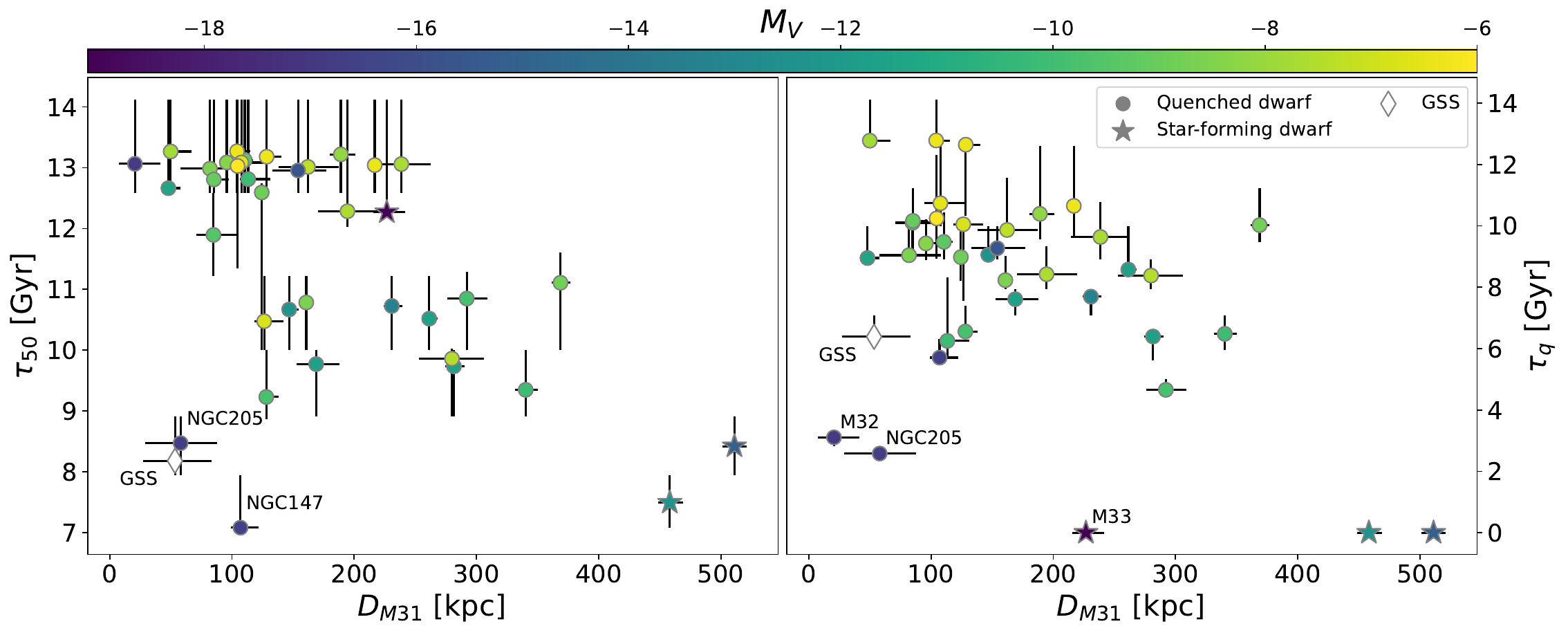}
    \caption{Median star formation epoch ($\tau_{50}$, left), and quenching epoch ($\tau_{q}$, right) as function of distance from M31 (taken from \citealt{Savino22}). The symbols are color-coded by the value of absolute luminosity. Star forming galaxies are shown as star symbols. Notable outliers from the general trend are highlighted by their name.}
    \label{Fig:Tq_DM31}
\end{figure*}

\subsection{Trends with Luminosity}
\label{Sec:Mv}
Fig.~\ref{Fig:Tq_Mv} shows the values of $\tau_{50}$ (i.e., the lookback time when 50\% of the total star formation is reached) and $\tau_q$, as a function of absolute luminosity (from \citealt{Savino22}) for our sample of 36 dwarf galaxies. Shown on the plot are also the measurements for M33 (white star), the GSS (dotted line), and the M31 halo (dashed line).

We find a clear correlation between $M_V$ and both $\tau_{50}$ and $\tau_q$ such that more luminous galaxies build their stellar mass over longer periods and quench at later times. The Pearson correlation coefficients are 0.46 for the trend with $\tau_{50}$ and 0.76 for the trend with $\tau_q$. Earlier studies of the 6 M31 ISLAndS satellites \citep{Skillman17} did not detect any significant trend, despite having nearly identical SFHs to ours. This highlights the increased statistical power of our larger sample.

The mixed degree of spatial coverage provided by HST complicates interpretation of these trends.
Given that all of our targets are at similar distances, there is a strong correlation between the galaxy's luminosity and $f_{\star}$. In the presence of significant age gradients, the stellar ages inferred for the bright galaxies in our sample would be biased young (due to the central position of our ACS fields), compared to the total SFH of the system \citep{Graus19}.

Despite this observational bias, there are two reasons to expect that this trend is driven by physics. First, such a correlation between galaxy luminosity and star formation duration is also seen in MW satellites \citep[e.g.,][]{Weisz14a,Skillman17} for which spatial selection effects are more complicated than a simple luminosity scaling. Second, the fact that more massive satellites tend to have more extended star formation is a well-established expectation that is widely seen in simulated galaxies \citep[e.g.,][]{Simpson18,Digby19,Garrison-kimmel19b,Applebaum21,Joshi21,Engler23}.

Interestingly, there are a few galaxies that deviate from this trend: M32, M33, and NGC185 do not fall on the same $\tau_{50}-M_V$ sequence traced by the other galaxies. It is likely that observational effects (e.g., spatial coverage issues outlined in \S~\ref{sec:Coverage}) play a non-negligible role for these systems. For example, significant stellar population gradients are observed in NGC~185 and M32 \citep[e.g.,][see also \citealt{Geha15} for a discussion on the SFH of NGC~185]{Worthey04,Rose05,Coelho09,Crnojevic14,Vargas14}. The situation is even more complex in M33, where our fields probe only a few lines of sight in this morphologically complex galaxy. In particular, our mass-weighted SFH mean is strongly driven by the central fields, that have older stars and much higher stellar densities. Additionally, our CMDs for M32 and M33 are substantially affected by crowding. This limits the photometric depth that we can reach to above the oMSTO. In turn, this makes the determination of the oldest ages (and in turn, $\tau_{50}$) less robust.

In spite of these issues, it is interesting to note   that these three galaxies are morphologically peculiar compared to the rest of the sample (1 spiral galaxy and 2 dwarf ellipticals). It is tempting therefore to speculate on the potential meaning of such old ages, should they be confirmed by future analyses. For instance, it has been suggested that M32 is the compact remnant of a much larger system that was recently accreted by M31 \citep[e.g.,][]{D'Souza18}. In this scenario, M32 itself would be more analogous to a bulge, rather than a genuine satellite galaxy, which would explain the very old median stellar age.

Finally, it is interesting, albeit quite speculative, to compare how the star formation timescales of the GSS and of the M31 Halo field compare to the satellite sample. Both $\tau_{50}$ and $\tau_q$ of these systems are compatible with those of satellites with $-17\lesssim M_V \lesssim -10$, corresponding to stellar masses of $10^6-10^9 M_{\odot}$. Taken at face value, these results could imply that both system progenitors had characteristics of moderately luminous systems, akin to bright classical dwarfs.

However, the interpretation is likely more complicated. Aside from the pencil beam nature of our ACS fields, the process of accretion itself can complicate direct comparisons between existing dwarf galaxies and stellar halo and/or stellar streams (e.g., the GSS). When a galaxy is accreted, star formation is effectively truncated, meaning that the resulting SFH will be biased towards older ages than if the system were not accreted in the first place. In fact, there is evidence that the GSS progenitor was a relatively massive system, with the stellar mass of the progenitor estimated to be $10^9-10^{10} M_{\odot}$ \citep[e.g.,][]{Fardal06,D'Souza18, Hammer18}.  The discovery of a metallicity gradient in the GSS \citep{Escala21} further implicates a progenitor galaxy massive enough to maintain such a gradient. Furthermore, given that a large fraction of the MW halo was formed by a single accreted system with stellar mass of $10^8-10^9 M_{\odot}$ \citep{Helmi18,Belokurov18,Mackereth20,Lane23}, it seems plausible that the M31 halo could have formed in a similar manner. This is also in accordance with expectations from CDM models, which predict that the stellar budget in haloes of present-day $L^*$ galaxies is dominated by a small number of massive progenitors \citep[e.g.,][]{Deason16}. A deep large area survey of M31's halo, e.g., with Roman, would provide more stringent constraints on this scenario.

\subsection{Trends with M31 Distance}
\label{Sec:DM31}
Fig.~\ref{Fig:Tq_DM31} shows the values of $\tau_{50}$ and $\tau_q$ as function of deprojected distance from M31, $D_{M31}$.  Here, we use the uniformly measured RR Lyrae-based distance from \citealt{Savino22}). 

We find that there is a clear trend between distance and both $\tau_{50}$ and $\tau_q$. The Pearson correlation coefficients are -0.52 for the trend with $\tau_{50}$ and -0.47 for the trend with $\tau_q$.  On average, satellites that are currently closer to M31 formed their stellar mass more rapidly and quenched at earlier epochs.

There are several outliers to these trends. M32, M33, NGC147, NGC205, and the GSS have substantially extended SFHs and late quenching, but are located close to M31. These targets are among the most luminous satellites of the M31 system. In fact, they represent the extreme examples of a more general dependence on luminosity. From Fig.~\ref{Fig:Tq_DM31}, it is evident that, at fixed $D_{M31}$, more luminous galaxies tend to have later quenching epochs, which is consistent with the results of \S\ref{Sec:Mv}. Given this scatter term introduced by the $\tau_q$-$M_V$ dependence, and given that satellite orbital motion can cause large variations of $D_{M31}$ over time, it is remarkable that a relatively clean correlation is observed in Fig.~\ref{Fig:Tq_DM31}.

This finding is similar to that of \citet{VandenBergh74}, who postulated a similar trend between satellite galactocentric distance and SFH duration among MW satellites. Even in M31, this was tentatively suggested with the smaller ISLAndS sample \citep{Skillman17}. However, these previous studies were limited by small samples.  Our paper firmly establishes this trend from a large and homogeneous sample of SFH measurements.

\begin{figure*}[t]
    \centering 
    \includegraphics[width=0.8\textwidth]{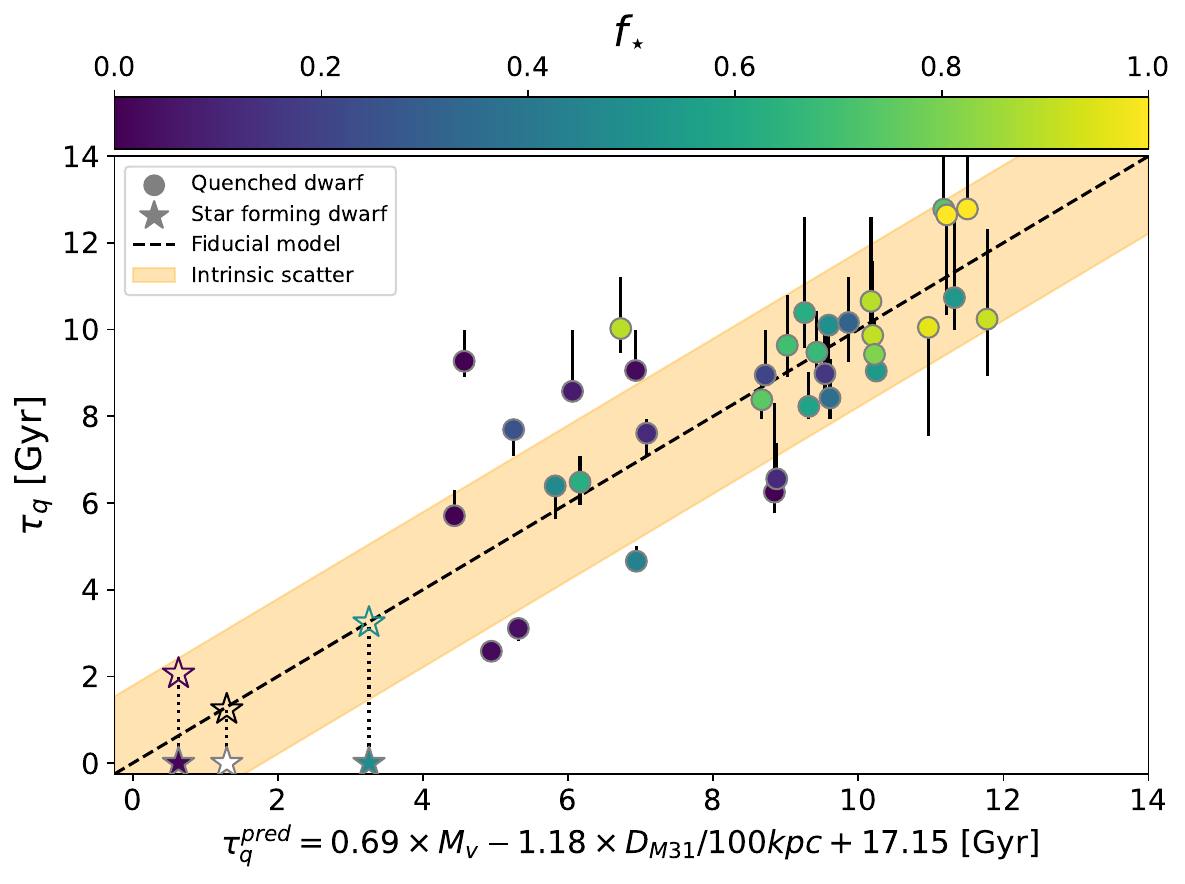}
    \caption{Comparison between the measured values of $\tau_q$ (y axis) and the values $\tau_{q}^{pred}$ predicted from eq.~\ref{Eq:FP} (x axis). The dashed line shows our fiducial model, while the shaded region shows the intrinsic scatter term of our model. The symbols are color-coded by the value of $f_{\star}$ (reported in Table\ref{Tab:Sample}). M33 is shown as a white symbol. Star forming galaxies are shown as star symbols. For the star forming galaxies, the empty star shows how the comparison would have looked like if $\tau_q\coloneqq\tau_{90}$ had been used, rather than $\tau_q\coloneqq0$.}
    \label{Fig:FP}
\end{figure*} 

The physical driver of this correlation is not entirely clear.  It may be the manifestation of accretion-driven quenching on the satellite population of M31. Some simulations suggest a correlation is expected between the typical orbital energy of satellites in $L^*$ halos and their infall epoch into the host halo \citep[e.g.,][]{Rocha12,Fillingham19}. Although an imperfect tracer of orbital energy, present-day galactocentric distance is also expected to loosely correlate with infall epoch (see, for example, Fig.~1 of \citealt{Rocha12}, or Fig.5 of \citealt{Santistevan23}). In this scenario, satellites at small $D_{M31}$ values would have been, on average, accreted by M31 at earlier times than satellites at large $D_{M31}$.  This would explain their older quenching epochs through early environmental perturbations. Satellites closer to M31 are also likely to have experienced more intense environmental effects e.g., through stronger tidal fields and/or more efficient ram pressure stripping, which also would contribute to more rapid quenching \citep[e.g.,][]{Samuel22,Samuel23}. 

Qualitatively, this scenario fits with other observed trends among low-mass galaxies in group environments, in that the morphological type, gas content, and star formation activity of dwarfs are strongly influenced by their distance to the massive host \citep[e.g.,][]{Blitz00,Grebel03,Karachentsev14,Putman21,Geha24}. In this regard, our results may be interpreted as an extension of these long-known environmental trends, suggesting that proximity to $L^*$ hosts not only plays a role in determining the star-forming/quenched status of low-mass satellites but also influences how early their star formation quenching is likely to occur.

Clearly, a limitation of this interpretation is that present-day galactocentric radius is not a perfect tracer of infall time or of the strength of environmental disturbances. Orbital parameters such as pericenter, apocenter, or orbital energy are likely to be better correlated with $\tau_{q}$, but this is not yet a feasible test, as only 5 satellites in our sample currently have published HST-based proper motions \citep[M33, NGC~147, NGC~185, \A{III}, and \A{VII};][]{VanderMarel19,Sohn20,Warfield23,Casetti-Dinescu24}. 

The outliers in Fig.~\ref{Fig:Tq_DM31} underscore the drawbacks of using only galactocentric distance. These are all massive stellar systems with evidence for recent or on-going star formation. While these galaxies are currently at small $D_{M31}$, it is likely that they have been accreted at recent times (e.g., the GSS is likely to have fallen in 1-2 Gyr ago;  \citealt{Fardal06,Fardal07,D'Souza18,McConnachie18,Dey23}). \citet{D'Souza18} also suggest that M32 could be associated with the same accretion event. 

From proper motion measurements, it seems that M33 is on a highly radial orbit and is either at its first or second pericenter passage \citep[e.g.,][]{Patel17,VanderMarel19}. NGC~205 has clear signs of tidal perturbations \citep[e.g.,][]{Choi02}, which could also indicate a relatively recent accretion. Overall, this interpretation is in line with cosmological expectations that, in $L^*$ hosts, massive satellites at small galactocentric radii usually fell in at later times \citep[e.g.,][]{Boylan-Kolchin11,Patel17,Santistevan23}.

\begin{figure*}
    \centering
    \plotone{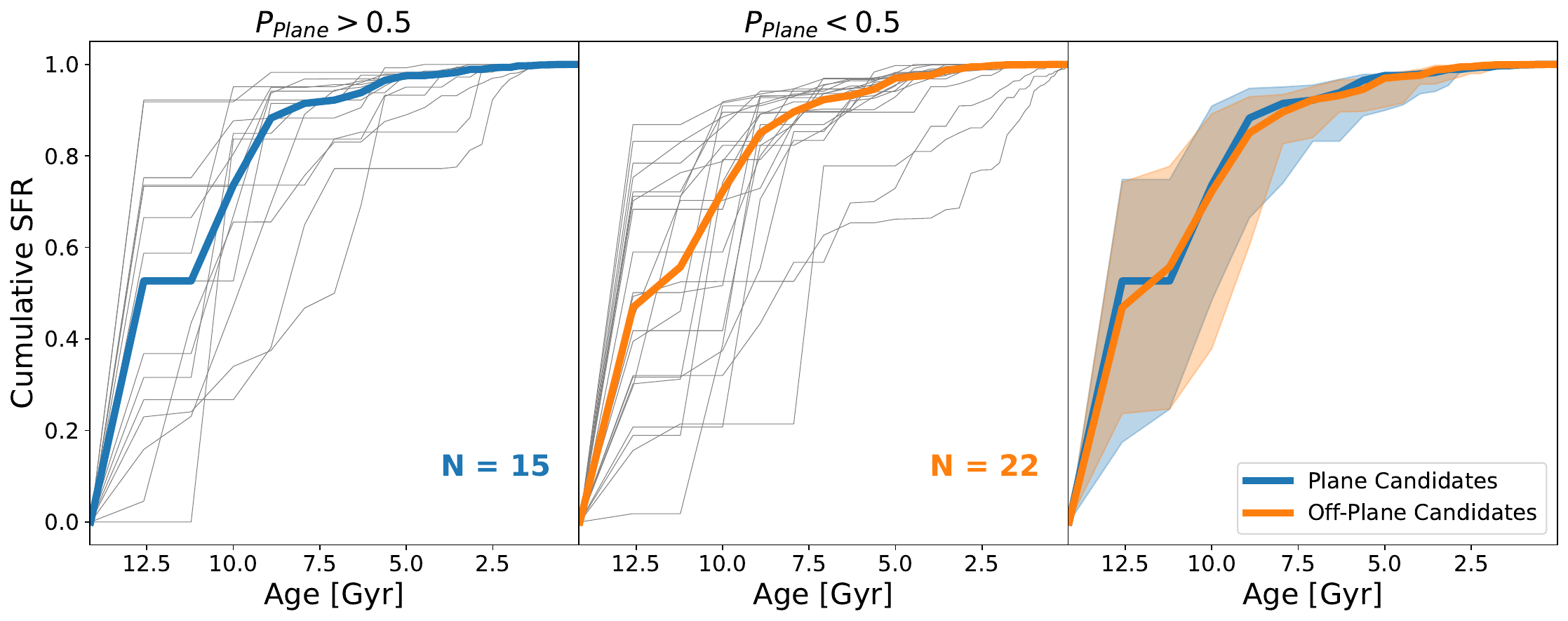}
    \caption{Median SFH for the GPoA candidate members (blue line, left panel) and out-of-plane candidates (orange line, middle panel). The grey lines show the SFH of individual galaxies. The right panel shows a direct comparison between the median SFH of the two samples. The shaded regions delimit that 68\% scatter around the median. Plane membership probabilities are taken from \citet{Savino22}.}
    \label{Fig:Plane}
\end{figure*}

\subsection{A Fundamental Predictor of Satellite Quenching?}
Motivated by clear correlations described in \S~\ref{Sec:Mv} and \ref{Sec:DM31}, we consider how well the quenching epoch in the M31 satellites can be predicted from only galactocentric distance and satellite luminosity (i.e., with no knowledge of past orbital properties).

We explore this question by fitting a linear model of the form:
\begin{equation}
    \tau_q=a\times M_V + b\times D_{M31} + c + \mathcal{N}(0, rms)
    \label{Eq:bivariate}
\end{equation}
to our M31 satellite data (our primary sample of 36 galaxies plus M33). We explore the parameter space of this model using the Markov chain Monte Carlo (MCMC) sampler \texttt{emcee} \citep{Foreman-Mackey13}. We use 16 walkers, a Gaussian likelihood function, and flat uninformative priors on the model parameters. We define convergence when the length of the MCMC chains is at least 50 times the autocorrelation length. Our fit converges on
\begin{align}
    \tau_q \,\, [Gyr]&= 0.69 (\pm 0.09) \times M_V - 1.18 (\pm 0.27) \times \frac{D_{M31}}{100~{\rm kpc}}\nonumber \\
    &+ 17.15 (\pm 1.14),
    \label{Eq:FP}
\end{align}
where the coefficients and uncertainties correspond to the median value and 68\% interval of our posterior distribution, respectively. The intrinsic $rms$ scatter inferred from the model is 1.8~Gyr. As a comparison, considering $M_V$ or $D_{M31}$ alone as independent variable, we obtain a scatter of 2.3 and 3.9 Gyr, respectively.

Fig.~\ref{Fig:FP} shows a comparison between the $\tau_q$ measured from our SFHs and the $\tau_{q}^{pred}$ predicted by eq.~\ref{Eq:FP}. Our model matches $\tau_q$ for nearly the entire sample, including the three star-forming galaxies (the agreement for the latter is even better if $\tau_q = \tau_{90}$ is used, instead of $\tau_q = 0$). Of our 37 sample galaxies, 11 fall slightly outside our $1\sigma$ scatter region, perfectly in line with statistical expectations.  All but one system (NGC~185) agree with this relation within $2\sigma$. The Pearson's correlation coefficient for this bi-variate relation is 0.85, which is a significant improvement compared to correlations of $\tau_q$ with $M_V$ or $D_{M31}$ alone.

Our finding in Fig.~\ref{Fig:FP} means that, within the M31 system, the position and the luminosity of a satellite can determine its quenching epoch to within less than 2~Gyr. As discussed in \S~\ref{Sec:Mv} and \ref{Sec:DM31}, this is likely interpreted in terms of environmental perturbations. At fixed luminosity, satellites closer to M31 are likely to have experienced stronger environmental perturbations at earlier epochs, therefore quenching at older times. At fixed distance to M31, more luminous galaxies are more resilient to quenching due to, e.g., a deeper potential well, resulting in more extended SFHs \citep[e.g.,][]{Samuel22,Samuel23,Santistevan23}.

It is remarkable that such a tight relation exists despite many effects that could introduce scatter into Fig.~\ref{Fig:FP}.  Fig.~\ref{Fig:FP} shows that most of the intrinsic scatter in our model is driven by those galaxies with the lowest values of $f_{\star}$ (darker symbols). This could be linked to the spatial coverage issues outlined in \S~\ref{sec:Coverage}, which primarily affect galaxies with higher luminosities. Additionally, galactocentric distance changes, due to orbital motion, smear the correlation along the $D_{M31}$ direction. This is particularly important in the inner halo where orbital timescales are shorter.

For these reasons, the 1.8~Gyr scatter should be interpreted as an upper limit on the correlation spread between satellite luminosity, orbital properties, and star formation quenching. When proper motions (e.g., with HST/JWST follow-up) and wide-field deep imaging (e.g., with Roman) become available for most M31 satellites, with the use of orbital integrals of motion (i.e., energy) and whole-galaxy SFHs, we expect to find an even tighter relationship.

Also interesting will be to investigate how the faint, mostly undiscovered, UFD population of M31 behaves with respect to eq.~\ref{Eq:FP}, especially those that will be discovered at large galactocentric distances. Our sample contains 6 relatively bright UFD galaxies. As discussed in \citet{Savino23}, these galaxies are likely to have been impacted by cosmic reionization but they were not ultimately quenched by it. The fact that these 6 galaxies are well described by the model of eq.~\ref{Eq:FP} might support the fact that their final quenching was caused by environmental effects. If fainter UFDs in the M31 system are true reionization fossils, we would expect them to significantly deviate from eq.~\ref{Eq:FP}. In particular, those UFDs in the outer halo of M31 are expected to have ancient quenching epochs, in spite of the much younger $\tau_q$ predicted by our model.

\begin{table}

\caption{Sample of dwarf galaxies, associated with the MW, used in this paper. We list absolute magnitudes and distance from the MW. References for the magnitudes, distances, and SFHs are also given.}
\hspace{-13mm}
\begin{threeparttable}
\begin{tabular}{llll}
\toprule
Galaxy ID & $M_V$ & $D_{MW}$ & Ref.\\
&mag &kpc&\\
\toprule
Bootes~I&$-6.02\pm0.25$&$64.0^{+2.5}_{-2.4}$&13, 4, 8\\
Canes Venatici~I&$-8.80\pm0.06$&$210.6^{+5.9}_{-5.7}$&13, 5, 10\\
Carina&$-9.43\pm0.05$&$107.2^{+5.5}_{-5.2}$&13, 11, 10\\
Draco&$-8.71\pm0.05$&$81.5^{+1.5}_{-1.5}$&13, 18, 10\\
Eridanus~II&$-7.21\pm0.09$&$371.7^{+8.6}_{-8.4}$&13, 15, 17\\
Fornax&$-13.22\pm0.1$&$144.6^{+3.2}_{-3.1}$&13, 16, 10\\
Leo~I&$-11.78\pm0.28$&$262.1^{+9.7}_{-9.3}$&13, 9, 10\\
Leo~II&$-9.74\pm0.04$&$235.9^{+14.4}_{-13.6}$&13, 3, 10\\
Leo~T&$-7.60\pm0.14$&$418.0^{+15.5}_{-14.9}$&13, 14, 10\\
NGC~6822&$-15.2\pm0.2$&$563.1^{+13.3}_{-13.0}$&7, 14, 10\\
Phoenix&$-9.9\pm0.4$&$409.2^{+41.5}_{-22.0}$&7, 6, 10\\
Sculptor&$-10.82\pm0.14$&$84.1^{+1.6}_{-1.5}$&13, 12, 10\\
Sagittarius& $-13.5\pm0.3$&$18.4^{+1.7}_{-1.9}$&7, 2, 10\\
Ursa Minor&$-9.03\pm0.05$&$77.8^{+3.6}_{-3.4}$&13, 1, 10\\
\toprule
\end{tabular}
\end{threeparttable}
\begin{tablenotes}
\leftskip=0pt
\raggedright
\item \hspace{-8mm} References: (1) \citet{Carrera02}; (2) \citet{Monaco04};\\ (3) \citet{Bellazzini05}; (4) \citet{Dallora06};\\ (5) \citet{Kuehn08}; (6) \citet{Battaglia12b};\\ (7) \citet{McConnachie12}; (8) \citet{Brown14};\\ (9) \citet{Stetson14}; (10) \citet{Weisz14a};\\ (11) \citet{Karczmarek15}; (12) \citet{Martinez-Vazquez15};\\ (13) \citet{Munoz18}; (14) \citet{Higgs21};\\ (15) \citet{Martinez-Vazquez21}; (16) \citet{Oakes22};\\ (17) \citet{Weisz23b}; (18) \citet{Bhardwaj24}
\end{tablenotes}
\label{Tab:MW}
\end{table}

\subsection{The Great Plane of Andromeda}
We use our data to search for differences in the SFHs of satellites that are reported to be members of the Great Plane of Andromeda \citep[GPoA, e.g.,][]{Koch06,Conn13,Ibata13,Pawlowski18,Santos-Santos20,Savino22} and those that are not. For this experiment, we select a plane-member sample using the plane model and the membership probability calculated in \citet{Savino22}. In particular we define as a plane-member candidate every galaxy with membership probability, $P_{Plane} > 0.5$, and off-plane candidates those galaxies with $P_{Plane} < 0.5$. 

Fig.~\ref{Fig:Plane} shows the cumulative SFHs of the plane and off-plane samples (grey lines). For the two samples, we also calculate the median SFH (thick line). We find that both plane and  off-plane samples contain galaxies with a range of SFHs, from mostly old to those with nearly constant SFHs. The right panel overplots the median SFHs from on- and off-plane samples, revealing they are virtually identical.

Given that the significance and stability of satellite planes are still debated \citep[e.g.,][]{Fernando17,Fernando18,Boylan-Kolchin21,Pawlowski21a,Pawlowski21b,Samuel21,Sawala23,Pham23}, the similarity of SFHs in the plane and off-plane satellites of M31 provides additional perspective beyond positions and velocities. Had the typical SFHs of on- and off-plane satellites been systematically different from each other, such result would have provided support to the GPoA being a physically distinct structure. The similarity of the SFHs in the two samples is not sufficient to rule out the significance of the GPoA, especially in light of its intriguing degree of kinematical coherence \citep[e.g.,][]{Ibata13,Sohn20}\footnote{This is primarily observed through radial velocity measurements, although two putative plane members, NGC~147 and NGC~185 have published proper motions \citep{Sohn20} that seem to support a certain degree of coorbitation.}. However, if future proper motion measurements of a larger sample of M31 satellites reveal that the GPoA is indeed a long-lived corotating structure, our result indicates that its formation origin has left no appreciable imprint in the SFH of its constituent galaxies.

 A popular hypothesis for the origin of the GPoA is that the plane is the result of accretion of a relatively large galaxy with its own satellite system or of a small galaxy group \citep[e.g.,][]{Ibata13,Hammer13,Ibata14,Angus16}. It is not clear whether this scenario would result in distinct star formation patterns among the plane members but any model that made such a prediction would be in tension with the results of Fig.~\ref{Fig:Plane}. On the other hand, scenarios in which the planar configuration is dynamically induced in a subset of long-term M31 satellites might be more compatible with the data.

\begin{figure*}
    \centering
    \plotone{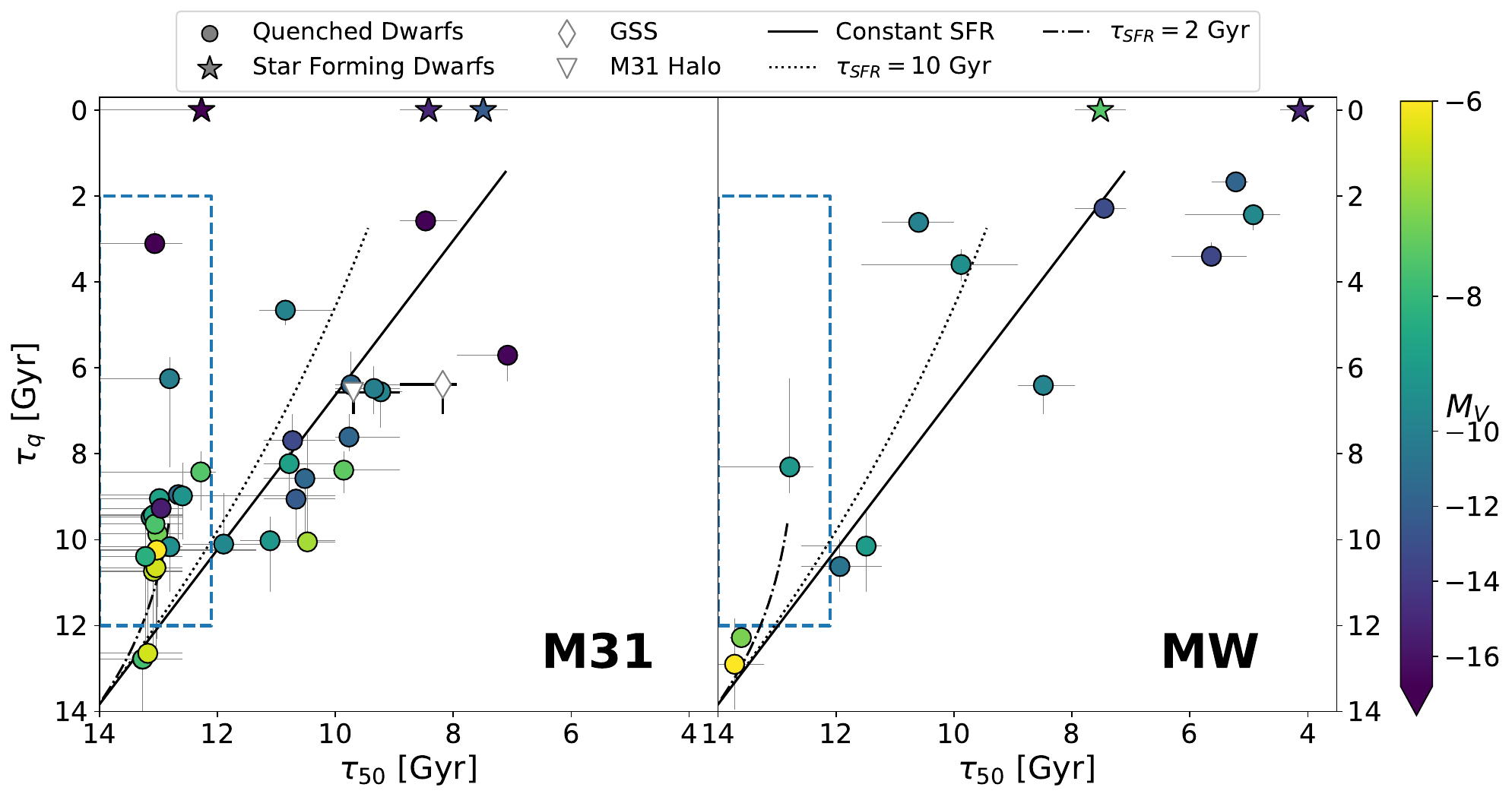}
    \caption{Distribution of $\tau_{50}$ vs $\tau_{q}$ for the dwarf galaxies associated with M31 (left) and the MW (right). Only galaxies with $M_V<-6$ are shown in this plot. The symbols are color-coded by absolute magnitude. Quenched galaxies are shown as circles while star-forming galaxies are shown as stars. The position of the GSS field and the M31 Halo field are also reported in these plots (white diamond and white triangle, respectively). Over-plotted are also loci corresponding to constant star formation (solid line) and exponentially declining star formation with timescales of 10~Gyr (dotted line) and 2~Gyr (dash-dotted line). The tracks correspond to sequences of increasingly younger star formation truncation. The blue dashed line delimits the region of old $\tau_{50}$ and intermediate-age $\tau_q$, where roughly half of the M31 sample is located.}
    \label{Fig:T50Tq}
\end{figure*}

\begin{figure}
    \centering
    \subfloat
        {\includegraphics[width=\linewidth]{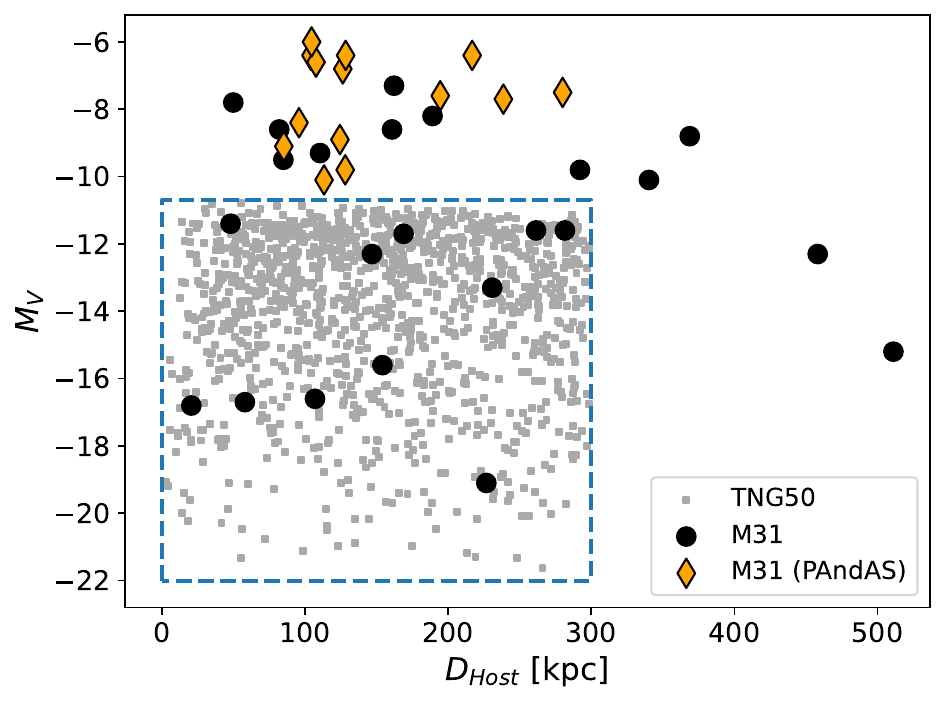}}\quad 
    \subfloat
        {\includegraphics[width=\linewidth]{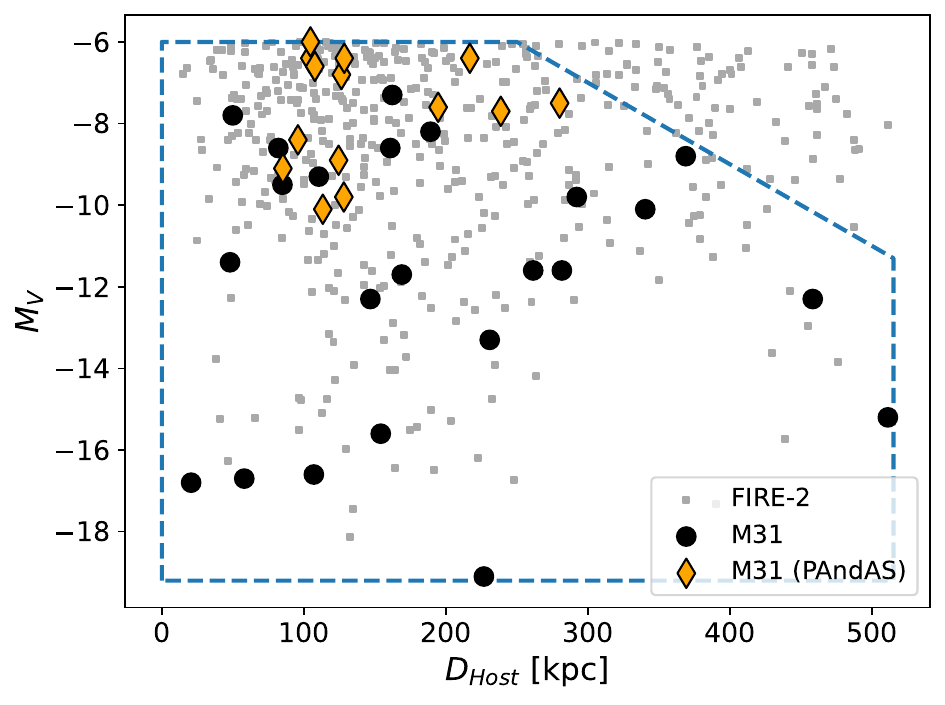}}
    \caption{Distribution of our M31 satellite sample in the galactocentric distance vs absolute luminosity parameter space. Satellites discovered by PAndAS are highlighted as orange diamonds. The 579 simulated satellites from TNG50 (top panel) and the 466 simulated satellites from FIRE-2 (bottom panel) are shown as gray squares. The region enclosed by the blue dashed line shows the parameter range in which the observation vs simulation comparison is performed. }

    \label{Fig:Selection}
\end{figure}

\begin{table*}
\caption{Fiducial model coefficients for eq.~\ref{Eq:bivariate}, when fit to the different satellite samples of this paper. The corresponding satellite sample size is also reported.}
\begin{tabular}{lrrrrr}
\toprule
Sample & N & a & b & c & rms\\
&&&&&Gyr\\
\toprule
M31 (full)&37&$0.69\pm0.09$&$-1.18\pm0.27$&$17.15\pm1.14$&$1.79^{+0.26}_{-0.21}$\\
M31 (TNG50 Matched)&11&$0.92\pm0.28$&$-0.36\pm0.89$&$19.89^{+4.73}_{-4.62}$&$2.28^{+0.81}_{-0.52}$\\
MW (M31 Matched) &14&$0.93\pm0.44$&$-0.84\pm0.68$&$16.62^{+4.58}_{-4.54}$&$3.98^{+1.12}_{-0.76}$\\
TNG50&579&$1.19\pm0.03$&$-0.87\pm0.10$&$24.25\pm0.53$&$2.04\pm0.06$\\
FIRE-2 (M31 Matched)&394&$1.16\pm0.04$&$-0.30\pm0.11$&$20.09\pm0.37$&$2.12\pm0.06$\\
\toprule
\end{tabular}
\label{Tab:Planes}
\end{table*}

\begin{figure*}
    \centering
    \plotone{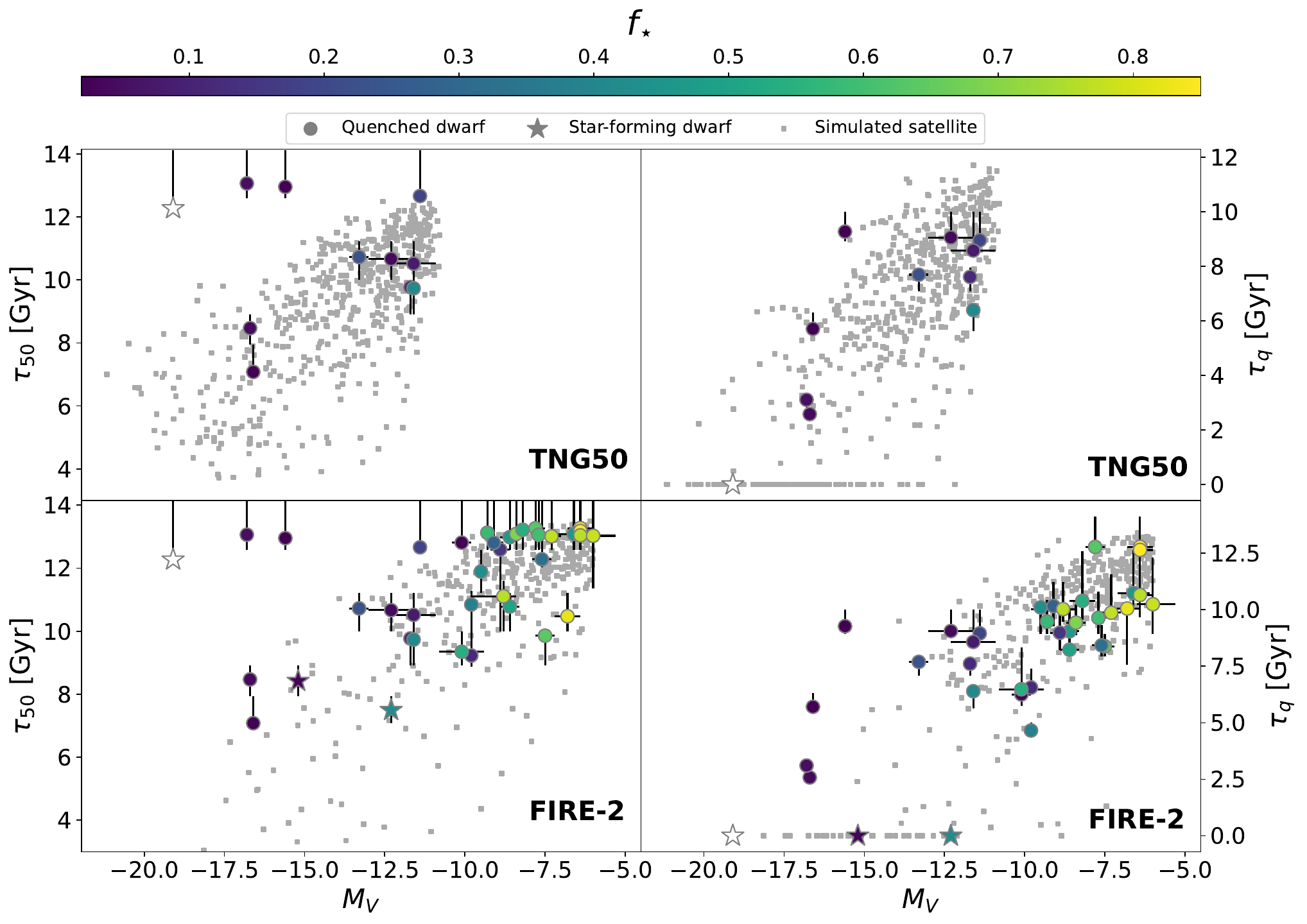}
    \caption{Trends of $\tau_{50}$ (left) and $\tau_{q}$ (right) with absolute luminosity in our M31 satellite sample. Circles represent quenched dwarfs, while stars represent star-forming dwarfs. Squares show 579 simulated satellite dwarfs from the TNG50 simulation \citep[top]{Engler23} and 394 simulated dwarfs from the FIRE-2 simulation \citep[bottom]{Wetzel23}. The M31 dwarfs are color-coded by value of $f_{\star}$. In the top panels, only the M31 satellites matched to the TNG50 selection ($M_V<-10.7$, $D_{Host}<300$~kpc) are shown.}
    \label{Fig:Sims_Lum}
\end{figure*}

\subsection{Comparison with the Milky Way}
We now compare the star formation timescales, and their trends with satellite properties, in the M31 and MW systems.  To do this, we measure $\tau_{50}$ and $\tau_q$ (as defined in \S~\ref{Sec:Quenching}) of MW satellites from the literature SFHs of \citet{Weisz14a}. As with the M31 satellites, we set $\tau_q = 0$ for the star forming dwarfs (NGC~6822 and Leo~T). We only consider MW satellites with $M_V\leq-6$ and within 500~kpc from the MW, to mirror the properties of our M31 sample. We supplement this sample with the SFH of Boo~{\sc I} from \citet{Brown14} and that of Eri~{\sc II} from \citet{Weisz23b}. We also adopt absolute luminosities and galactocentric distances from the literature. The resulting set of 14 MW satellites, listed in Table~\ref{Tab:MW}, represents the almost complete set of known dwarfs (with $M_V\leq-6$) in the proximity of the MW, with the exceptions being the Sextans dSph and the Magellanic Clouds. 

We find that the star formation timescales in the satellites of the MW follow similar trends with luminosity and galactocentric distances to what observed in the M31 system, albeit at a lower statistical significance due to higher intrinsic scatter and the smaller sample size. Fitting the same model of eq.~\ref{Eq:bivariate} to the MW satellites, we obtain:
\begin{align}
    \tau^{MW}_q \,\, [Gyr]&= 0.93 (\pm 0.44) \times M_V - 0.84 (\pm 0.68) \nonumber \\
    & \times \frac{D_{MW}}{100~{\rm kpc}} + 16.62^{+4.58}_{-4.54},
    \label{Eq:FP_MW}
\end{align}
with an intrinsic scatter of 3.98 Gyr. The large uncertainties in this fit are mainly driven by the limited number of data points. Within the uncertainties, this result is broadly compatible with the trend reported in eq.~\ref{Eq:FP}.

As a second comparison, figure~\ref{Fig:T50Tq} shows the distribution of the M31 and the MW satellite systems in the $\tau_{50}$-$\tau_q$ parameter space. There are several interesting trends in this comparison. First, is that M31 hosts a much larger number of satellites in this luminosity range. This is likely due at least in part to the higher halo mass of M31 \citep[e.g.,][]{Perrett02,Penarrubia14,Patel23}. Including the dwarfs not in our samples, there are 17 known galaxies with $M_V\leq-6$ within 500~kpc from the MW, whereas the number is 38 for M31\footnote{Although it is likely that the M31 satellite census is much less complete than in the MW, especially at low luminosities and large galactocentric radii \citep[e.g.,][]{DolivaDolinsky22,DolivaDolinsky23}.}. This abundance of low mass galaxies highlights the value of our SFH sample for the improvement of low-mass galaxy formation models.

\begin{figure*}
    \centering
    \plotone{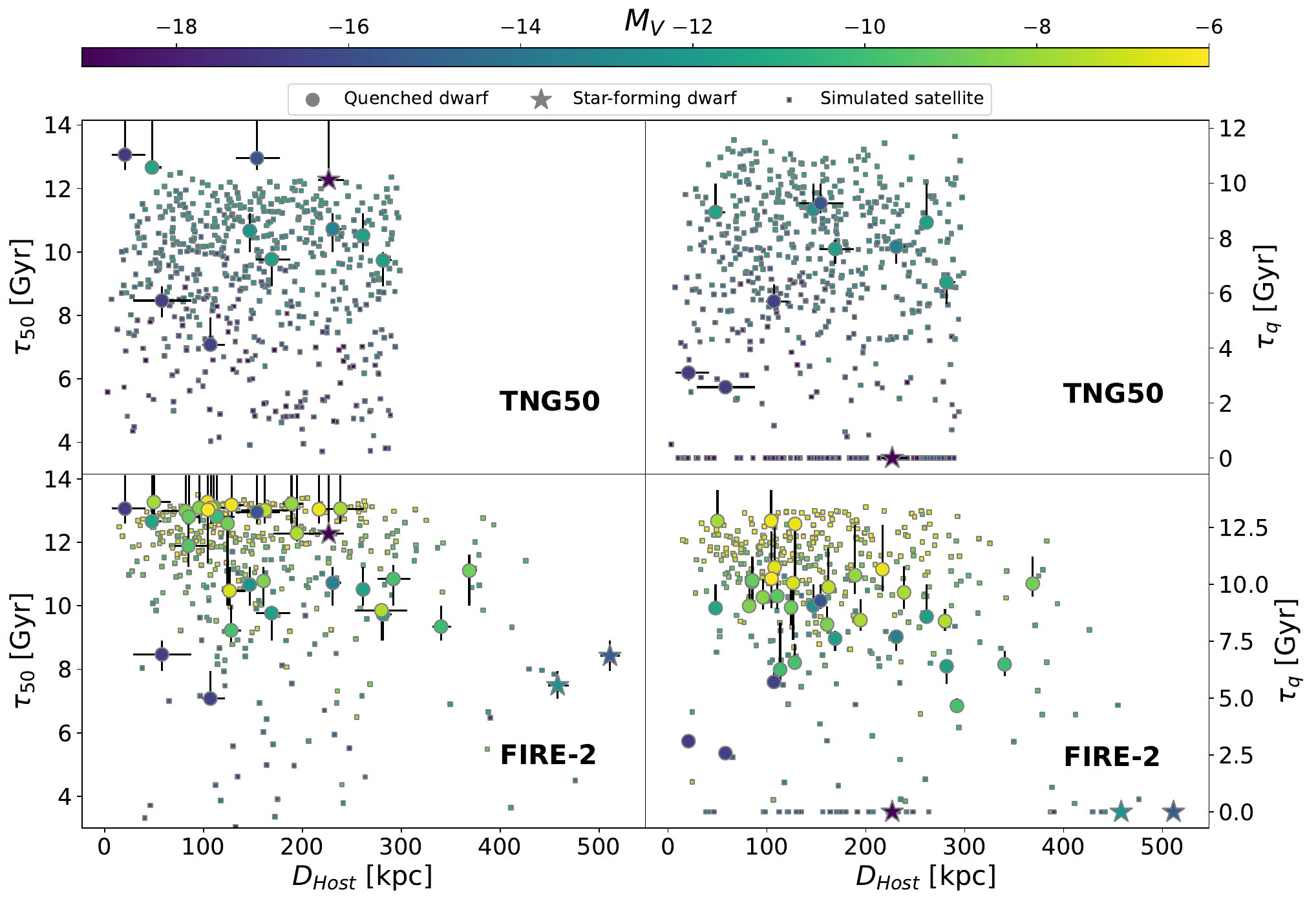}
    \caption{Trends of $\tau_{50}$ (left) and $\tau_{q}$ (right) with galactocentric distance in our M31 satellite sample. Circles represent quenched dwarfs, while stars represent star-forming dwarfs. Squares show 579 simulated satellite dwarfs from the TNG50 simulation \citep[top]{Engler23} and 394 simulated dwarfs from the FIRE-2 simulation \citep[bottom]{Wetzel23}. The M31 and simulated dwarfs are color-coded by absolute luminosity. In the top panels, only the M31 satellites matched to the TNG50 selection ($M_V<-10.7$, $D_{Host}<300$~kpc) are shown.}

    \label{Fig:Sims_Dist}
\end{figure*}

Second, the MW hosts a small number of satellites with moderate luminosities ($-10>M_V>-12$), extended SFHs ($\tau_{50}<6$ Gyr), and recent quenching ($\tau_q<3$ Gyr). While M31 hosts a few bright galaxies that quenched at such recent times (i.e., M32 and NGC~205) it does not have recently quenched galaxies at lower luminosities (except, possibly, Psc~{\sc I}). This was already noted in other studies of the M31 satellite population \citep[e.g.,][]{Skillman17,Weisz19}.

Finally, before their quenching, most of the satellites in the MW system are reasonably well-described by SFH parametrizations with constant (solid line in Fig.~\ref{Fig:T50Tq}) or slowly declining (dotted line) star formation rates. This is also the case for approximately half of the M31 satellites.

However, the other half of the M31 sample falls close to a track of rapidly declining star formation rate (dot-dashed line), with very old $\tau_{50}$ ($>12$~Gyr ago) and intermediate age $\tau_q$ ($8-10$~Gyr ago). While there is no obvious structural or morphological characteristic that sets this group of satellites apart from the rest of the M31 system, these galaxies seem to be characterized by lower typical luminosities (median $M_V$ of $-8.5$,  compared to $-10.4$ for the rest of the M31 system) and higher radial concentration (median $D_{M31}$ of 111~kpc, versus 168~kpc for the rest of the M31 system). However, this group of galaxies also contains higher luminosity, more peculiar objects, such as NGC~185 and M32.

\begin{figure}
    \centering
    \includegraphics[width=0.5\textwidth]{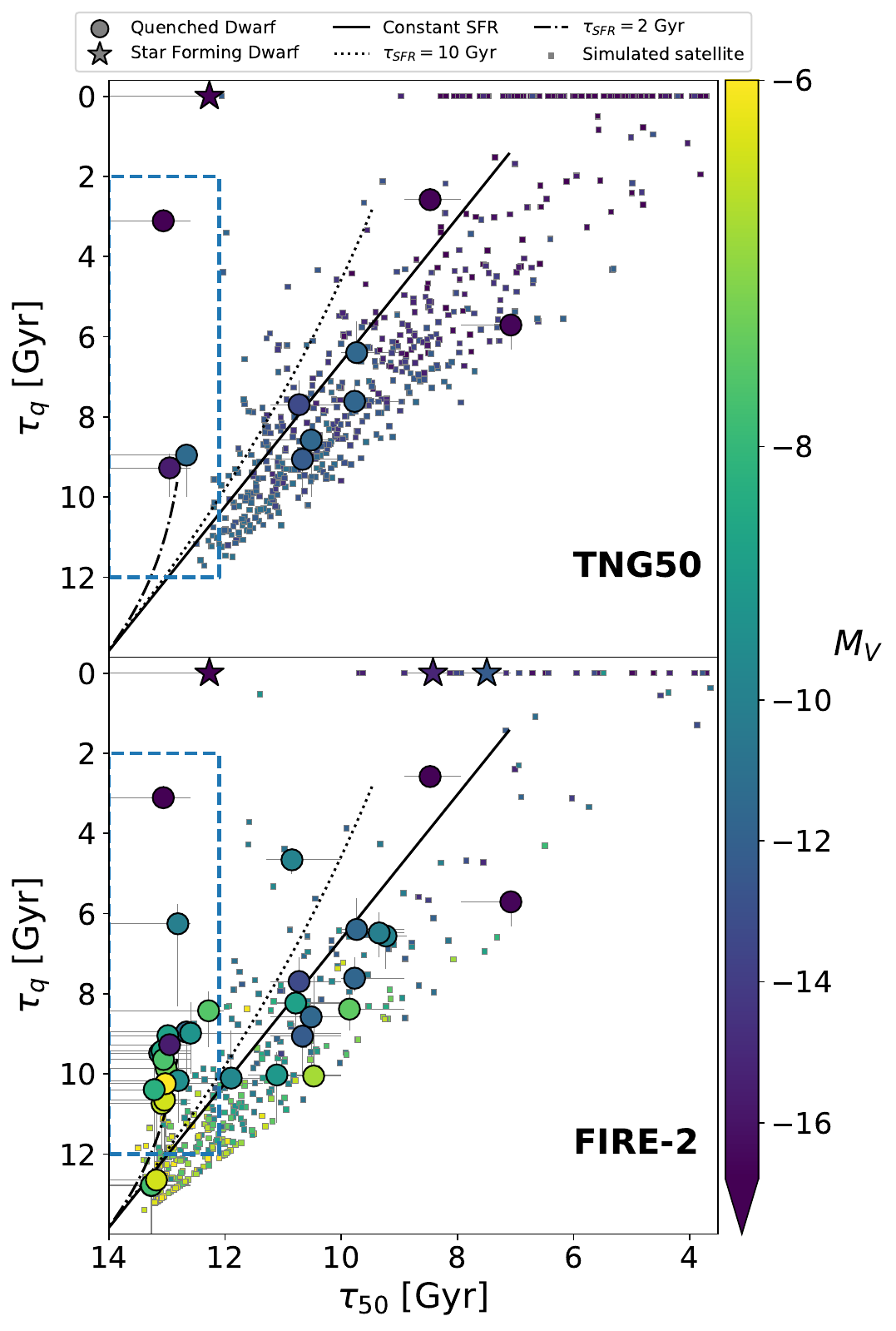}
    \caption{Values of $\tau_{50}$ and $\tau_{q}$ for our M31 satellite sample. Circles represent quenched dwarfs, while stars represent star-forming dwarfs. Squares show 579 simulated satellite dwarfs from the TNG50 simulation \citep[top]{Engler23} and 394 simulated dwarfs from the FIRE-2 simulation \citep[bottom]{Wetzel23}. Over-plotted are also loci corresponding to constant star formation (solid line) and exponentially declining star formation with timescales of 10~Gyr (dotted line) and 2~Gyr (dash-dotted line). The blue dashed line delimits the region of old $\tau_{50}$ and intermediate-age $\tau_q$, where roughly half of the M31 sample is located. In the top panel, only the M31 satellites matched to the TNG50 selection ($M_V<-10.7$, $D_{Host}<300$~kpc) are shown.}
    \label{Fig:Sims_2}
\end{figure}

Inspection of the cumulative SFHs in Fig.~\ref{Fig:SFHs} reveals that some of these galaxies (e.g., \A{V}, \A{XV}) experienced a strong initial episode of star formation and then a rapid decline, but sustained low levels of star formation for several Gyr, which leads to a relatively delayed $\tau_q$. These patterns may be seen even more clearly in the differential SFHs presented in Appendix~\ref{App:SFR}.

However, several other galaxies in this group (e.g., \A{I} or  \A{XXI}), do not have a rapidly declining SFH, but rather they show a strong initial star formation burst followed by a lull, and then re-ignition of high levels of star formation. Alternating phases of star formation and quiescence are in fact relatively common across the whole M31 satellite sample. Roughly 30\% of our M31 satellite sample show signs of this episodic star formation activity early on in their evolution. These galaxies are present in both M31 subpopulations of Fig.~\ref{Fig:T50Tq}.

Even within the MW system there are dwarfs that present evidence of intermittent star formation activity (e.g., Fornax, Carina, and Leo~{\sc I}; \citealt{deBoer14,Savino15,Ruiz-Lara21,Rusakov21}). However, these galaxies are located in a different region of parameter space. Aside from CVn~{\sc I}, the MW has no other satellites with ancient $\tau_{50}$ and intermediate-age $\tau_{q}$, similar to what we find in M31. Given how many of these satellites are observed in M31, and the size of the respective satellite populations, it is unlikely that this dearth in the MW is the effect of low-number statistics. Instead, it is very likely to be a true difference between the two systems that warrants a deeper investigation with theoretical models.

\subsection{Comparison with Simulations}
It is instructive to compare the satellite population trends we find in M31 with predictions from galaxy formation models. To this end, we use two sets of simulated satellite populations. The first, described in \citet{Pillepich24} and \citet{Engler23}, belongs to the TNG50 simulation \citep{Nelson19,Pillepich19}. In particular, we use 579 dwarf galaxies belonging to 60 M31-like ($10^{10.9}M_{\odot}<M_{\star}<10^{11.2}M_{\odot}$; $10^{12.1}M_{\odot}<M_{200c}<10^{12.9}M_{\odot}$) hosts. The second dataset consists of simulated satellites from the FIRE-2 suite of simulations \citep{Wetzel23}. The FIRE-2 cosmological zoom-in simulations of galaxy formation are part of the Feedback In Realistic Environments (FIRE) project, generated using the Gizmo code \citep{Hopkins15} and the FIRE-2 physics model \citep{Hopkins18}. The satellites we use belong to 14 host galaxies with $10^{10.3}M_{\odot}<M_{\star}<10^{11.1}M_{\odot}$ and $10^{11.9}M_{\odot}<M_{200m}<10^{12.4}M_{\odot}$ \citep{Wetzel16,Garrison-Kimmel17a,Garrison-kimmel17b,Hopkins18,Garrison-kimmel19b,Samuel20}. These halos host 466 dwarf satellites brighter than $M_{V}=-6.0$ and closer than $D_{Host}=515$~kpc (the approximate limits of our M31 sample). Because the simulated galaxies do not have the contamination issues discussed in \S~\ref{Sec:Quenching}, in both simulations we define $\tau_q = \tau_{90}$ at all magnitudes, except for those satellites with active star formation in the last 200~Myr, for which $\tau_q$ is set to 0.

It is important that the observed and simulated samples are selected to be as consistent to each other as possible. Figure~\ref{Fig:Selection} shows the distribution, in the luminosity versus galactocentric distance parameter space,  of our M31 satellite sample against the TNG50 and FIRE-2 satellites, where absolute luminosity of the FIRE-2 satellites is derived from their total stellar mass, assuming a mass-to-light ratio of 2.

As discussed in \citet{Pillepich24} and \citet{Engler23}, the TNG50 satellites have been selected to be within 300~kpc from their main host, and have absolute magnitudes that go down to $M_V<-10.7$, whereas our M31 sample contains both fainter satellites and satellites at larger distances. Therefore, when comparing to the TNG50 data, we apply the same selection criteria to our M31 sample, which results in a smaller sample of 11 galaxies. On the contrary, the FIRE-2 sample and the M31 sample span a very comparable range in terms of luminosities and host distances. The FIRE-2 sample contains a population of faint dwarfs at large galactocentric distances that are not present in the M31 sample. It is likely that this is a manifestation of detection biases in our M31 satellite sample. The faintest satellites in our sample have been discovered in the PAndAS dataset \citep{Martin06,Ibata07,McConnachie09,Martin09,Richardson11}. While the survey only extends to projected distances of $\sim150$ kpc from M31, the de-projected distances of these faint satellites are as high as $\sim 300$~kpc. Beyond this distance there a no known M31 satellites with $M_{V}>-8$. To mitigate this potential selection effect, we remove from the FIRE-2 sample faint satellites at large distances, as illustrated in Fig.~\ref{Fig:Selection}. With this selection, the sample is reduced from 466 to 394 simulated dwarfs.

Fig.~\ref{Fig:Sims_Lum} and Fig.~\ref{Fig:Sims_Dist} show the trends of $\tau_{50}$ and $\tau_{q}$ with absolute luminosity and galactocentric distance in our M31 sample and in the simulated satellites. Within the ranges of luminosities probed by the simulations, median age and quenching epoch show a similar trend with absolute magnitude as that observed around M31 (with the exception of M32, M33, and NGC205, whose caveat we discussed in \S~\ref{Sec:Mv}). As already discussed, the scaling of star formation duration with galaxy mass is a well-established prediction of formation models \citep[e.g.,][]{Simpson18,Digby19,Garrison-kimmel19b,Applebaum21,Joshi21,Engler23}. 

The comparison of satellite properties as a function of galactocentric distance is more nuanced. From Fig.~\ref{Fig:Sims_Dist}, it is clear that the quenching epochs in the simulations, especially in TNG50, are distributed much more evenly with galactocentric radius compared to the tight correlation we observe in M31. On the other hand, the simulated samples contain a much larger number of bright galaxies, and the aforementioned correlation between $\tau_q$ and $M_V$ might be playing a large role in increasing the scatter in the simulated sample.

At the faint end, the FIRE-2 sample contains a substantial number of faint galaxies that have large host distances ($D_{Host}\gtrsim200$kpc) and very early quenching ($\tau_q\gtrsim 10$~Gyr), which are not present in the M31 sample. As we illustrate in Fig.~\ref{Fig:Selection}, the FIRE-2 sample has been restricted to distances and luminosities spanned by the known M31 satellites, so it is unlikely that selection effects alone can justify this discrepancy.

Fitting the same bi-variate model of eq.~\ref{Eq:bivariate} to the TNG50 and FIRE-2 data, we obtain the fiducial solutions:
\begin{align}
    \tau^{TNG50}_q \,\, [Gyr]&= 1.19 (\pm 0.03) \times M_V - 0.87 (\pm 0.10)   \nonumber \\ 
    &\times \frac{D_{Host}}{100~\text{kpc}}  + 24.25 (\pm 0.53),
\label{eq:TNG50}
\end{align}
and
\begin{align}
    \tau^{FIRE}_q \,\, [Gyr]&= 1.16(\pm0.04) \times M_V - 0.30(\pm 0.11)   \nonumber \\ 
    &\times \frac{D_{Host}}{100~\text{kpc}}  + 20.09 (\pm 0.37),
\label{eq:FIRE}
\end{align}
with $rms^{TNG50}$ and $rms^{FIRE}$ of 2.04 and 2.12 Gyr, respectively. Interestingly, these intrinsic scatter terms are very close to what we measure in the M31 data. For ease of comparison, the fiducial model coefficients for the M31, MW, and simulated satellites are summarized in Table~\ref{Tab:Planes}.

Examining eq.~\ref{eq:TNG50} and eq.~\ref{eq:FIRE}, both simulation suits predict a slightly stronger dependence between satellite quenching epoch and absolute luminosity than observed in the M31 sample. As for the galactocentric distance term, a very weak correlation is observed in the FIRE-2 sample, while the TNG50 sample shows a similar trend as observed in M31, albeit with a shallower dependence. In principle, the TNG50 results should be compared to the subset of M31 satellites with the same selection criteria. However, due to the small sample size of such subsample, the resulting fit uncertainties are too large for a meaningful comparison (see Table~\ref{Tab:Planes}). Within recent studies of simulated MW/M31 analogs, claims of trends in the SFH duration with the host distance have been mixed \citep{Simpson18,Digby19,Joshi21,Engler23,Christensen24}. Reproducing the luminosity/distance scaling relations observed in M31 is a promising path forward to obtaining a high-fidelity population of simulated $L^*$ satellite systems.

Finally, Fig.~\ref{Fig:Sims_2} shows the M31 and simulated samples in the $\tau_{50}$-$\tau_{q}$ parameter space. The simulated satellites form a relatively tight sequence, close to the locus of constant star formation. This is the same region spanned by roughly half of the M31 satellites (and by most MW satellites). However, the other subpopulation of M31 satellites (blue dashed box in Fig.~\ref{Fig:Sims_2}) has very few counterparts in the FIRE-2 sample and virtually no analog in the TNG50 sample. 

We note that some galaxy formation models \citep[e.g.,][]{Kravtsov22} produce a low-mass galaxy population that lies in the same region of the $\tau_{50}$-$\tau_q$ space spanned by this second population of M31 satellites. In the \citet{Kravtsov22} models, these galaxies only emerge due to reionization feedback. It could therefore be that we are witnessing a population of galaxies that, while not quenched by reionization, had their SFH significantly altered due to heating from the cosmic ultraviolet background. A more in-depth comparison with formation models is warranted to assess whether, e.g., the stellar mass range and the other properties of this sample are compatible with such a scenario.

\section{Conclusions}
\label{sec:conclusions}

In this paper, we presented deep optical imaging from the Cycle 27 HST Treasury Survey of M31 Satellites (GO-15902; PI Weisz). By combining new and archival observations of $>1000$ orbits, we have assembled homogeneous photometric catalogs for 36 dwarf galaxies within 500~kpc from M31 (90\% of the currently known satellite system), along with 10 fields in the M31 halo, the GSS, and M33. 

Through detailed modeling of their CMDs we measured lifetime SFHs for the entire sample.  This large sample allows us to analyze star formation trends across the population of M31 satellites. We find that:

\begin{enumerate}
\item The median star formation epoch and the quenching epoch in the M31 satellites are correlated with galaxy luminosity, such that brighter satellites have more extended SFHs. This confirms similar trends observed in the MW system and is in accordance with expectations from galaxy formation models.

\item The median star formation epoch and quenching epoch also correlate with present-day distance from M31, with galaxies at larger radii having, on average, more extended SFHs. We interpret this trend in terms of infall-related quenching, with satellites at small radii being more likely to have been accreted at earlier times. A few notable outliers from this trend are massive stellar systems such as M32, M33, and NGC205, which supports other indications that these systems may have fallen in recently.

\item Combined, galaxy luminosity and present-day distance from M31 are enough to predict satellite quenching to within 1.8~Gyr. The existence of this tight correlation, despite the limitations in the data (e.g., field of view effects), highlights the fundamental connection between satellite halo mass, environmental history, and star formation duration.

\item There appears to be no meaningful difference between the typical SFH of satellites in the GPoA and the that of the rest of the satellite system. This result is not clearly aligned with a scenario in which a long-lived plane of satellites was formed through coherent accretion of a large number of dwarfs.

\item Notable differences between the satellite system of M31 and that of the MW are (a) the lack of low-mass, late quenching satellites in the former, and (b) around M31, the presence of a sizeable population of galaxies with prominent star formation at early times and intermediate-age quenching epochs. This type of galaxy is uncommon around the MW, which may be a reflection of different underlying accretion histories of the two hosts.

\item A comparison of our SFHs with TNG50 and FIRE-2 simulated satellite populations shows that both simulation suits: a) predict a slightly stronger dependence of satellite quenching epoch with absolute luminosity, compared to the M31 system; b) predict a shallower dependence of quenching epoch with present-day galactocentric distance, compared to the satellite population of M31; and c) do not contain a sizeable population of satellites with ancient $\tau_{50}$ and intermediate-age quenching, which is observed in M31.

\end{enumerate}

The results of this study represent a significant step forward in our understanding of the M31 satellite system. Although already valuable on their own, the SFHs presented in this paper will be even more informative when combined with other large homogeneous datasets, such as spectroscopic abundances and proper motions, which have been or are being collected for a large number of M31 low-mass satellites.


%
\begin{acknowledgments}
We thank the anonymous referee for helping us improve the quality of this paper. Support for this work was provided by NASA through grants GO-13768, GO-15476, GO-15902, AR-16159, GO-16273, and AR-17026 from the Space Telescope Science Institute, which is operated by AURA, Inc., under NASA contract NAS5-26555. MBK acknowledges support from NSF CAREER award AST-1752913, NSF grants AST-1910346, AST-2108962, and AST-2408247; NASA grant 80NSSC22K0827; HST-GO-16686, HST-AR-17028, and HST-AR-17043 from STScI; and from the Samuel T. and Fern Yanagisawa Regents Professorship in Astronomy at UT Austin. This research has made use of NASA’s Astrophysics Data System. This work has made use of the Local Volume Database: \url{https://github.com/apace7/local_volume_database }.
\end{acknowledgments}

\facilities{HST(ACS). All the data presented in this paper were obtained from the Mikulski Archive for Space Telescopes (MAST) at the Space Telescope Science Institute. The specific observations analyzed can be accessed via \dataset[DOI: 10.17909/ftep-8k64]{https://doi.org/10.17909/ftep-8k64}. The HLSP products related to this paper can be accessed via \dataset[DOI: 10.17909/2xyp-hh35]{https://doi.org/10.17909/2xyp-hh35}.}


\software{ This research made use of routines and modules from the following software packages: \texttt{Astropy} \citep{Astropy13,Astropy18}, \texttt{DOLPHOT} \citep{Dolphin16a}, \texttt{IPython} \citep{IPython},
\texttt{MATCH} \citep{Dolphin02}, \texttt{Matplotlib} \citep{Matplotlib}, \texttt{NumPy} \citep{Numpy}, \texttt{Pandas} \citep{Pandas}, and \texttt{SciPy} \citep{Scipy}.}



\appendix

\begin{figure*}
    \centering
    \includegraphics[width=\textwidth]{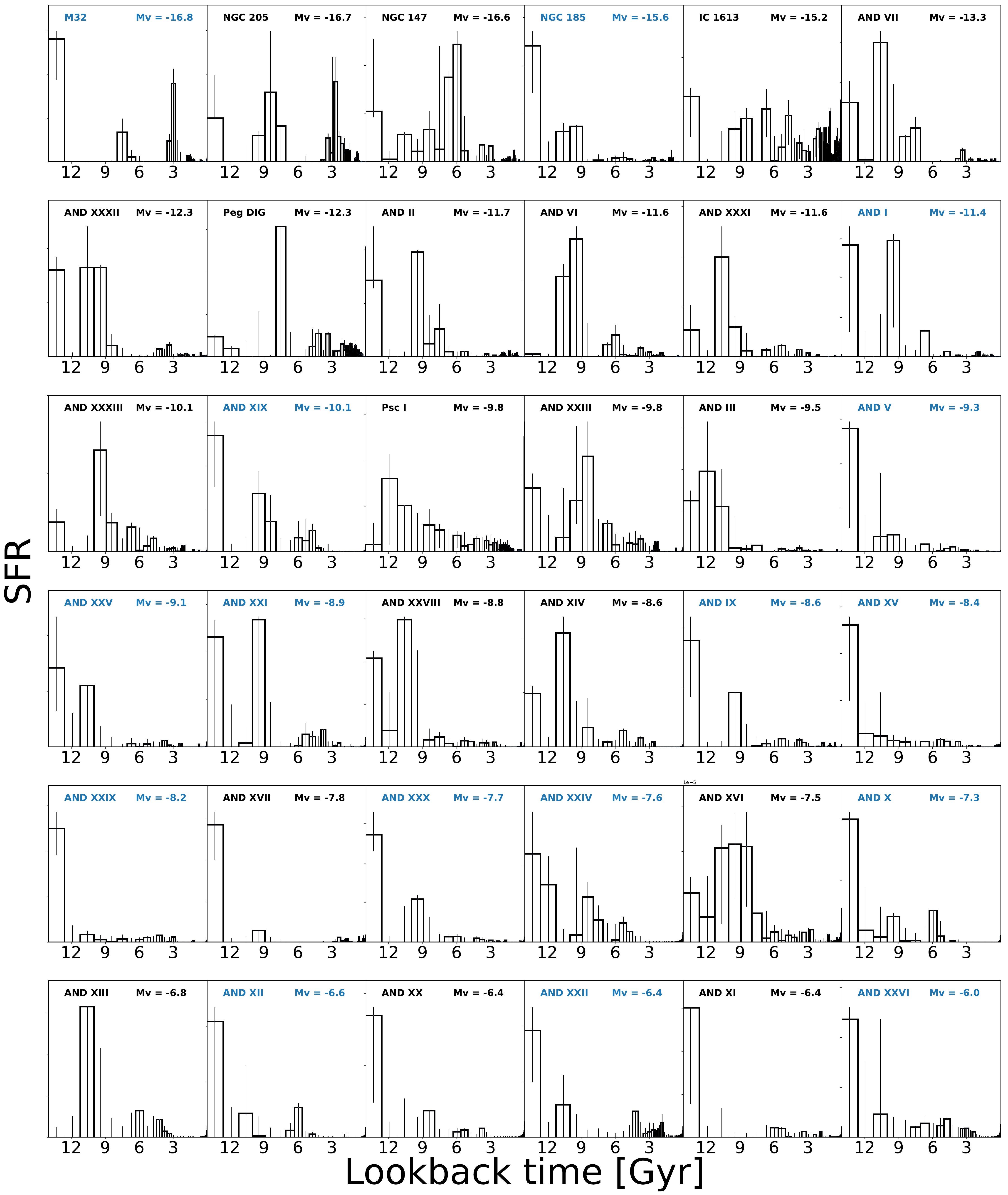}
    \caption{Differential SFHs for the 36 dwarf galaxies in our primary sample. Due to the different absolute amount of star formation in each dwarf, the y axis is scaled arbitrarily to highlight the SFH details. The galaxies belonging to the M31 subpopulation with old $\tau_{50}$ and intermediate-age $\tau_q$ are highlighted in blue.}
    \label{Fig:SFRs}
\end{figure*}

\begin{figure*}
    \centering
    \includegraphics[width=\textwidth]{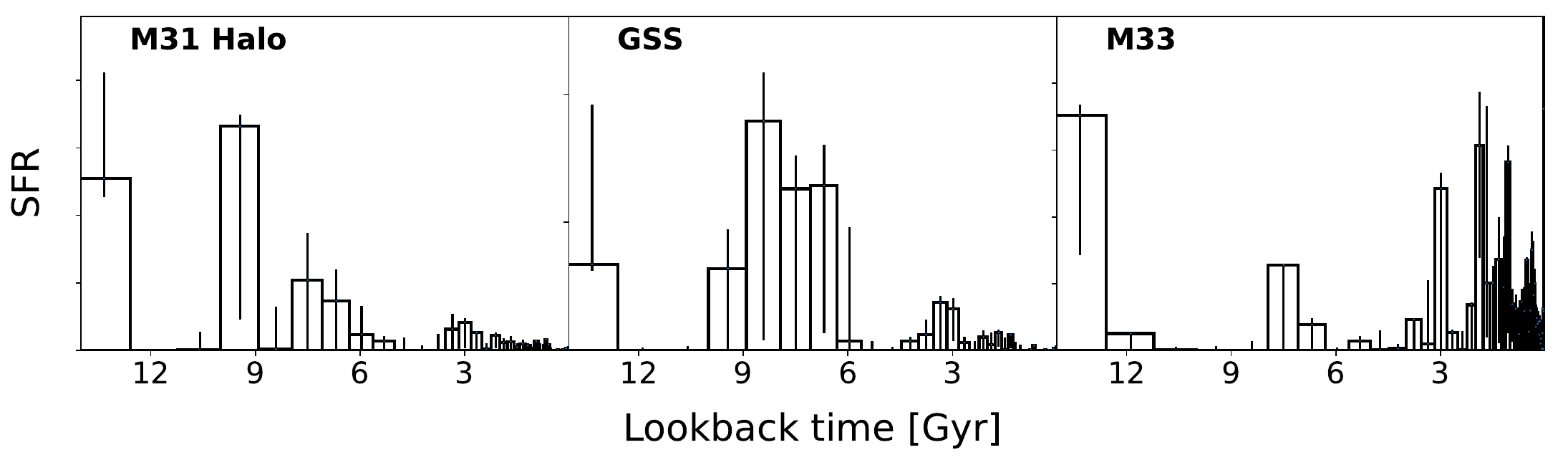}
    \caption{Differential SFHs for the M31 Halo field, the GSS field, and M33 (combined across the eight fields). Due to the different absolute amount of star formation, the y axis is scaled arbitrarily to highlight the SFH details.}
    \label{Fig:SFRs_Extra}
\end{figure*}

\section{High Level Science Products}
Here we provide a description of the publicly released data products of our program. We detail the different types of data products and a column-by-column description of their content. We publish data products for the 36 ACS fields in our primary dwarf galaxy sample, the 15 UVIS parallel fields for which we detect the target stellar population, and the 10 ACS fields belonging to our auxiliary targets. The data products can be retrieved on our MAST HLSP repository: \url{ https://archive.stsci.edu/hlsp/m31-satellites}.
\label{App:Products}

\subsection{Reference Drizzled Image}
For every field in this paper, we release the drizzled HST image used as a reference in the DOLPHOT reduction. This is contained in the \texttt{*drz.fits} file. This image has been used to perform astrometric alignment of the individual HST exposures and to set the absolute WCS reference frame for the detected sources. In most cases this image is a single-visit drizzled frame directly obtained from MAST, typically in the F814W band. In specific instances where a deeper reference image was required (e.g., in sparser fields) we have created a custom drizzled image by combining frames from multiple visits.
\subsection{Photometry}
For each field in this paper, we release two types of photometric catalogs. The first type, in the \texttt{phot-full} HLSP files, contains the full photometric catalog as obtained from DOLPHOT. This means that no selection is applied to the sources. This catalog corresponds to the ``raw" step, as outlined in Figure~\ref{Fig:Culling}. In this catalog we provide basic information about each source identified in our images, such as ID, position, VEGAMAG photometry in each band, and a set of DOLPHOT output metrics\footnote{For an explanation of DOLPHOT's output, refer to the official DOLPHOT webpage (\url {http://americano.dolphinsim.com/dolphot/}), or to the DOLPHOT tutorial produced by the ERS-1334 team (\url{https://dolphot-jwst.readthedocs.io/}.)}. We also provide multi-epoch photometry, in the form of VEGAMAG photometry and DOLPHOT output metrics for each individual exposure in our dataset. The file header contains information about each exposure's filter, reference MJD, and original MAST filename. A column-by-column breakdown of the \texttt{phot-full} files is provided in Table~\ref{tab:photfull}.

The second type of photometric catalog is contained in the \texttt{phot-clean} HLSP files. These catalogs contain a subset of our \texttt{phot-full} catalogs, selected to contain good-quality stellar sources. In particular, we apply the quality cuts described in \S~\ref{Sec:Culling} and the bright-star masks described in \S~\ref{Sec:Masks}. No radial spatial cuts are applied. The resulting catalogs correspond to the ``After Masking" step of Figure~\ref{Fig:Culling}. For each source, only the global photometry is provided. If the corresponding time-series are desired, they can be obtained from the \texttt{phot-full} catalogs, matching sources by the ID field. A column-by-column breakdown of the \texttt{phot-clean} files is provided in Table.~\ref{tab:photclean}.

\begin{table}[]
    \centering
    \begin{tabular}{ll}
    \toprule
    Column & Description\\
    \toprule
    ID & Unique source identifier within the field\\
    RA & Right Ascention [deg]\\
    Dec& Declination [deg]\\
    Rell& Elliptical radius from photometric center, in units of $r_h$\\
    X& X position in reference image [px]\\
    Y& Y position in reference image [px]\\
    Chi& DOLPHOT $\chi$ for global PSF fit\\
    SNR& DOLPHOT Signal-to-noise ratio for global PSF fit\\
    Sharp& DOLPHOT Sharpness for global PSF fit\\
    Round& DOLPHOT Roundness for global PSF fit \\
    Crowd& DOLPHOT Crowding for global PSF fit [mag]\\
    Type& DOLPHOT Object Type\\
    \toprule
    \multicolumn{2}{c}{For each filter}\\
    \toprule
    \textit{FilterName}& VEGAMAG magnitude in that filter [mag]\\
    \textit{FilterName}\_Photerr & DOLPHOT photometric uncertainty in that filter [mag]\\
    \textit{FilterName}\_Chi & DOLPHOT $\chi$ for the PSF fit in that filter\\
    \textit{FilterName}\_SNR& DOLPHOT  Signal-to-noise ratio  for the PSF fit in that filter\\
    \textit{FilterName}\_Sharp & DOLPHOT Sharpness for the PSF fit in that filter\\
    \textit{FilterName}\_Round& DOLPHOT  Roundness for the PSF fit in that filter\\
    \textit{FilterName}\_Crowd & DOLPHOT Crowding for the PSF fit in that filter [mag]\\
    \textit{FilterName}\_Flag& DOLPHOT error flag for the PSF fit in that filter\\
    \toprule
    \multicolumn{2}{c}{For each image}\\
    \toprule
    Im\textit{N}\_Mag& VEGAMAG magnitude for the PSF fit in image \textit{N}\\
    Im\textit{N}\_Photerr& DOLPHOT photometric uncertainty  for the PSF fit in image \textit{N} [mag]\\
    Im\textit{N}\_Chi& DOLPHOT $\chi$ for the PSF fit in image \textit{N}\\
    Im\textit{N}\_SNR& DOLPHOT Signal-to-noise ratio for the PSF fit in image \textit{N}\\
    Im\textit{N}\_Sharp& DOLPHOT Sharpness for the PSF fit in image \textit{N}\\
    Im\textit{N}\_Round& DOLPHOT Roundness for the PSF fit in image \textit{N}\\
    Im\textit{N}\_Crowd&DOLPHOT Crowding for the PSF fit in image \textit{N} [mag]\\
    Im\textit{N}\_Flag& DOLPHOT error flag for the PSF fit in image \textit{N}\\
    \toprule
    \end{tabular}
    \caption{Description of columns in the \texttt{phot-full} photometric catalogs. For each image, the MAST filename, the corresponding HST filter, and the MJD of the reference epoch are provided in the file header.}
    \label{tab:photfull}
\end{table}

\begin{table}[]
    \centering
    \begin{tabular}{ll}
    \toprule
    Column & Description\\
    \toprule
    ID & Unique source identifier within the field\\
    RA & Right Ascension [deg]\\
    Dec& Declination [deg]\\
    Rell& Elliptical radius from photometric center, in units of $r_h$\\
    X& X position in reference image [px]\\
    Y& Y position in reference image [px]\\
    Type& DOLPHOT Object Type\\
    \toprule
    \multicolumn{2}{c}{For each filter}\\
    \toprule
    \textit{FilterName}& VEGAMAG magnitude in that filter [mag]\\
    \textit{FilterName}\_Photerr & DOLPHOT photometric uncertainty in that filter [mag]\\
    \textit{FilterName}\_SNR& DOLPHOT  Signal-to-noise ratio  for the PSF fit in that filter\\
    \textit{FilterName}\_Sharp & DOLPHOT Sharpness for the PSF fit in that filter\\
    \textit{FilterName}\_Round& DOLPHOT  Roundness for the PSF fit in that filter\\
    \textit{FilterName}\_Crowd & DOLPHOT Crowding for the PSF fit in that filter [mag]\\
    \textit{FilterName}\_Flag& DOLPHOT error flag for the PSF fit in that filter\\
    \toprule

    \end{tabular}
    \caption{Description of columns in the \texttt{phot-clean} photometric catalogs. }
    \label{tab:photclean}
\end{table}

\subsection{Artificial Star Tests}
As detailed in \S~\ref{Sec:ASTs}, we perform and release AST catalogs for our photometry. Our current release only contains AST catalogs for the ACS fields. The ASTs for the UVIS parallels will be produced and made available in a future data release. The catalogs contain a list of input properties for each artificial star injected (coordinates, image position, magnitudes) and a list of output properties, recovered from the DOLPHOT run. The output consists of recovered source position, measured magnitudes, and a set of DOLPHOT photometry metrics. In any given band, an output magnitude entry of 99.999 indicates a non-detection in that specific band. For each ACS field, we release two distinct AST catalogs. The first, provided in the \texttt{ast-full} file, contains ASTs that are consistent with the \texttt{phot-full} catalogs, i.e. with no selection citeria applied after the DOLPHOT run. The second set, provided in the \texttt{ast-clean} file, contains ASTs that have been processed with the exact selection criteria as the \texttt{phot-clean} catalogs. Artificial stars that fail the \texttt{phot-clean} selection criteria are marked as non-detections. The column-by-column content of the AST files is detailed in Table.~\ref{tab:astcat}. 

\begin{table}[]
    \centering
    \begin{tabular}{ll}
    \toprule
    Column & Description\\
    \toprule
    RA & Input Right Ascension [deg]\\
    Dec& Input Declination [deg]\\
    Rell& Input elliptical radius from photometric center, in units of $r_h$\\
    Xi& Input X position in reference image [px]\\
    Yi& Input Y position in reference image [px]\\
    Xo& Output X position in reference image [px]\\
    Yo& Output Y position in reference image [px]\\
    Type& Output DOLPHOT Object Type\\
    \toprule
    \multicolumn{2}{c}{For each filter}\\
    \toprule
    \textit{FilterName}\_i& Input VEGAMAG magnitude in that filter [mag]\\
    \textit{FilterName}\_o& Output VEGAMAG magnitude in that filter [mag]\\
    SNR\_\textit{FilterName}& Output DOLPHOT Signal-to-noise ratio in that filter\\
    Sharp\_\textit{FilterName}& Output DOLPHOT Sharpness in that filter\\
    Round\_\textit{FilterName}& Output DOLPHOT Roundness in that filter\\
    Crowd\_\textit{FilterName}& Output DOLPHOT Crowding in that filter [mag]\\
    \toprule

    \end{tabular}
    \caption{Description of columns in the \texttt{ast-full} and \texttt{ast-clean} AST catalogs. }
    \label{tab:astcat}
\end{table}
\subsection{RR Lyrae Variables}
Our data release also contains catalogs of RR Lyrae variable stars identified in our ACS fields. The methodology behind the identification and analysis of the RR Lyrae variables, as well as a description of the source properties, is detailed in \citet{Savino22}. The RR Lyrae catalogs are provided for each ACS field in our primary sample of 36 dwarfs, as well as for the M31 halo field, the GSS field, and one field in M33 (the D2 pointing). The \texttt{rrl} HLSP file contains the RR Lyrae stars used in \citet{Savino22} for determining the distance to our targets. The catalog contains source ID, source position, as well as pulsation properties (period, amplitude, intensity-averaged magnitude) and their related uncertainties. The RR Lyrae light curves can be obtained from the \texttt{phot-full} catalog, matching sources by the ID field. A column-by-column breakdown of the \texttt{rrl} files is provided in Table.~\ref{tab:rrl}.

A small number of RR Lyrae candidates were discarded by \citet{Savino22}, due to low-quality light-curve fits. If an ACS fields contains any of these sources, they are listed in a corresponding \texttt{rrl-candidates} file. These files, whose content is detailed in Table~\ref{tab:rrl-candidates}, contain source ID and position, as well as a rough estimate of the photometric variability in each band.

\begin{table}[]
    \centering
    \begin{tabular}{ll}
    \toprule
    Column & Description\\
    \toprule
    ID & Unique source identifier within the field\\
    RA & Right Ascension [deg]\\
    Dec& Declination [deg]\\
    P & Pulsation period [d]\\
    dP- & Lower uncertainty on period [d]\\
    dP+ & Upper uncertainty on period [d]\\
    Type & RR Lyrae type: either RRab or RRc\\
    \toprule
    \multicolumn{2}{c}{For each filter}\\
    \toprule
    \textit{FilterName}\_Mag& Intensity-average mean magnitude in that filter [mag]\\
    \textit{FilterName}\_dMag-& Lower uncertainty in the mean magnitude [mag]\\
    \textit{FilterName}\_dMag+& Upper uncertainty in the mean magnitude [mag]\\
    \textit{FilterName}\_Amp& Pulsation amplitude in that filter [mag]\\
    \textit{FilterName}\_dAmp-& Lower uncertainty in the pulsation amplitude [mag]\\
    \textit{FilterName}\_dAmp+& Upper uncertainty in the pulsation amplitude [mag]\\
    
    \toprule

    \end{tabular}
    \caption{Description of columns in the \texttt{rrl} RR Lyrae catalogs. }
    \label{tab:rrl}
\end{table}

\begin{table}[]
    \centering
    \begin{tabular}{ll}
    \toprule
    Column & Description\\
    \toprule
    ID & Unique source identifier within the field\\
    RA & Right Ascension [deg]\\
    Dec& Declination [deg]\\
    \toprule
    \multicolumn{2}{c}{For each filter}\\
    \toprule
    Delta\_\textit{FilterName}& Maximum observed magnitude variation in that filter [mag]\\
    \toprule

    \end{tabular}
    \caption{Description of columns in the \texttt{rrl-candidates} RR Lyrae catalogs. }
    \label{tab:rrl-candidates}
\end{table}
\subsection{Star Formation Histories}
The SFHs measured in this paper are provided in our HLSP repository, in the \texttt{*sfh.fits} files. We publish the SFHs measured from the 36 ACS fields in our primary dwarf galaxy sample, as well as SFHs measured from the 10 auxiliary fields. For M33, we also provide the combined SFH solution, obtained from the unweighted sum of the solutions for the 8 individual ACS fields. Each entry in the file represents a time bin of the SFH solution, for which we provide the minimum and maximum age spanned by the bin, the average star formation rate in that bin, and the cumulative SFH at the end of that bin. We also provide the associated statistical and total (statistical plus systematic) uncertainties. The column-by-column breakdown of the \texttt{sfh} files is provided in Table~\ref{tab:sfhcat}.

\begin{table}[]
    \centering
    \begin{tabular}{ll}
    \toprule
    Column & Description\\
    \toprule
    tmax& Oldest lookback time in the SFH bin [Gyr]\\
    tmin&Youngest lookback time in the SFH bin [Gyr]\\
    SFR&Average star formation rate in the SFH bin [$M_{\odot}/yr$]\\
    dSFR-\_Stat&Lower statistical uncertainty on SFR [$M_{\odot}/yr$]\\
    dSFR+\_Stat&Upper statistical uncertainty on SFR [$M_{\odot}/yr$]\\
    dSFR-\_Sys&Lower total uncertainty on SFR [$M_{\odot}/yr$]\\
    dSFR+\_Sys&Upper total uncertainty on SFR [$M_{\odot}/yr$]\\
    Cum&Cumulative SFH (normalized) at the end of the SFH bin\\
    dCum-\_Stat&Lower statistical uncertainty on cumulative SFH\\
    dCum+\_Stat&Upper statistical uncertainty on cumulative SFH\\
    dCum-\_Sys&Lower total uncertainty on cumulative SFH\\
    dCum+\_Sys&Upper total uncertainty on cumulative SFH\\
    \toprule

    \end{tabular}
    \caption{Description of columns in the \texttt{sfh} SFH files. }
    \label{tab:sfhcat}
\end{table}

\section{Star Formation Rates}
\label{App:SFR}

Here we show the differential version of our SFH, i.e., the star formation rate as  a function of time, for our primary sample of 36 dwarfs (Fig.~\ref{Fig:SFRs}) and our three auxiliary targets (Fig.~\ref{Fig:SFRs_Extra}). As discussed in \citet{Savino23}, differential SFHs require caution in the interpretation of the uncertainties, as significant covariance exists between the star formation rates in neighboring time bins \citep[e.g.,][]{Dolphin02}. This, in turn, means that the significance of the SFH features is higher than it could appear from the size of the error bars alone. This is the reason why cumulative SFHs are preferred in the main manuscript.

Nevertheless, differential SFHs provide an intuitive way to appreciate the relative amount of star formation in the galaxy at any point in time. In particular, it can be appreciated how certain M31 satellites have distinct phases of star formation (e.g., \A{I}, \A{XXI}), while others are characterized by a main star formation episode followed by a rapid decline (e.g., \A{V}, \A{XV}).

\section{Individual Fields in M33}
\label{App:M33}

\begin{figure*}
    \centering
    \includegraphics[width=\textwidth]{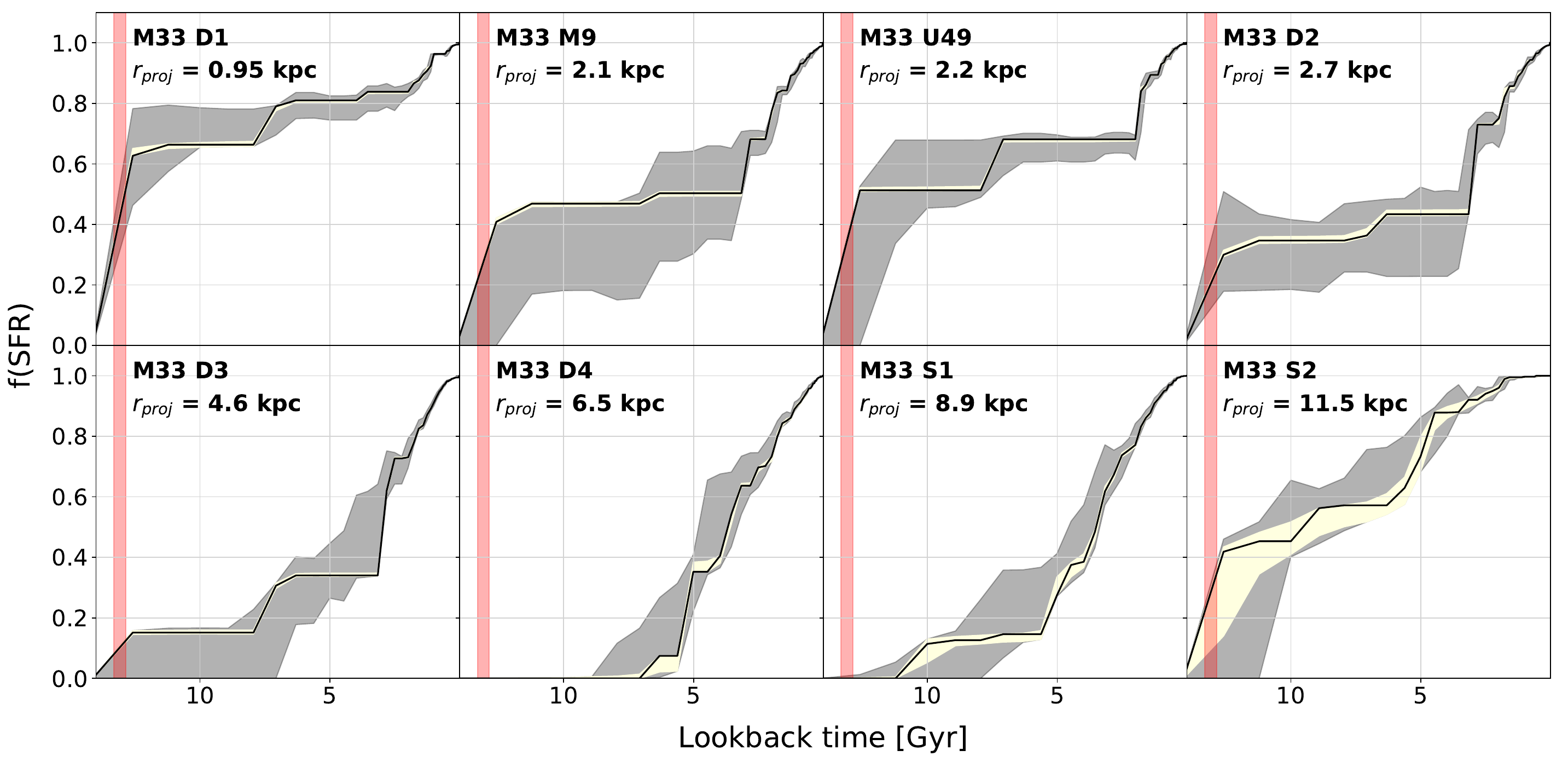}
    \caption{Cumulative SFHs for the eight ACS fields belonging to M33. Lines and colors are the same as in Figure~\ref{Fig:SFHs}. The projected distance from M33 is also reported for each field.}
    \label{Fig:M33}
\end{figure*}

Here we provide the cumulative SFHs for the individual ACS fields we analyze in M33 (Fig.~\ref{Fig:M33}). The fields are arranged in order of increasing projected distance from the center of M33. This makes immediately clear the inside-out star formation experienced by the main body of M33 and the subsequent turnover to older ages in the most external field (for more details see discussions in, e.g., \citealt{Barker07b}, \citealt{Williams09}, \citealt{Barker11}, \citealt{Bernard12}). Also provided are the values of $\tau_{50}$, $\tau_{80}$, and $\tau_{90}$ measured in each field, along with the value of internal extinction $dA_V$ used in the CMD fitting (Table~\ref{tab:M33}).

\begin{table}
    \centering
    \caption{Star formation timescales measured in the eight individual M33 fields, along with the amount of internal reddening used in the CMD fitting.}
    \begin{tabular}{lrrrr}
    \toprule
    Field & $dA_V$ & $\tau_{50}$ & $\tau_{80}$ & $\tau_{90}$\\
    & mag &Gyr & Gyr& Gyr\\
    \toprule
        D1 & 0.9&$12.9_{-0.3~(-0.8)}^{+1.2~(+1.2)}$&$6.7_{-0.4~(-4.4)}^{+0.4~(+0.4)}$&$1.4_{-0.1~(-0.2)}^{+0.1~(+0.1)}$\\
        
        D2 & 0.4&$3.1_{-0.3~(-0.3)}^{+0.1~(+9.5)}$&$1.8_{-0.1~(-0.2)}^{+0.2~(+0.2)}$&$1.1_{-0.0~(-0.1)}^{+0.1~(+0.2)}$\\
        
        D3 & 0.2&$3.0_{-0.1~(-0.1)}^{+0.2~(+1.5)}$&$1.7_{-0.1~(-0.1)}^{+0.1~(+0.2)}$&$1.1_{-0.1~(-0.1)}^{+0.1~(+0.1)}$\\
        
        D4 & 0.3&$3.7_{-0.1~(-0.4)}^{+0.3~(+1.1)}$&$1.8_{-0.0~(-0.2)}^{+0.2~(+0.3)}$&$1.0_{-0.0~(-0.1)}^{+0.1~(+0.2)}$\\
        
        M9 & 0.3&$6.4_{-3.2~(-3.2)}^{+0.7~(+0.8)}$&$1.9_{-0.1~(-0.3)}^{+0.1~(+0.1)}$&$1.1_{-0.1~(-0.2)}^{+0.0~(+0.1)}$\\
        
        U49 & 0.9&$12.6_{-0.0~(-4.8)}^{+1.5~(+1.5)}$&$1.8_{-0.1~(-0.2)}^{+0.2~(+0.2)}$&$1.1_{-0.1~(-0.1)}^{+0.0~(+0.5)}$\\
        
        S1 & 0.1&$3.5_{-0.3~(-0.3)}^{+0.0~(+1.1)}$&$1.9_{-0.1~(-0.1)}^{+0.1~(+0.3)}$&$1.3_{-0.0~(-0.1)}^{+0.1~(+0.1)}$\\
        
        S2 & 0&$9.5_{-1.5~(-1.9)}^{+1.1~(+2.1)}$&$4.8_{-0.3~(-0.8)}^{+0.3~(+0.9)}$&$3.4_{-0.3~(-0.5)}^{+0.5~(+1.1)}$\\
    \toprule
    \end{tabular}
    \label{tab:M33}
\end{table}

\section{Comparison with Literature SFHs}
\label{App:Lit}
\begin{figure*}
    \centering
    \includegraphics[width=\textwidth]{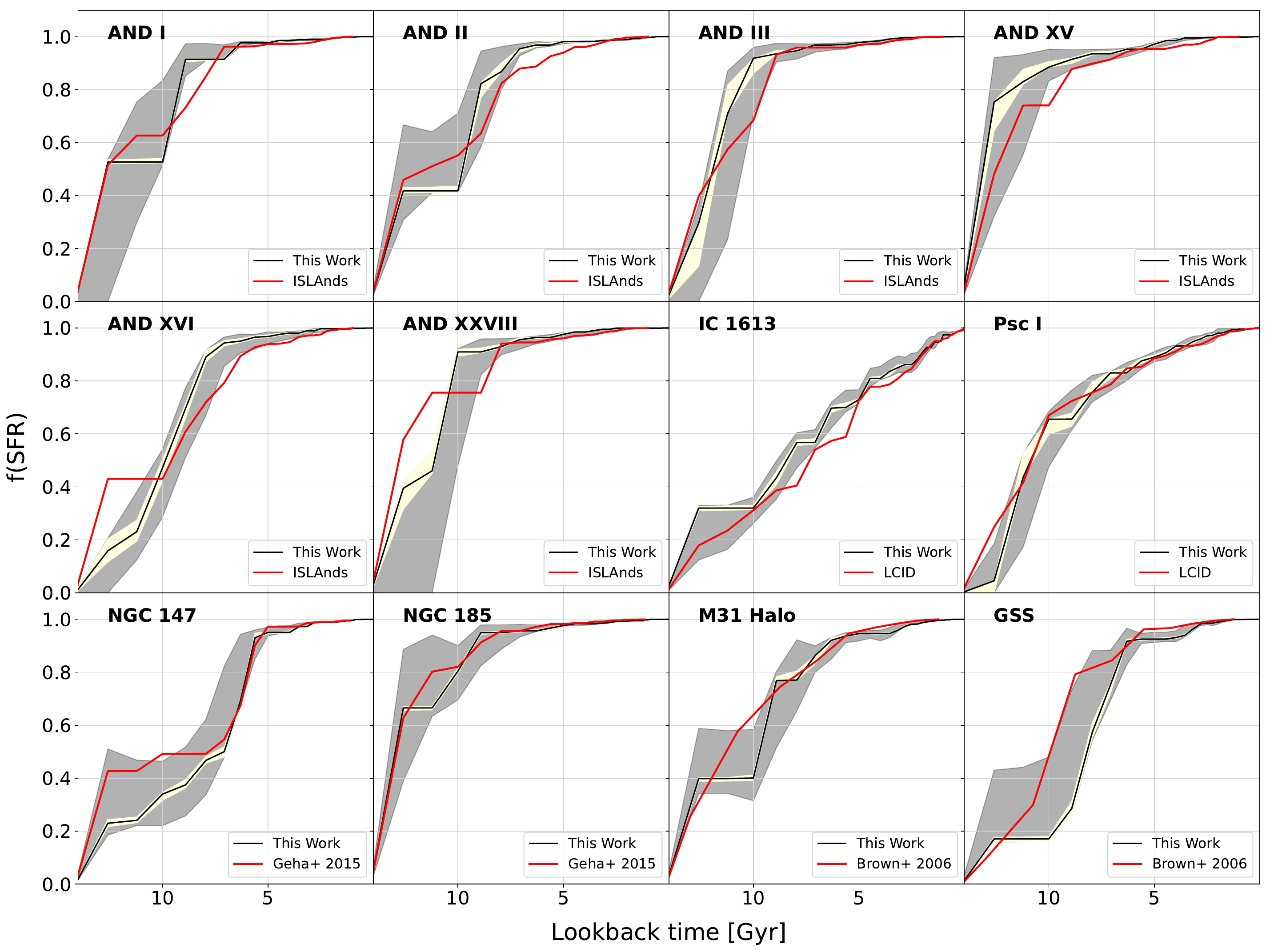}
    \caption{Comparison with literature SFHs. The black line is the best fit SFH from this paper, while the red line is the literature SFH from ISLAndS \citep{Skillman17}, LCID \citep{Gallart15}, \citet{Geha15}, and \citet{Brown06}. The yellow and grey regions show the statistical and systematic uncertainties from our fit, respectively.}
    \label{Fig:SFHs_Lit}
\end{figure*}

Here we compare our SFH solutions to a set of literature SFHs, derived from the same ACS pointings of our study (Fig.~\ref{Fig:SFHs_Lit}). We compare the SFHs of \A{I}, \A{II}, \A{III}, \A{XV}, \A{XVI}, and \A{XXVIII} to the results of the ISLAndS program \citep{Skillman17}; our SFHs for Psc~{\sc I} and IC~1613 to the solutions of the LCID program \citep{Hidalgo11,Skillman14,Gallart15}; the SFHs of NGC~147 and NGC~185 to the solutions of \citet{Geha15}; and the SFHs of the M31 halo and the GSS to the solutions of \citet{Brown06}.

The comparison of Fig.~\ref{Fig:SFHs_Lit} shows that our solutions and literature results are comparable, with the differences between the two sets of SFHs being mostly of the order of the systematic uncertainties. The largest differences occur for \A{XVI} and \A{XXVIII}, for which the ISLAndS SFH contain $\sim$ 20\% more stars with $t>11$~Gyr, compared to our solutions. In other targets, such as \A{XV}, NGC~147 and the GSS, we note systematic differences in the age of the oldest star formation. Such differences, however, fall within the range of possible solutions that are generated by our systematic uncertainty estimation procedure (grey region). Overall, the agreement of our SFHs with previous determinations is very good, considering that we have used different stellar models, and different values of distance and extinction in our CMD fitting.

\section{Trends with $\tau_{80}$ and $\tau_{90}$}
\label{App:T8090}
\begin{figure*}
    \centering
    \includegraphics[width=\textwidth]{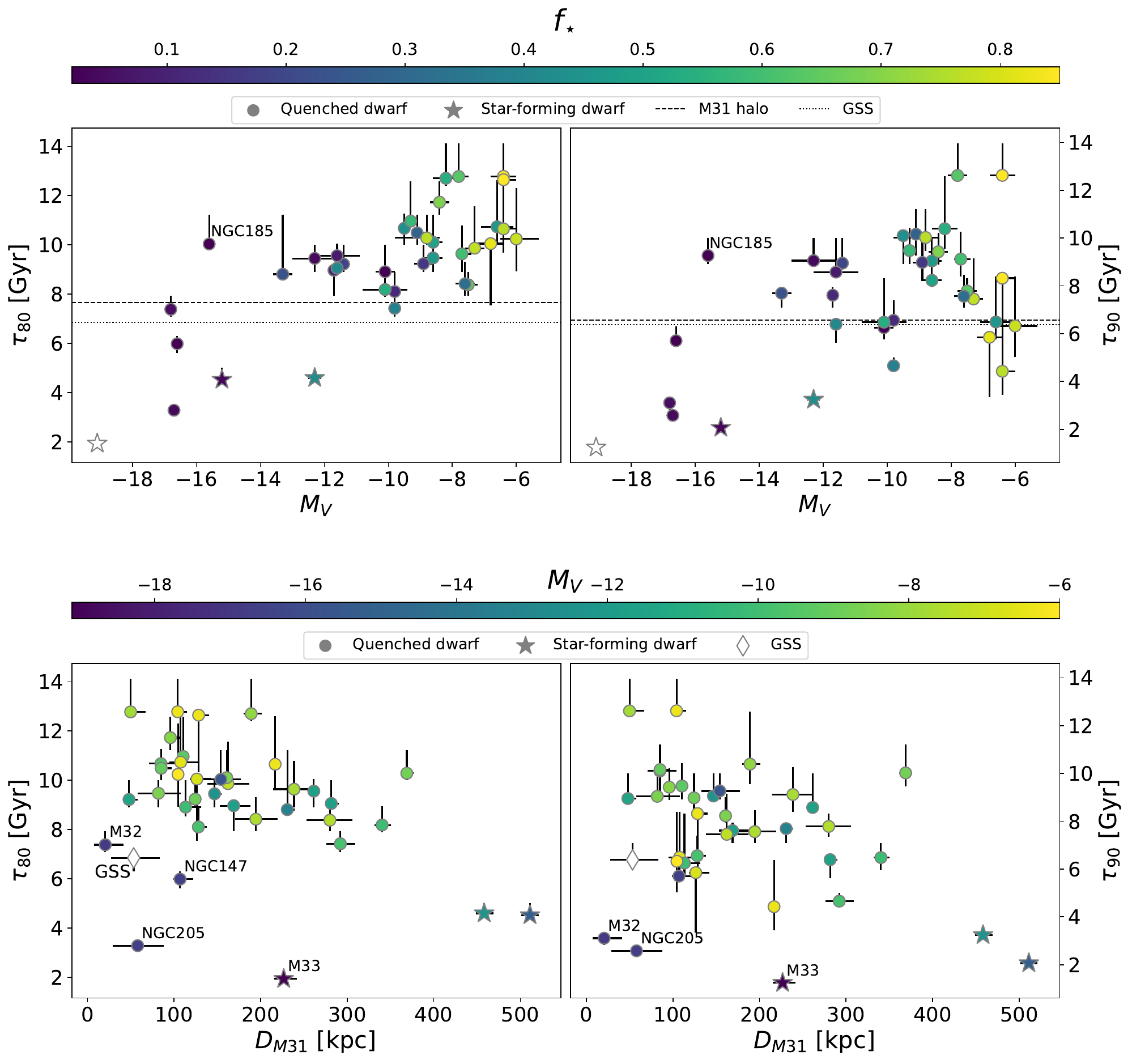}
    \caption{Trends of $\tau_{80}$ (left) and $\tau_{90}$ (right) with absolute luminosity (top) and galactocentric distance (bottom). The symbols and colors of the top panels are the same as Fig.~\ref{Fig:Tq_Mv}, while those of the bottom panels are the same as Fig.~\ref{Fig:Tq_DM31}.}
    \label{Fig:T80_Mv}
\end{figure*}

\begin{figure*}
    \centering
    \plotone{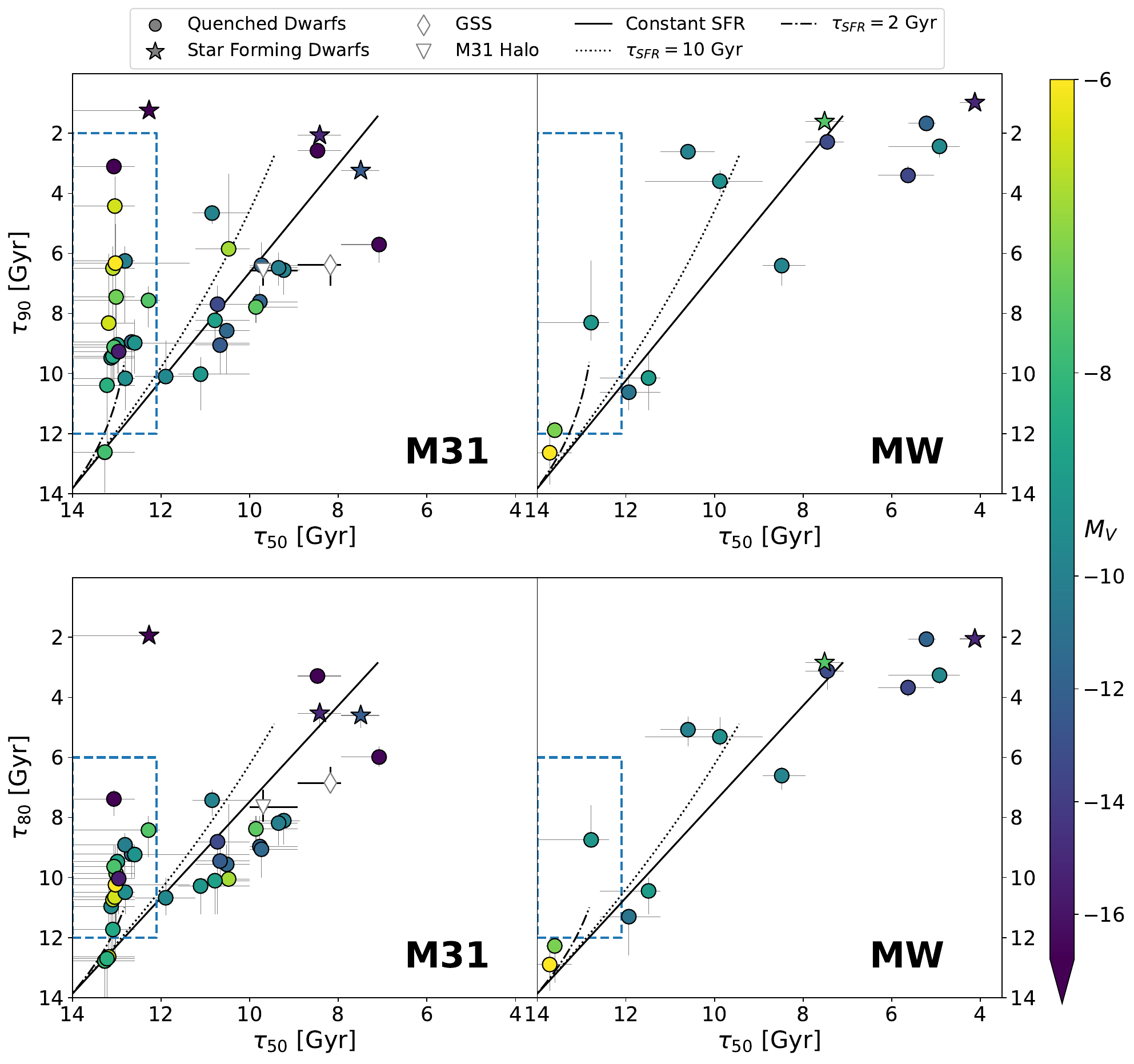}
    \caption{Distribution of $\tau_{50}$ vs $\tau_{90}$ (top panels) and $\tau_{50}$ vs $\tau_{80}$ (bottom panels) for the dwarf galaxies associated with M31 (left) and the MW (right). Only galaxies with $M_V<-6$ are shown in this plot. The symbols and colors in this plot are the same as in Fig.~\ref{Fig:T50Tq}.}
    \label{Fig:T50T90}
\end{figure*}

Here we report the same SFH trends with galaxy luminosity and galactocentric distance of Fig.~\ref{Fig:Tq_Mv} and \ref{Fig:Tq_DM31}, except we use $\tau_{90}$ or $\tau_{80}$ instead of $\tau_q$ (Fig.~\ref{Fig:T80_Mv}). We also show the comparison with the MW satellite system in terms of $\tau_{50}$ versus $\tau_{80}$ and $\tau_{50}$ versus $\tau_{90}$ (Fig.~\ref{Fig:T50T90}). Overall, the same results we discuss in the main text can be appreciated here when using different metrics to quantify quenching. The main qualitative difference is that the values of $\tau_{90}$ for low-luminosity galaxies ($M_V\gtrsim -8$) are substantially more recent than the $\tau_q$ we used in the main text. This makes the low-luminosity satellites behave as outliers in the trends with satellite luminosity and galactocentric distance. As we have discussed in \S~\ref{Sec:Quenching}, in such low-mass galaxies it is challenging to interpret $\tau_{90}$ in terms of genuine star formation, due to the stronger concerns about CMD contamination. 

\clearpage

\bibliography{Bibliography}
\bibliographystyle{aasjournal}



\end{document}